\documentclass[pdflatex,sn-nature,iicol]{sn-jnl}% Style for submissions to Nature Portfolio journals
\usepackage{graphicx}%
\usepackage{multirow}%
\usepackage{amsmath,amssymb,amsfonts}%
\usepackage{amsthm}%
\usepackage{mathrsfs}%
\usepackage[title]{appendix}%
\usepackage{xcolor}%
\usepackage{textcomp}%
\usepackage{manyfoot}%
\usepackage{booktabs}%
\usepackage{algorithm}%
\usepackage{algorithmicx}%
\usepackage{algpseudocode}%
\usepackage{listings}%
\usepackage{bbm} % added
\usepackage{aas_macros} % added - reference errors due to undefined journals
\usepackage{anyfontsize} % added - to remove font size errors
\usepackage{bookmark} % added - to remove paragraph hierarchy warnings associated with hyperref
\usepackage{etoolbox}   
\makeatletter
\patchcmd{\@maketitle}{\artauthors}{{\artauthors}}{}{}
\makeatother

\begin{document}

\title[Article Title]{Nature versus nurture in galaxy formation: the effect of environment on star formation with causal machine learning}


\author*[1]{\fnm{Sunil} \sur{Mucesh}}\email{sunil.mucesh.18@ucl.ac.uk}
\author[2]{\fnm{William G.} \sur{Hartley}}
\author[1, 3]{\fnm{Ciar\'an M.} \sur{Gilligan-Lee}}
\author[1]{\fnm{Ofer} \sur{Lahav}}

\affil*[1]{\orgdiv{Department of Physics \& Astronomy}, \orgname{University College London}, \orgaddress{\street{Gower Street}, \city{London}, \postcode{WC1E 6BT}, \country{UK}}}

\affil[2]{\orgdiv{Department of Astronomy}, \orgname{University of Geneva}, \orgaddress{\street{ch. d’Ecogia 16}, \city{CH-1290 Versoix}, \country{Switzerland}}}

\affil[3]{\orgname{Spotify}, \city{London}, \country{UK}}

\abstract{
Understanding how galaxies form and evolve is at the heart of modern astronomy. With the advent of large-scale surveys and simulations, remarkable progress has been made in the last few decades. Despite this, the physical processes behind the phenomena, and particularly their importance, remain far from known, as correlations have primarily been established rather than the underlying causality. We address this challenge by applying the causal inference framework. Specifically, we tackle the fundamental open question of whether galaxy formation and evolution depends more on nature (i.e., internal processes) or nurture (i.e., external processes), by estimating the causal effect of environment on star-formation rate (SFR) in the IllustrisTNG simulations. To do so, we develop a comprehensive causal model and employ cutting-edge techniques from epidemiology to overcome the long-standing problem of disentangling nature and nurture. We find that the causal effect is negative and substantial, with environment suppressing the SFR by a maximal factor of $\sim 100$. While the overall effect at $z=0$ is negative, in the early universe, environment is discovered to have a positive impact, boosting star formation by a factor of $\sim 10$ at $z \sim 1$ and by even greater amounts at higher redshifts. Furthermore, we show that: (i) nature also plays an important role, as ignoring it underestimates the causal effect in intermediate-density environments by a factor of $\sim 2$, (ii) controlling for the stellar mass at a snapshot in time, as is common in the literature, is not only insufficient to disentangle nature and nurture but actually has an adverse effect, though (iii) stellar mass is an adequate proxy of the effects of nature. Finally, this work may prove a useful blueprint for extracting causal insights in other fields that deal with dynamical systems with closed feedback loops, such as the Earth's climate.
}




\maketitle

\section{Introduction}
\label{sec:introduction}
In the local universe, there are two distinct broad populations of galaxies: the red sequence of massive, red, early-type, quiescent galaxies, and the blue sequence of less-massive, blue, late-type, star-forming galaxies. These two populations are unevenly distributed, with the red sequence mainly found in groups and clusters and the blue sequence located in relative isolation in the field. The fundamental question is then: is environment responsible for the bimodality? 

Correlations between galaxy properties and environment are well established. For example, the morphology–density \citep{dressler_1980}, colour–density \citep{kodama_2001}, and star-formation rate (SFR)–density \citep{gomez_2003} relations show a shift from late-type to early-type morphologies \citep{hubble_1931, zwicky_1937, morgan_1961, abell_1965, oemler_1974, davis_1976, postman_1984, whitmore_1991, santiago_1992, whitmore_1993, hermit_1996, guzzo_1997, dominguez_2001, giuricin_2001, treu_2003, goto_2003}, an increase in the fraction of red galaxies \citep{wilmer_1998, brown_2000, pimbblet_2002, zehavi_2002, hogg_2004, blanton_2005, martinez_2006}, and a decline in the star-formation activity of galaxies \citep{balogh_1997, balogh_1998, hashimoto_1998, poggianti_1999, balogh_2000, couch_2001, postman_2001, carter_2001, lewis_2002, balogh_2004a, tanaka_2004, rines_2005} with increasing environmental density, respectively. However, ``correlation does not imply causation''. Notably, stellar mass is also correlated with most galaxy properties \citep{mcgaugh_1997, blanton_2003, kauffmann_2003a, kauffmann_2003b, baldry_2004a, hogg_2004} and environment \citep{balogh_2001, hogg_2003, mo_2004, croton_2005, hoyle_2005, blanton_2005}. Consequently, does galaxy formation and evolution depend on internal processes (i.e., `nature') that scale with stellar mass or external processes (i.e., `nurture') associated with environment? Resolving this nature versus nurture debate \citep{irwin_1995} is a big open challenge in astronomy.

The current consensus is that both nature and nurture influence the evolution of galaxies. To isolate the effect of environment, previous studies have controlled for stellar mass, typically by binning galaxies. Here, most have found galaxy properties to still depend on environment \citep{kauffmann_2004, balogh_2004b, baldry_2004b, baldry_2006, weinmann_2006, bamford_2009, skibba_2009}. However, binning galaxies according to their stellar mass does not suffice to disentangle nature and nurture \citep{lucia_2012}. 

The dilemma (as in the original debate in biology) is that nurture likely influences nature and vice versa in a feedback loop, as galaxies interact with their environments and co-evolve over time. Thus, the distinction between nature and nurture is blurred. Furthermore, given the above, the evolutionary history of the galaxy population is likely necessary to separate the two components. However, observational data in the past were limited predominantly to the local universe due to the limitation of the then-available surveys. The advent of deep surveys has enabled the tracing of the galaxy population over time, but the data is still snapshots of galaxies, which ranks the lowest in a hierarchy of designs for evidence concerning causality \citep{vanderweele_2016}. Consequently, the problem remains unsolved, and the individual causal effects of nature and nurture are not yet known.

We address this challenge by applying, for the first time to our knowledge, the causal inference framework in the overall context of galaxy formation and evolution. Causal inference is concerned with the identification and quantification of causal relationships (see Appendix \ref{sec:causal_inference} for a brief overview). With significant advancements made in the last few decades (\citenum{lee_2017, lee_2020, dhir_2020, lee_2022, jeunen_2022, zeitler_2023, goffrier_2023}; also see \citenum{pearl_2010}, for a review), the field is now well established and has been applied to answer crucial questions in economics \citep{angrist_1991, card_1993, cengiz_2019}, political science \citep{kam_2008}, education \citep{angrist_1999, carlsson_2015}, policy \citep{ghosh_2018}, public health \citep{doll_1950, chay_2003, clark_2013, desouza_2022}, and more recently, quantum mechanics \cite{allen_2017} and quantum cryptography \cite{lee_2018, lee_2019}. Notably, causal inference methods are now starting to be adopted in astronomy, with applications to exoplanets \citep{scholkopf_2015, wang_2016}, and in the past few years, galaxy formation and evolution \citep{pasquato_2019, pang_2021, pasquato_2023, jin_2024}. We combine causal inference with machine learning (ML) to handle high-dimensional and non-linear data. The tools of ML have increasingly been applied to problems in astronomy (\citenum{mucesh_2021}; also see \citenum{baron_2019, fluke_2020}, for recent reviews), including causal insights into galaxy evolution \citep{teimoorinia_2016, bluck_2019, bluck_2020a, bluck_2020b, bluck_2022, brownson_2022, piotrowska_2022, mcgibbon_2022}. However, overall, ML on its own cannot infer causality as it is correlation-based.

Causal machine learning \citep[Causal ML; see][for a review]{kaddour_2022} is an emerging field, but it has already been successful, e.g., in improving the accuracy of medical diagnosis (\citenum{richens_2020, perov_2020, reynaud_2022, vlontzos_2023}; also see \citenum{sanchez_2022}, for a review of causal ML for healthcare). In this work, we apply causal ML to disentangle the roles of nature and nurture. Specifically, we establish the causal nature of the SFR–density relation in the IllustrisTNG simulations \citep{pillepich_2018a, springel_2018, nelson_2018, naiman_2018, marinacci_2018, nelson_2019}. We estimate the overall causal effect of environment on the SFR at $z = 0$ and at different redshifts out to $z \sim 3$ to determine how the role of environment has changed over cosmic time. Furthermore, we answer the fundamental questions:

\begin{enumerate}
    \item Is stellar mass an adequate proxy of the effects of nature?
    \item Is controlling for the stellar mass at a snapshot in time sufficient to disentangle nature and nurture and estimate the causal effect of environment?
    \item Is nature important? Specifically, is galaxy formation and evolution top-down determined by environment with no reverse influence of nature?
\end{enumerate}

\section{Causal model of galaxy formation and evolution}
\label{sec:causal_model}
The gold standard for estimating causal effects is a randomised control trial (RCT; \citenum{chalmers_1981}), but this is impossible in our setting. In the absence of experimentation, causal inference necessitates identifying and nullifying any biases present in observational data with expert knowledge and \textit{apriori} assumptions about the structure of the data-generating process (DGP) in the form of a causal structure, i.e., a model capturing the causal relationships between variables (referred
to as causal model for simplicity from here). Consequently, to estimate the causal effect of environment on SFR, we construct a causal model of galaxy formation and evolution.

We assume the cold dark matter (CDM) paradigm, in which galaxies form and evolve in dark matter haloes \citep{white_1978, efstathiou_1983, blumenthal_1984}. To build our causal model, we review established theories of galaxy formation and evolution, and in particular ideas from semi-analytic modelling (SAM; \citenum{white_1991, cole_1991, kauffmann_1993, cole_1994, kauffmann_1999, somerville_1999, springel_2001, hatton_2003, springel_2005, kang_2005, lu_2011, benson_2012, herniques_2015}; also see \citenum{baugh_2005, benson_2010}, for reviews), and express them as causal graphs \citep{wright_1921}. Briefly, we found galaxy formation and evolution to depend on both nature and nurture, with halo mass representing nature and environment related to nurture. See Appendix \ref{sec:galaxy_formation_and_evolution} for a full description of the construction of the causal model.

\begin{figure*}
    \centering
    \includegraphics[height=\textheight]{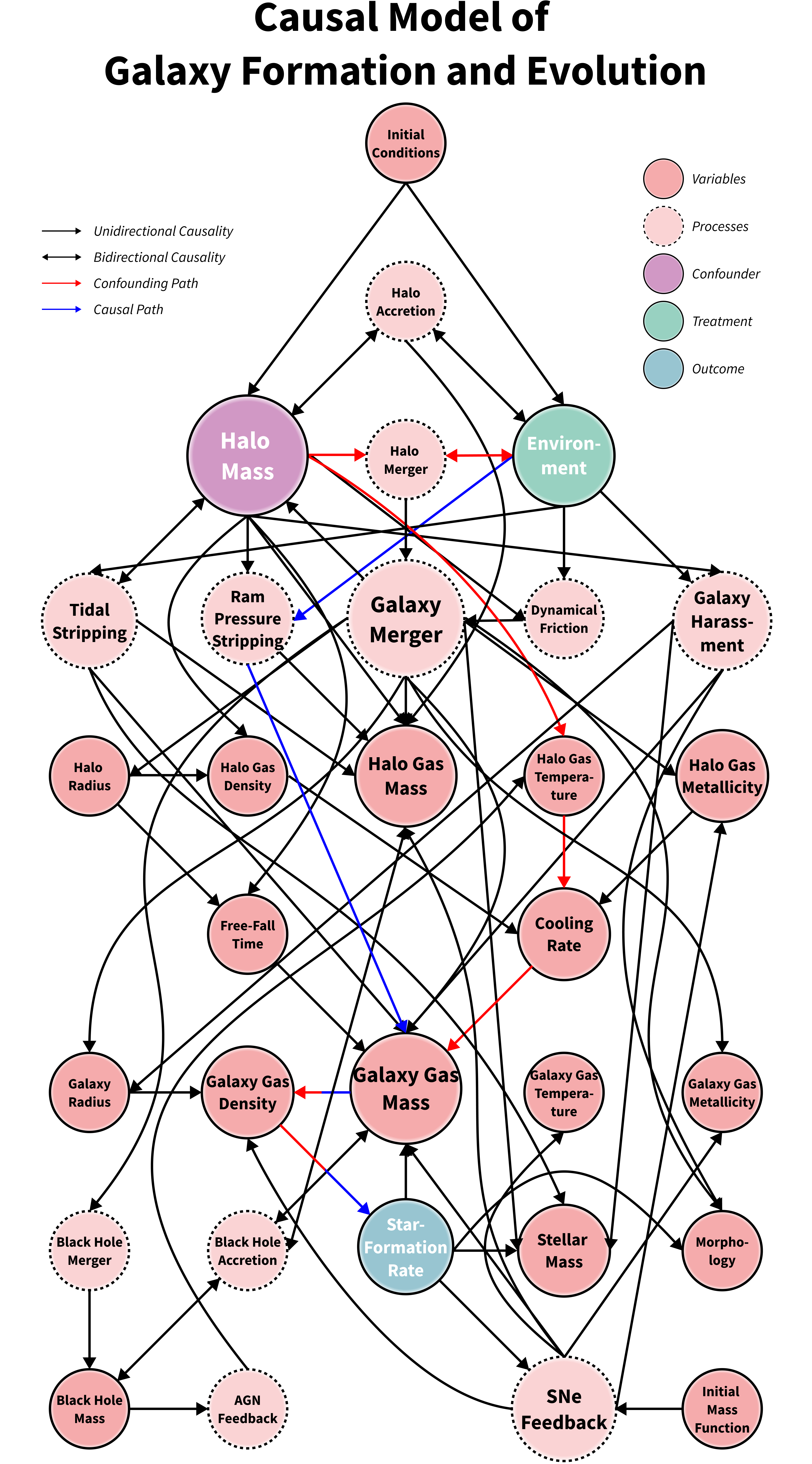}
    \caption{Causal model of galaxy formation and evolution (technically, a causal graph of the causal structure). The nodes are variables or processes (size indicates number of connections), and the edges communicate the causes. The confounder, treatment, and outcome (light purple, green, and blue nodes, respectively) are halo mass, environment, and star-formation rate (SFR), respectively. The blue and red arrows show example causal and confounding paths, respectively. The bicoloured arrows indicates the influence of both confounder and treatment. The naming convention is as follows: any variables associated with the halo and galaxy are preceded by them, respectively. Furthermore, halo refers to the dark matter halo that hosts a galaxy, and host halo refers to the parent dark matter halo that hosts other haloes. As such, halo refers to both distinct haloes and subhaloes. Note that only the connections between variables from the construction of the model (Appendix \ref{sec:galaxy_formation_and_evolution}) are shown.}
    \label{fig:causal_model_galaxy_formation_and_evolution}
\vspace{-89.75pt}
\end{figure*}

Fig. \ref{fig:causal_model_galaxy_formation_and_evolution} shows the causal model of galaxy formation and evolution (in the form of a causal graph). The causal model is complex and richly interconnected, with multiple interactions and feedback loops between the variables. Notably, halo mass (i.e., nature) and environment (i.e., nurture) are interacting with each other through accretion and mergers. Furthermore, internal processes driven by halo mass are affected by environment, and external processes linked to environment are dependent on halo mass. In other words, nurture influences nature and vice versa, and the effects of one depend on the other, i.e., nature and nurture are heavily intertwined. In this context, we define the nature versus nurture debate in this Article.

\begin{figure*}
    \centering
    \includegraphics[scale=0.3]{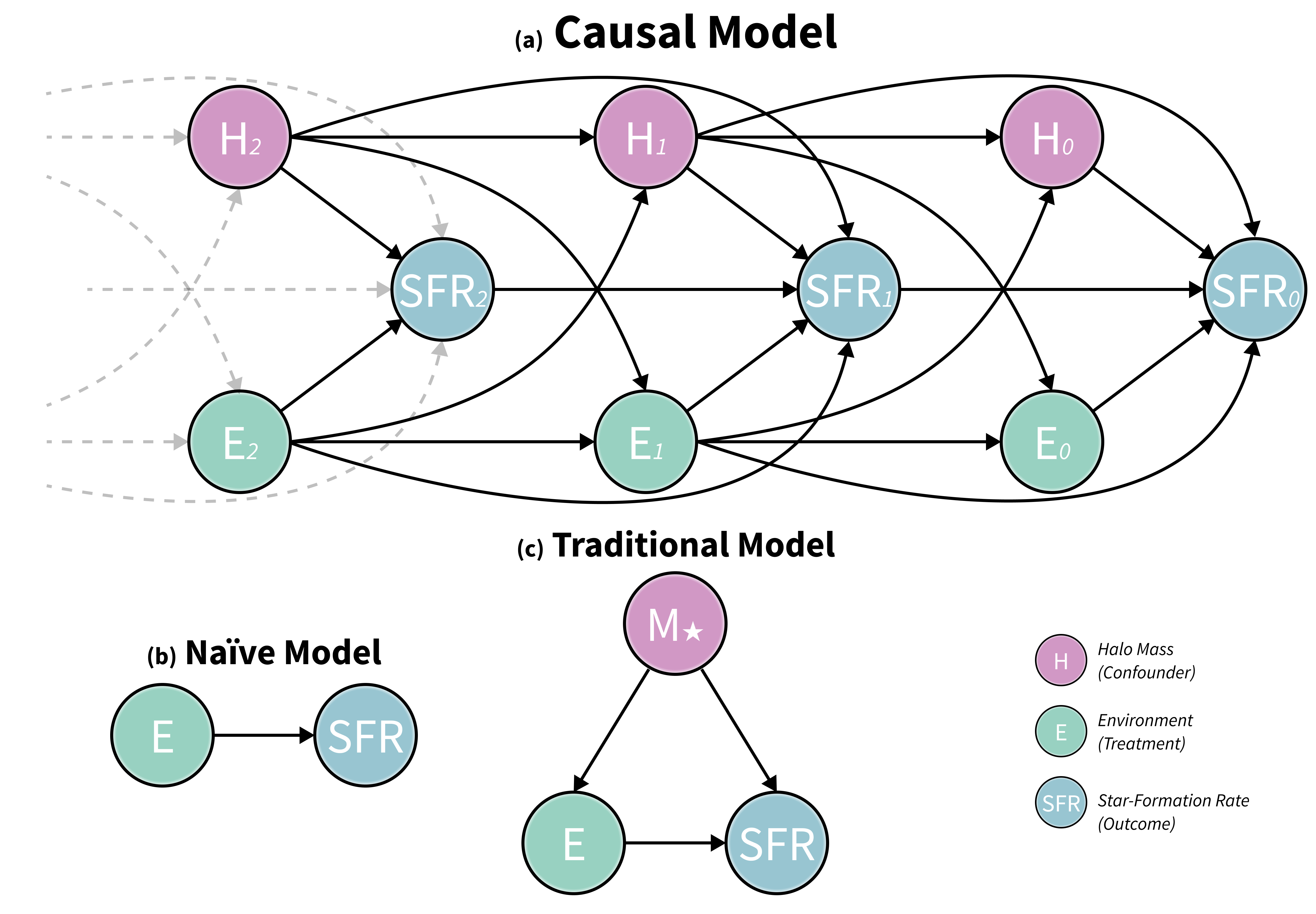}
    \caption{Causal directed acyclic graph (DAG) for determining the causal effect of environment on star-formation rate. (a) It is constructed by carefully tracing the causal chains between halo mass ($H$; confounder), environment ($E$; treatment), and star-formation rate ($SFR$; outcome) and unravelling the feedback loops over time, in the causal model of galaxy formation and evolution (Fig. \ref{fig:causal_model_galaxy_formation_and_evolution}). The subscripts indicate time, increasing from left to right (with zero marking the present). The causal chains of variables between the fundamental quantities are condensed for visual clarity, but more importantly, the adjustment set of only halo mass is both sufficient and necessary to estimate the causal effect of environment on SFR, according to d-separation. This DAG is the causal model. (b) Basic DAG for the relationship between environment and SFR whereby the raw correlation is the unbiased causal effect. (c) DAG of the model implicitly assumed in the literature when controlling for the stellar mass ($M_{\star}$) at a snapshot in time to disentangle nature and nurture. These naïve and traditional models are compared to our physics-informed causal model in Section \ref{sec:model_comparison}.}
    \label{fig:causal_model}
\end{figure*}

The presence of feedback loops in the causal model blurs the notion of causality. If two variables cause each other in a cycle, then what is the causal effect of one thing on another? Fundamentally, this fuzziness emerges due to a lack of a causal arrow of time, i.e., a cause precedes its effect. Thus, we unravel the feedback loops over time to make it easy to interpret and estimate causal effects, transforming the directed cyclic graph (DCG) into a directed acyclic graph (DAG).

\subsection{Causal DAG}
We start by carefully going through the many paths between environment and SFR to identify any potential biases that may distort the causal effect (example causal path shown in blue in Fig. \ref{fig:causal_model_galaxy_formation_and_evolution}). Here, we find that halo mass is a fundamental common cause \citep{reichenbach_1956} or confounder (ignoring initial conditions, which is shown as a token) of environment (i.e., the treatment) and SFR (i.e., the outcome) through various processes (example confounding path shown in red). This is an issue because if halo mass influences the environment (of a galaxy) and SFR, then what is the causal effect of environment versus that of halo mass? The presence of this ``confounding bias'' (Fig. \ref{fig:confounding_bias}) means that the correlation between environment and SFR does not translate to the causal effect. To determine the causal effect of environment on SFR, the influence of halo mass must be adjusted for. We carefully trace the causal chains and unravel the feedback loops present in the causal model to construct the DAG shown in Fig. \ref{fig:causal_model}a, which we refer to also as ``causal model'' from here. In the causal model, the causal chains of variables connecting the fundamental quantities have been condensed for visual clarity. More importantly, the adjustment set of only halo mass is both sufficient and necessary to estimate the causal effect on environment on SFR, according to the d-separation criterion \cite{pearl_2009a}. The figure also shows the naïve (b) and traditional (c) models, which we compare to our physics-informed causal model in Section \ref{sec:model_comparison}.

The initial haloes and their environments emerge from the initial conditions in the early universe. Subsequently, they interact and co-evolve over time in a feedback loop, influencing galaxies in the process. The structure of the causal model is as follows:

\begin{description}
    \item[$E_{k} \rightarrow SFR_{k}$] Environment affects the SFR as: (i) ram-pressure stripping (RPS) and tidal stripping deplete the fuel necessary for star formation by removing the cold galaxy and hot halo gases and (ii) galaxy harassment impacts the galaxy gas density by impulsively heating the cold gas.
    \item[$H_{k} \rightarrow SFR_{k}$] Halo mass dictates the amount and density of the cold gas for star formation as it indirectly influences the cooling rate and free-fall time via the halo gas temperature and density. Furthermore, it determines the susceptibility of a galaxy to environmental processes, which affects the SFR.    
    \item[$SFR_{k} \rightarrow SFR_{k+1}$] Intrinsically, the act of forming stars consumes gas, thus impacting the future SFR. Additionally, feedback as a consequence of star formation actively affects the SFR through the expulsion of hot halo and cold galaxy gases and suppression of the cooling process.
    \item[$H_{k-1}, E_{k-1} \rightarrow H_{k}, E_{k}$] Halo mass and environment are determined by the halo accretion and merger rates, which ultimately depend on the previous halo mass and environment. Also, environmental processes such as tidal stripping affect the halo mass.
    \item[$H_{k-1}, E_{k-1} \rightarrow SFR_{k}$] Accretion and mergers alter many halo and galaxy properties besides halo mass, which all converge on the SFR. In other words, there is a direct lagged effect of the previous halo mass and environment on the current SFR.
\end{description}

For an in-depth understanding of and reasoning behind the structure of the causal model, see Appendix \ref{sec:galaxy_formation_and_evolution}. There are many causal effects in the causal model, then what is the causal effect of environment on SFR? We provide an explicit definition next.

\subsection{Causal effects}
\label{sec:causal_effects}
Environment affects galaxies over a period of time, so it is a time-varying treatment rather than a time-fixed treatment. There are many causal effects one can compute when the treatment varies in time, each providing an answer to a specific type of decision-making task. We define two types of causal effects:

\begin{enumerate}
    \item The marginal causal effect is the effect of a single treatment $T_{k}$ on outcome $Y_{k}$.
    \item The joint causal effect is the effect of multiple treatments or treatment history $\bar{T}_{k}$ on outcome $Y_{k}$, where $\bar{T}_{k} = [T_{0}, T_{1},...,T_{k}]$.
\end{enumerate}

The marginal effect of $E_{k}$ represents the short-term environmental impact, while the joint effect of $\bar{E}_{k}$ captures the long-term impact on SFR. We focus on the joint effect as we are interested in the overall impact and refer to it as the causal effect. We define the environmental history,

\begin{equation}
    \bar{E}_{j} = \frac{1}{N} \sum_{k=0}^{j} E_{k},
    \label{eq:environmental_history}
\end{equation}

\noindent
where $N$ is the number of treatments and $j=k$ (see \citenum{hagedoorn_2021}, for more complex characterisations). We note that $j$ is used instead of $k$ for mathematical correctness in this equation and where necessary, but not in text for consistency. The joint effect of $\bar{E}_{k}$ represents the impact of average environment, which we de facto mean by the causal effect of environment.

As found above, there is confounding bias due to halo mass that must be negated to estimate the true causal effect of environment on SFR. First, consider the marginal effect of $E_{0}$ on $SFR_{0}$. There are two confounders, $H_{1}$ and $E_{1}$, as they directly cause the treatment $E_{0}$, and directly and indirectly (via $SFR_{1}$ and $H_{0}$) cause the outcome $SFR_{0}$. Thus, we need to adjust for them to estimate the causal effect of the `current' environment on the `current' SFR. Intuitively, if the `previous' halo mass and environment affect the current SFR, then their roles must be negated to determine the impact of only the current environment. We can achieve this by conditioning on them.

Now, consider the joint effect of $E_{0}$, $E_{1}$, and $E_{2}$ on $SFR_{0}$. Once again, $H_{1}$ is a confounder of the causal effect of $E_{0}$ on $SFR_{0}$, so we must adjust for it. However, if we condition on the confounder, the causal effect of $E_{2}$ on $SFR_{0}$ that flows through it becomes blocked. As a result, the causal effect of environment on the SFR would be underestimated (in the positive or negative direction). In other words, if we attempt to eliminate confounding bias, then we introduce over-adjustment bias, and if we do not, the causal effect remains biased. More precisely, we cannot estimate the joint causal effect with the conditional approach because there is time-varying confounding and treatment-confounder feedback: $H_{k}$ is a time-varying confounder since it causes outcome $SFR_{k}$ and the subsequent treatment $E_{k+1}$, and there is treatment-confounder feedback as $H_{k}$ is affected by the previous treatment $E_{k-1}$, i.e., the previous environment affects the current halo mass which then affects the subsequent environment in a cycle. The key takeaway is that simply adjusting for the halo mass and environmental histories is insufficient to disentangle nature and nurture.

In summary, estimating the causal effect of environment is difficult due to the interdependence of nature and nurture. The causal effects of nature and nurture are intertwined as the causal effect of environment partially flows through halo mass and vice versa. As a result, it is challenging to isolate the effect of one from the other without introducing bias. Conditional approaches cannot adequately separate the causal effects even if given all the necessary data. So, the challenge is not only of data but also methodological. We employ a marginal approach in the form of the Robins' generalised method \citep[g-method; see][for an overview]{naimi_2017}, inverse probability weighting (IPW) of marginal structural models (MSMs; \citenum{robins_2000}), to disentangle nature and nurture. The method has been applied to conceptually similar problems in other fields \citep{fewell_2004, mortimer_2005, mansournia_2012, nandi_2012, li_2016}, e.g., to study the effects of neighbourhood poverty on alcohol use \citep{cerda_2010}. See Appendix \ref{sec:ipw_of_msms} for a description of the method and definitions of the conditional and marginal approaches.

We combine IPW of MSMs with the random forest (RF; \citenum{breiman_2001}) algorithm and devise and implement an overall two-step estimation process on the IllustrisTNG simulations, specifically the TNG100-1 run. The dataset consists of $18629$ galaxies traced over cosmic time with merger trees from $z \sim 6$ to the present day $z=0$, and our environment proxy is the three-dimensional (3D) 10th nearest neighbour density. See Section \ref{sec:method} for a description of the galaxy sample and the estimation process. We estimate the overall causal effect of environment on the SFR at $z=0$ and at different redshifts back to $z \sim 3$ (with a baseline at $z \sim 6$) to understand the role of environment over time. See Appendix \ref{sec:validation} for validity of the results.

\section{Results}
\label{sec:results}

\subsection{Overall causal effect of environment}
\begin{figure*}
    \centering
        \includegraphics[width=\textwidth]{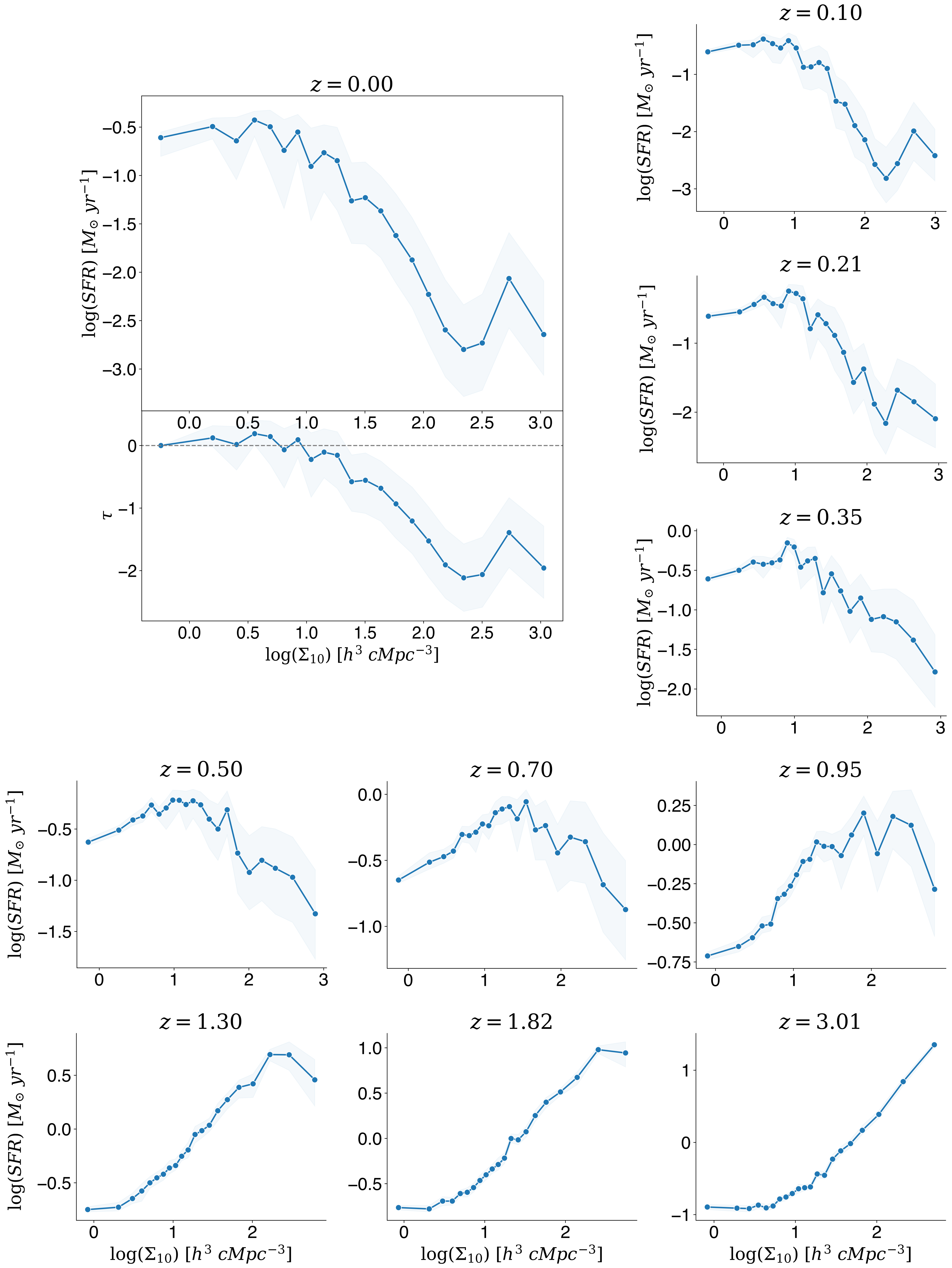}
    \caption{Causal dose-response curves (CDRCs) of the causal effect of environment on the star-formation rate (i.e., causal SFR–density relations) at $z=0$ and at different redshifts going back to $z \sim 3$, assuming the causal model (Fig. \ref{fig:causal_model}a). Specifically, they represent the average SFR of galaxies if they inhabited, on average, the specific density environment (10th nearest neighbour density) over time. The bottom panel of $z=0$ shows the average causal effects $\tau$ of different density environments (comparing to the lowest-density environment). The shaded regions represent the $68 \%$ confidence interval, estimated with bootstrapping.}
    \label{fig:cdrc_current}
\vspace{-55.3305pt}
\end{figure*}

Fig. \ref{fig:cdrc_current} shows the causal dose-response curves (CDRCs) of the causal effect of environment on the SFR (i.e., causal SFR–density relations) at $z=0$ and at different redshifts going back to $z \sim 3$. The CDRC denotes the average response in the population if all units were subject to treatment $T=t$. By comparing any two points on the curve, one can determine the mean change in the outcome if all units received, for example, treatment $T=t_{a}$ instead of $T=t_{b}$. This difference in outcome is the average causal effect $\tau$ (ACE; \citenum{holland_1986}). Here, the CDRCs represent the average SFR of galaxies if they inhabited, on average, the specific density environment over time. The bottom panel of $z=0$ shows the ACEs of different density environments (comparing to the lowest-density environment, which we define as our baseline).

Focusing on $z=0$, the CDRC is relatively flat up to $\log(\Sigma_{10}) \sim 1$ and therefore the causal effect of environment is negligible. Subsequently, the CDRC trends downwards and the causal effect becomes negative as the average SFR decreases with increasing environmental density until $\log(\Sigma_{10}) \sim 2.5$. At even higher densities the impact of environment saturates, or possibly even weakens. We return to this point in Section \ref{sec:causal_vs_naive_model}. In brief, environment does not appear to influence the SFR at low densities, but at intermediate-to-high densities, it has a negative effect. However, the causal effect is not the strongest in the densest environments. Furthermore, there is a characteristic density ($\log(\Sigma_{10}) \sim 1$) beyond which environment starts playing a role. This `break' in the SFR–density relation has previously been remarked upon by \cite{lewis_2002} and \cite{gomez_2003}, who in fact reported the same value (albeit in projected 2D density). These authors identified the density to be typical of environments a few virial radii away from cluster centres, thus representing something of a transition from the field to more circum-cluster and group-like environments. Above all, the overall causal effect is negative and substantial, with environment suppressing the average SFR by a maximal factor of $\sim 100$.

\subsection{Role of environment over time}
The causal effect is at its most substantial at $z=0$ and weaker in the recent past, implying that the impact of environment accumulates over time. By $z=0.7$ the negative trend seems to start flattening, and somewhat surprisingly, fully reverses by $z=0.95$. Beyond $z \sim 1$ the average SFR continues to increase with increasing environmental density, so the causal effect of environment on the SFR is positive. Furthermore, the effect is significant and becomes stronger with redshift, rising from a factor of $\sim 10$ to over $100$ by $z \sim 3$.

The downtrend we find at low redshifts is consistent with observational studies, and typically interpreted as a result of processes in our causal model such as tidal stripping, RPS, dynamical friction, and galaxy harassment (see Fig. \ref{fig:causal_model_galaxy_formation_and_evolution}). But the uptrend at high redshifts is unexpected given that these same physical processes should operate in dense environments regardless of cosmic epoch. Having said that, studies have been inconsistent in their findings, and there remains an active debate on whether the SFR–density relation, as observed in the local universe, exists at intermediate ($z \sim 1$) to high redshifts ($z > 1$). Some studies have found that the relation persists in the early universe \citep{patel_2009, muzzin_2012, quadri_2012, chartab_2020}, others have evidenced a flattening \citep{feruglio_2010, grutzbauch_2011, scoville_2013, ziparo_2014, darvish_2016}, while others yet have noted a reversal \citep{elbaz_2007, cooper_2008, tran_2010, popesso_2011, santos_2015, lemaux_2022, shi_2024}. The lack of consistency between the studies is due to a multitude of reasons, beginning with the ambiguity around the SFR–density relation itself. The term is loosely used in the literature and is polysemous, as studies have analysed the specific star-formation rate ($SFR/M_{\star}$)–density, colour–density, and star-forming/quiescent fraction–density relations under the SFR–density relation moniker. These quantities, while related, are fundamentally different and cannot be compared. Another point for contention is the data or lack thereof. Deep surveys observe a small patch of the sky, so such studies are often susceptible to cosmic variance. Lastly, since there is no universal definition of environment \citep{muldrew_2012}, the choice of measure is possibly responsible for part of the disagreements.

It is difficult to draw parallels between the literature for all the aforementioned reasons, but the most critical aspect is that the analyses have been largely statistical rather than causal in nature. Regardless, both observational \citep{elbaz_2007, lemaux_2022} and simulation studies \citep{tonnesen_2014, hwang_2019} have found SFR–density reversals (albeit at a weaker level) independent of the stellar mass correlation, thus supporting a positive impact of environment. Additionally, \cite{hwang_2019} and \cite{lemaux_2022} observed the uptrend strengthening with redshift, as we do here. The consistency of our results, with \cite{hwang_2019} in particular, given that they also used the TNG simulations (though the larger volume TNG300 instead of TNG100), is strong evidence for the reliability of our method and study.

Assuming that IllustrisTNG provides a reasonable reflection of reality, a possible explanation for the positive causal effect of environment in the early universe is that denser environments have a larger reservoir of material from which to form stars. As a result, galaxies in such environments are able to accrete more gas more quickly and sustain higher SFRs. This seemingly simple and obvious statement belies the complexity around how we choose to compare galaxy populations occupying different environments and what we mean by a galaxy's `nature'. We will elaborate this point in the next section. Naturally, high-density environments at low redshift also contain a greater abundance of matter than lower-density regions. The key difference is that in the late universe a far greater fraction of the gas is maintained in the form of hot gas, inaccessible for star formation. The above, combined with the absence of negative processes in the early universe typically associated with environment, as groups and clusters are still forming, can explain the positive effect.

Another potential cause is major mergers, which are more likely in dense regions at high redshifts \citep{fevre_2000, kampczyk_2007, kartaltepe_2007, lotz_2011} when the velocities are not too extreme. An influx of cold gas in a gas-rich galaxy merger can trigger a galaxy-wide starburst (though clearly this channel will not explain the entire factor $\sim 100$ enhancement measured at the highest redshift). Nonetheless, for a definitive answer, the causal effect of individual environmental processes must be estimated. The critical work has been done with the construction of a comprehensive causal model (Fig. \ref{fig:causal_model_galaxy_formation_and_evolution}). Using this model, one can extend our analysis to estimate the causal effects of different processes \citep[see][for work in this direction]{smethurst_2017}.

We stress that the positive trend is not just a consequence of the fact that massive galaxies are preferentially found in denser environments. Rather, what we are seeing is a reflection of the well-established galaxy ``downsizing'', in which the most massive galaxies form early and quickly, and barely evolve thereafter (\citenum{cowie_1996, heavens_2004, kodama_2004, jimenez_2005, juneau_2005, thomas_2005, bauer_2005, bell_2005, nelan_2005, feulner_2005, bundy_2006, drory_2008, vergani_2008, mortlock_2011}; also see \citenum{fontanot_2009}, for a detailed discussion). Our measurements explain downsizing as a consequence of environment-driven accelerated evolution of galaxies in dense environments rather than high-stellar mass galaxies directly causing their own high SFRs and subsequent quenching.

\subsection{Model comparison}
\label{sec:model_comparison}
We now place our results into a wider context by comparing our causal model to others. We answer the following critical questions: (i) is nature important? Specifically, is galaxy formation and evolution top-down dominated by environment with no reverse influence of halo mass? (ii) is controlling for the stellar mass at a snapshot in time sufficient to disentangle nature and nurture and estimate the causal effect of environment? and (iii) is stellar mass an adequate proxy of nature?

\paragraph{Naïve model}
We first consider the possibility of no confounding. In our causal model, halo mass is the confounder as it causes environment and SFR. However, in order to answer the first of our key questions it is useful to consider a universe in which this is not the case, one in which environment completely dominates the evolution of a galaxy and halo mass has no reverse influence on environment. In this scenario, the raw correlation between environment and SFR is the unbiased causal effect of environment, and one does not need to adjust for nature. We represent this as the naïve model (Fig. \ref{fig:causal_model}b).

\begin{figure*}
    \centering
    \includegraphics[width=\textwidth]{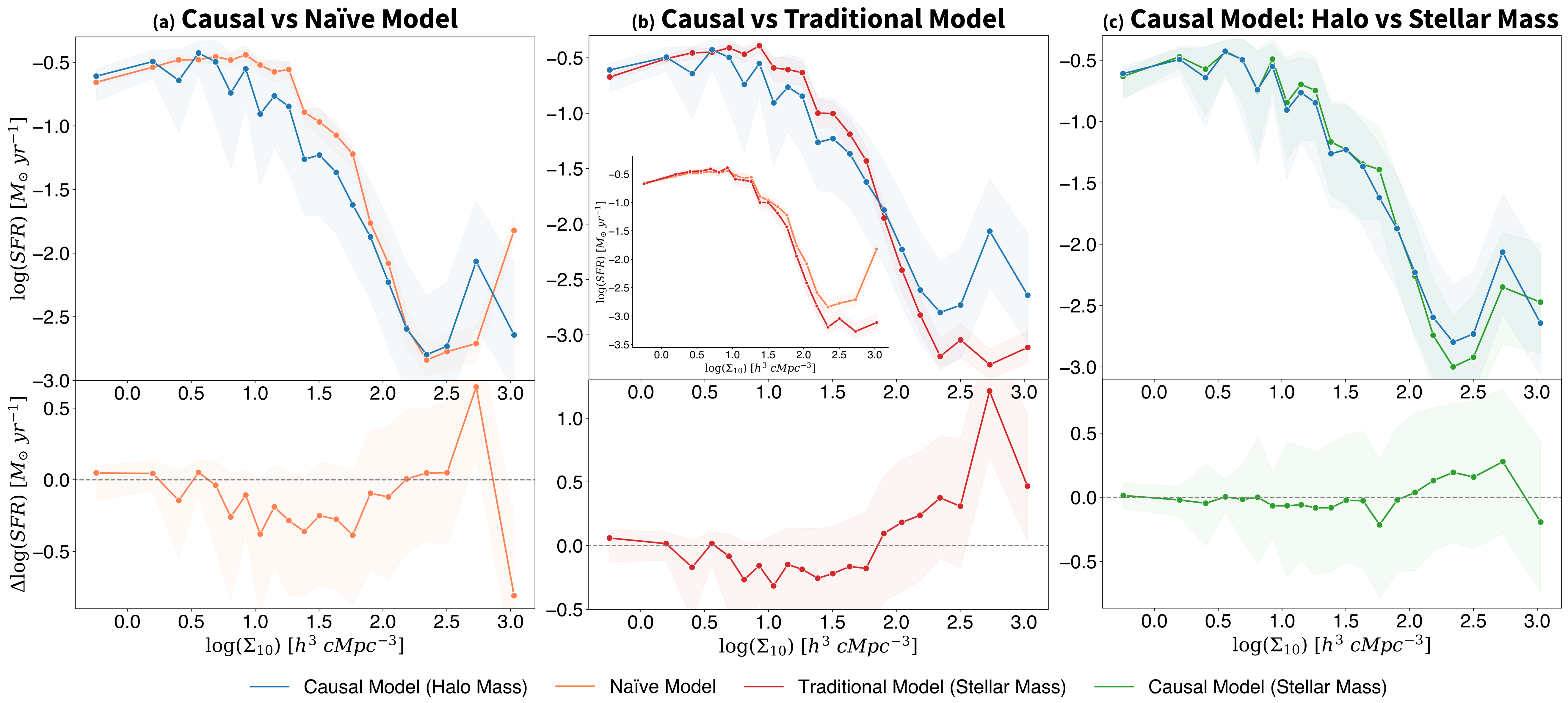}
    \caption{Causal dose-response curves (CDRCs) of the causal effect of environment on the star-formation rate (i.e., causal SFR–density relations) at $z=0$ of the (a) naïve (orange) and (b) traditional (red) models, compared to our physics-informed causal model (blue). The naïve model (Fig. \ref{fig:causal_model}b) assumes no confounding and that the raw correlation between environment and SFR is the unbiased causal effect. The traditional model (Fig. \ref{fig:causal_model}c) is the model implicitly assumed in the literature when controlling for the stellar mass at a snapshot in time to disentangle nature and nurture. (c) Causal model (stellar mass; green) is the causal model (Fig. \ref{fig:causal_model}a) but with stellar mass as the time-varying confounder instead of halo mass. The inset of the centre panel compares the naïve and traditional models. The CDRCs represent the average SFR of galaxies at $z=0$ if they inhabited, on average, the specific density environment (10th nearest neighbour density) over time. Comparing the CDRC at a given density (`treatment') to a baseline, chosen to be the lowest-density environment (`no treatment'), therefore reveals the causal effect of that environment on SFR. The bottom panels shows the difference in the average SFRs between the models. The shaded regions represent the $68 \%$ confidence interval, estimated with bootstrapping.}
    \label{fig:cdrc_joint_effect_outcome_current_model_comparison}
\end{figure*}

\paragraph{Traditional model}
Next, we mimic previous studies, that is we adjust for the stellar mass at a snapshot in time to estimate the causal effect of environment. The method supposes the causal model illustrated in Fig. \ref{fig:causal_model}c, in which stellar mass ($M_{\star}$) is the confounder. Obviously, stellar mass does not cause environment and SFR, but this is nevertheless the model implicit in the literature. In fact, stellar mass may be an effect of both, as per our causal model, so adjusting for the variable could induce selection bias (Fig. \ref{fig:selection_bias}) instead of negating confounding bias (see ``Processing steps'' in Section \ref{sec:data} for a more detailed discussion). Furthermore, the model fails to capture dynamic systems with feedback loops that galaxies ultimately are. The lack of a framework to infer causality, combined with the shortcomings of observational data, has meant that controlling for the stellar mass at a snapshot in time has been the only practical and logical option (given that stellar mass should correlate with internal processes) to disentangle nature and nurture, even though it has been demonstrated by \citep{lucia_2012} that this is insufficient. We refer to this model as the traditional model.

\paragraph{Causal model (stellar mass)}
Finally, we assume our physics-informed causal model (Fig. \ref{fig:causal_model}a), but on this occasion, we substitute stellar mass for halo mass as the time-varying confounder. Our goal is to answer the question: is stellar mass a suitable proxy of nature (halo mass) given a valid causal model and method to disentangle nature and nurture? This is important because halo mass (not host halo mass) is challenging to infer observationally. Fig. \ref{fig:cdrc_joint_effect_outcome_current_model_comparison} shows the CDRCs of the causal effect of environment on the SFR at $z=0$ for the aforementioned models, compared to the causal model (labelled as ``causal model (halo mass)''). The bottom panels show the difference in the average SFRs between the models. We focus on $z=0$ as we are primarily interested in the difference in the overall causal effect.

\subsubsection{Is nature important?}
%\subsubsection{Causal vs Naïve Model}
\label{sec:causal_vs_naive_model}
The CDRCs of the naïve and causal models are different (Fig. \ref{fig:cdrc_joint_effect_outcome_current_model_comparison}a), which implies that the halo mass of a galaxy influences the environment it inhabits. Breaking it down further, the CDRC of the naïve model is above the causal model's up to $\log(\Sigma_{10}) \sim 2$, which signifies that halo mass has a positive effect on the SFR because, post adjustment, the average SFR is lower. The probable explanation is that at low-to-intermediate densities, a larger halo is able to accrete more gas from its surroundings which, up to a certain extent, translates to enhanced star formation. Between $\log(\Sigma_{10}) \sim 2-2.5$, there is no discernible difference between the two models, so halo mass has no effect on environment in this density regime (assuming it impacts the SFR). In the host halo mass–environment distribution in \ref{fig:galaxy_properties_distributions}, $\log(\Sigma_{10}) \gtrsim 2$ in average density probably corresponds to large group and cluster host haloes, where environment is indeed believed to dictate galaxy evolution. Naturally therefore, the causal effect of environment is the largest in this density regime.

Beyond $\log(\Sigma_{10}) \sim 2.5$ the models diverge again, but it is difficult to conclude whether halo mass impacts environment due to the noise and limited data. Nonetheless, a possible explanation may lie in the fact that the density regime (primarily) corresponds to central regions of massive groups and clusters, given that the ratio of the number of central galaxies to satellites is comparatively larger than in the prior density range, as can be observed in \ref{fig:galaxy_properties_distributions}. The centre of a host halo is a special location, where a galaxy is not subject to some of the environmental processes that satellite galaxies are, and may also potentially be able to have some influence. This also explains the weaker environmental impact on SFR in the density regime. Also, central galaxies at the bottom of the potential well may be recipients of gas for star formation in the case of cool-core clusters. The dichotomy between central and satellite galaxies could be explored within our causal framework by using an \emph{effect modification} and adapting IPW of MSMs to estimate the separate causal effects of environment on these two object classes. However, doing so is rather involved and we leave this for future work. 

In conclusion, nature is important and galaxy formation and evolution is not simply top-down determined by environment. The environment affects halo mass, but halo mass has an influence on the environment a galaxy inhabits as well. Ignoring the role of nature by not adjusting for halo mass leads to the causal effect in intermediate-density environments being underestimated by a factor of $\sim 2$. A potential future study would be to estimate the causal effect of halo mass on SFR to finally answer the long-standing question: which is more important, nature or nurture? 

\subsubsection{Does controlling for stellar mass disentangle nature and nurture?}
%\subsubsection{Causal vs Traditional Model}
\label{sec:causal_vs_traditional_model}
The CDRC of the traditional model is largely identical to the naïve model's except in the high-density regime (inset of Fig. \ref{fig:cdrc_joint_effect_outcome_current_model_comparison}b), so adjusting for the stellar mass at the single redshift of observations has mostly had no effect. It has not disentangled the effects of nature and nurture and is therefore insufficient to estimate the causal effect of environment. In fact, the traditional model overall deviates further from the assumed truth (i.e., the causal model) than the naïve model, so adjusting for the stellar mass at a snapshot in time actually has an adverse effect. The failure of the traditional model is also indicated by the skewed weight distributions in Fig. \ref{fig:weights_distributions_traditional_model}, with mean values farther from one compared to the causal model's (Fig. \ref{fig:weight_distributions_causal_model}). 

We remark that the traditional model also predicts a positive causal effect of environment at high redshifts (Fig. \ref{fig:cdrc_current_traditional_model}), as opposed to a flattening observed by many previous observational studies. The uptrend is weaker compared to the causal model, but it is present nonetheless, especially at $z = 1.82$ and $z \sim 3$. The result further confirms that the positive trend observed with the causal model is not simply due to massive galaxies in denser environments that form more stars because we adjust for this fact by controlling for stellar mass with the traditional model.

\subsubsection{Is stellar mass an adequate proxy of nature?}
%\subsubsection{Causal Model: Halo vs Stellar Mass}
\label{sec:halo_vs_stellar_mass}
While there are minor differences (primarily at high densities), the CDRCs of causal model (halo mass) and causal model (stellar mass) are similar overall, as further highlighted by the minimal deviations of the residual curve (Fig. \ref{fig:cdrc_joint_effect_outcome_current_model_comparison}c). Consequently, we conclude that stellar mass is an adequate proxy of halo mass and thus nature. This finding is important because unlike a galaxy's halo mass, its stellar mass can be readily inferred observationally, and the ultimate goal of this study is to estimate the causal effect of environment on galaxies in the real universe. Evidently, the lack of evolutionary histories of galaxies observationally, or more specifically, the halo/stellar mass and environmental histories, is a major hurdle. On this front, star-formation histories (and thus stellar mass histories) can be recovered by modelling and fitting spectral energy distributions \citep[see][for a recent review]{conroy_2013}, and recently, \cite{sarpa_2022} employed the extended Fast Action Minimization (eFAM) method to reconstruct the environmental history of galaxies. Consequently, it is feasible to estimate the real causal effect (on star formation at least).

\section{Conclusions}
\label{sec:conclusion}
Our understanding of galaxy formation and evolution has advanced significantly in the past few decades as an overall picture has been pieced together. Having said that, the underlying processes responsible for the observed phenomena, and especially their importance, are still not fully known, as establishing causality from correlations has proven challenging. Here, simulations have provided some insights, but the causal effect itself is intractable given the high degree of complexity. In this Article, we have addressed this by adopting the causal inference framework. In particular, we have tackled the long-standing problem of disentangling the roles of nature (i.e., internal processes) and nurture (i.e., external processes) on the formation and evolution of galaxies and estimated the causal effect of environment on star-formation rate (SFR) in the IllustrisTNG simulations. To achieve this, we developed a comprehensive causal model from the ground up and employed advanced techniques from epidemiology.

We have found that the causal effect is negative and substantial, with environment suppressing the SFR by a maximal factor of $\sim 100$. However, while the overall effect at $z=0$ is negative, we have discovered that in the early universe ($z \gtrsim 1$), environment had a positive impact. Furthermore, the causal effect is significant, with environment boosting star formation by a factor of $\sim 10$ at $z \sim 1$ and by even greater amounts at higher redshifts. This challenges the consensus that environment always acts to inhibit star formation. Our results require a broader view of what we mean by the impact of environment, and are an expression of the well-known phenomenon of galaxy ``downsizing''. We also reveal that:

\begin{enumerate}
    \item Nature (associated with halo mass) is important. Specifically, galaxy formation and evolution is not top-down determined by environment as the environment affects halo mass, but halo mass has an influence on the environment a galaxy inhabits as well. Ignoring the role of nature leads to the causal effect in intermediate-density environments being underestimated by a factor of $\sim 2$.
    \item Controlling for the stellar mass at a snapshot in time, as is common in the literature, does not disentangle nature and nurture. Not only is it insufficient to estimate the causal effect of environment, but it actually has an adverse effect. We remark that the causal effect estimated this way at high redshifts is still positive, though reduced in magnitude. Overall, snapshot studies are inadequate, and the evolutionary history of galaxies is required.
    \item Nevertheless, stellar mass is a sufficient proxy of the effects of nature (i.e., halo mass), assuming our causal model is valid and given stellar mass history and method to disentangle nature and nurture.
\end{enumerate}

With the introduction of a theoretical framework to infer causality and a potential solution to the challenging nature–nurture problem, this work paves the way towards unravelling some of the biggest questions in galaxy formation and evolution, such as: what drives galaxy quenching, are environmental processes responsible for the morphological transformations of galaxies, what is the impact of supermassive black holes (SMBHs) on their host galaxies, and the question at the heart of this Article, which is more important: nature or nurture? The timing is opportune as the field is entering the era of ``Big Data'' with the next generation of large-scale surveys coming online, such as Euclid \citep{euclid_2011}, the Rubin Observatory Legacy Survey of Space and Time (LSST; \citenum{lsst_2009}), and the Nancy Grace Roman Space Telescope (Roman Space Telescope; \citenum{roman_2015}). These surveys have the power to measure the correlations to exquisite precision, and thus hold great promise in teasing out subtle patterns that are currently hidden. However, a causal framework, such as the one we have presented, will be necessary to determine the underlying mechanisms and their effects to truly understand how galaxies have formed and evolved.

\section{Methods}
\label{sec:method}

\subsection{Galaxy sample}
\label{sec:data}
According to the causal model and method, we require the evolutionary history of galaxies to disentangle nature and nurture. Such data is not readily available observationally, so we rely on simulations, which enable the tracing of galaxies over time using merger trees. In particular, we utilise the IllustrisTNG simulations \citep{pillepich_2018a, springel_2018, nelson_2018, naiman_2018, marinacci_2018, nelson_2019}.

\paragraph{IllustrisTNG}
IllustrisTNG (hereafter TNG) is a suite of cosmological, gravo-magnetohydrodynamical (MHD) simulations run with the moving-mesh code \texttt{AREPO} \citep{springel_2010}. TNG adopts a flat $\Lambda \text{CDM}$ cosmology with \cite{planck_2016} cosmological parameters ($\Omega_{\Lambda, 0}=0.6911, \Omega_{m, 0}=0.3089, \Omega_{b, 0}=0.0486, \sigma_{8}=0.8159, n_{s}=0.9667, \ \text{and} \ h=0.6774$). The simulations start at $z=127$ from initial conditions created with the Zeldovich approximation \citep{zeldovich_1970} and the \texttt{N-GenIC} code \citep{springel_2015}. There are $100$ snapshots of each simulation, approximately equally spaced in cosmic time from $z \sim 20$ to the present day $z=0$. For our analysis, we use the highest resolution run of the TNG100 simulation, TNG100-1. The simulation is initialised with $1820^{3}$ dark matter and gas particles of mass resolutions $7.5 \times 10^{6} \text{M}_{\odot}$ and $1.4 \times 10^{6} \text{M}_{\odot}$, respectively.

\paragraph{Processing steps}
First, we select a sample of galaxies to trace over time as follows. We start with the group catalogue at $z=0$, which contains $6291349$ friends-of-friend (FoF; \citenum{davis_1985}) haloes and $4371211$ \texttt{SUBFIND} \citep{springel_2001, dolag_2009} subhaloes. We match each subhalo to its FoF halo using the \texttt{SubhaloGrNr} field, which results in $3430706$ FoF haloes (i.e., more than half of FoF haloes do not have any subhaloes). There are some subhaloes of non-cosmological origin, which means they have not formed due to the process of structure formation and collapse and are likely fragments or clumps rather than bonafide galaxies \citep{nelson_2019}. We discard these objects by setting \texttt{SubhaloFlag} $=1$. Finally, we remove any subhaloes with no detectable dark matter and retain all galaxies with stellar mass $M_{\star} > 10^{9} \text{M}_{\odot}$, resulting in $20935$ galaxies. We chose the relatively high stellar mass cut to trace galaxies further back in time. 

There is a possibility that applying the stellar mass cut introduces selection bias since stellar mass may be a common effect of both environment and SFR. One can interpret this as survivor bias (a form of selection bias) since we are selecting galaxies that made it to the stellar mass at the `end' of the galaxy formation and evolution process. In the model, there is a causal connection between SFR and stellar mass, as well as environment and stellar mass, but significantly only the former is direct, while the latter is indirect through SFR. There is a `direct' connection via galaxy harassment and tidal stripping between environment and stellar mass, but it is likely to be weak in comparison and not universal. Consequently, we reason that selecting galaxies based on their stellar mass at $z=0$ does not bias our analysis. Indeed, preliminary tests supported this as different stellar mass cuts were applied, and it was found that while the amplitude of the SFR–density relation changed, the shape did not. In other words, the causal effect remained unmodified. 

We track the selected galaxies back in time with \texttt{SUBLINK} \citep{rodriguez_2015} merger trees. The merger tree of a subhalo can have many branches if its progenitors have undergone mergers. We follow the main progenitor branch (MPB), which traces the most massive progenitor at each point in time. For the analysis, we use $11$ approximately equally-spaced snapshots in cosmic time from $z \sim 6$ to $z=0$ (i.e., snapshots $13, 25, 35, 43, 51, 59, 67, 75, 83, 91, \text{and} \ 99$). We decided upon $z \sim 6$ as the baseline because it was the furthest we could trace most galaxies back in time. The average timespan between each snapshot is $\sim 1.3$ Gyr. After all the preprocessing steps, the galaxy sample contains $18629$ galaxies. 

\paragraph{Environment and measurement choices}
There are different measures available of halo and galaxy properties. In our analysis, we use the quantities derived by summing all particles/cells bound to a subhalo associated with the particular property. The choice of measure does not impact our results because our questions are causal rather than statistical in nature. For the same reason, we stick with instantaneous SFRs measured in the simulations instead of using time-averaged SFRs that better match SFRs estimated observationally with various tracers. Due to the finite numerical resolution of the simulation, the instantaneous SFRs of galaxies are unresolvable if they are below the minimum value of $\log(SFR) \sim -4$ for TNG100 \citep{donnari_2019}. The SFRs of such galaxies are labelled zero, which could cause numerical issues when estimating the causal effect. Following \cite{donnari_2019}, we resolve the problem by randomly assigning an SFR value between $-4$ and $-5$. 

There are many definitions of environment in the literature \citep[see][for a review]{muldrew_2012}, but the most popular are nearest-neighbour-based and fixed-aperture-based measures. The former best probe the `local environment', while the latter the `large-scale environment'. Simply put, there is no universal definition of environment, and the most suitable method is scale dependent \citep{muldrew_2012}. As we are interested in the impact of the local environment, our environment proxy is the $N$th nearest neighbour density,

\begin{equation}
    \Sigma_{N} = \frac{N}{(4 \pi/3) r_{N}^{3}},
\end{equation}

\noindent
where $r_{N}$ is the three-dimensional (3D) distance to the $N$th nearest neighbour from the galaxy in question. Specifically, we compute densities at the 10th nearest neighbour, which is a popular choice in the literature \citep{lewis_2002, pimbblet_2002, cassata_2007, sobral_2011}. We note that a rough analysis was performed with a range of nearest neighbours from $N=3-64$. Below $N \leq 7$, the SFR–density relation was found to be noisy and flat, and thereafter, it became more stable and negative with increasing $N$. These preliminary results indicate that the causal effect of environment varies at different scales. Most subhaloes in the simulation are small and dark (i.e., do not possess any galaxies), so considering the entire population would result in a noisy density measure. For a more informative estimate, we remove such subhaloes by applying the same cuts as when selecting the galaxy sample, but we do not apply the stellar mass cut at $10^9 \text{M}_{\odot}$. Instead, we drop subhaloes with no detectable stars. We remark that we considered using host halo mass as a proxy since many environmental processes are either a consequence of it or scale with it. However, we ultimately decided against it because it is not a fine-grained measure of environment and its effects, and due to conceptual as well as practical issues.

Following our naming convention, FoF haloes and subhaloes are host haloes and haloes in our causal model respectively, so we refer to them and the associated properties accordingly from here onwards. \ref{fig:galaxy_properties_distributions} shows the relationships between fundamental halo and galaxy properties, such as host halo mass, halo mass, stellar mass, and SFR, as well as the average environmental density of the galaxy sample at $z=0$. A clear positive correlation can be observed between host halo mass and the 10th nearest neighbour density, which suggests that the latter is a suitable measure of environment, at least to the first order \citep[see][for a comparison between different environment measures and host halo mass]{haas_2012}. As expected, the SFR overall decreases with increasing environmental density.

\subsection{Estimation process}
\label{sec:estimation}
We apply inverse probability weighting (IPW) of marginal structural models (MSMs; \citenum{robins_2000}) to disentangle nature and nurture and estimate the causal effect of environment on SFR. IPW is a statistical technique to adjust for confounding and estimate causal effects \citep[see][for an overview]{chesnaye_2022}. As the name suggests, the basic premise is to weight each unit according to the inverse of their probability of receiving treatment, i.e., the propensity score (PS; \citenum{rosenbaum_1983}), to create a pseudo-population in which the treatment is independent of confounders. Overall, IPW of MSMs involves: (i) estimating weights to adjust for biases and (ii) fitting a MSM using the weights to estimate causal effects (see Appendix \ref{sec:ipw_of_msms} for a more detailed description and explanation of the method). Weights can be estimated directly from data if the treatment and confounders are binary or categorical variables and there are a limited number of them. In our case, environment and halo mass are inherently continuous variables, and we suffer from the curse of dimensionality with $11$ total snapshots, so direct estimation is not feasible. Additionally, we cannot fit a parametric MSM to estimate causal effects since the causal relationship between environment and SFR is unknown. We apply ML for both tasks, which allows us handle the high-dimensional data and model the potentially non-linear relationships between halo mass, environment, and SFR. Specifically, we employ the random forest (RF; \citenum{breiman_2001}) algorithm as it has already been shown to perform the best (out of the algorithms compared) in estimating propensity scores \citep{cannas_2019}.

\paragraph{Overview}
We define two types of models:

\begin{itemize}
    \item \textit{Weighting model} – the input features and target variable are the confounders and treatment, respectively. The model output is the expectation of treatment given confounders, $\mathbb{E}[T|X]$. 
    \item \textit{Outcome model} – the input feature and target variable are the treatment and outcome, respectively. The model output is the expectation of outcome given treatment, $\mathbb{E}[Y|T]$.
\end{itemize}

As the names suggest, the weighting and outcome models estimate weights and MSMs, respectively. We devise the following multi-step estimation process:

\begin{enumerate}
    \item Train a weighting model.
    \item Predict the treatment of each unit with the weighting model to estimate the propensity score.
    \item Repeat the above steps, but now to estimate the numerator.
    \item Construct weights.
    \item Train an outcome model with each unit weighted.
    \item Predict treatment outcomes with the outcome model to estimate causal effects.
\end{enumerate}

\paragraph{Application}
In the causal model, the previous halo mass $H_{k-1}$ and environment $E_{k-1}$ affect the current environment $E_{k}$ and star-formation rate $SFR_{k}$. Therefore, it is necessary to adjust for them at each time point to eliminate confounding bias. We go a step beyond and adjust for the entire previous halo mass and environmental histories ($\bar{H}_{k-1}$ and $\bar{E}_{k-1}$) as a precautionary measure to account for any direct lagged effects of halo mass and environment that may hypothetically exist from further back in time. Taking everything into consideration leads us to the weights at time point $k(=j)$,

\begin{equation}
    w_{j} = \prod_{k=0}^{j} \frac{f(E_{k}|\bar{E}_{k-1})}{f(E_{k}|\bar{E}_{k-1}, \bar{H}_{k-1})},
    \label{eq:stabilized_weights_time_varying_specific}
\end{equation}

\noindent
adapted from the general form for time-varying treatments (equation \ref{eq:stabilized_weights_time_varying_general}; \citenum{robins_2000}). The numerator is the conditional probability density function (PDF) of the current environment given the previous environmental history, and the denominator is the conditional PDF of the current environment given the previous environmental and halo mass histories.

We estimate the propensity score (i.e., the denominator) as follows. We train a weighting model with the previous halo mass and environmental histories as the inputs and the current environment as the target. Once trained, we predict the current environment with the model, which is the $\mathbb{E}[E_{k}|\bar{E}_{k-1}, \bar{H}_{k-1}]$. Subsequently, we construct a normal distribution with the mean set to the prediction and the standard deviation equal to that of the residuals. Finally, we evaluate the density function at the true value. 

We estimate the numerator following the same process. In the weighting model, we input the previous environmental history, with the target once again the current environment. The model output is the $\mathbb{E}[E_{k}|\bar{E}_{k-1}]$. We train separate weighting models to estimate the weights at each time point. The exception being $z \sim 6$ because there is no prior confounding by default as it is the baseline snapshot. The weights at the redshift are equal to one. We note that we have assumed the conditional PDFs follow the normal distribution to estimate the densities. Also, we trained and predicted on the same dataset because our goal is causal inference, not prediction. 

Finally, we construct the weights according to equation (\ref{eq:stabilized_weights_time_varying_specific}) and incorporate them in outcome models to estimate the causal effect of environment. As our environment proxy, the 10th nearest neighbour density, is a continuous variable, we estimate causal dose-response curves (CDRCs) rather than single causal effects. For this, we define a grid of $21$ treatment values between the 1st and 99th percentiles of the treatment distribution and predict with outcome models. 

We first focus on $z=0$ to understand the overall effect. We train an outcome model with the average environment $\bar{E}_0$ and the star-formation rate $SFR_{0}$ at $z=0$ as the input and target, respectively. Once trained, we predict the final SFR at different average environments to estimate the CDRC. The weights applied are the product of weights at all redshifts, and the model prediction is the $\mathbb{E}[SFR_{0}|\bar{E}_{0}]$. Next, we extend our analysis to all the redshifts to determine how the role of environment has changed over time. We bootstrapped the entire estimation process to obtain confidence intervals around the CDRCs. This resulted in weighting models predicting a few extreme weights, probably due to the limited sample size. Given that they could drastically skew the causal effect, we trimmed the weights at the 1st and 99th percentiles. $1000$ bootstrap samples were used.

We utilised the \texttt{scikit-learn} \citep{scikit-learn} ML Python package to train the RF models, specifically the \texttt{RANDOMFORESTREGRESSOR} module. In regards to hyperparameter tuning, we kept the defaults and only coarsely tuned $\texttt{min\_samples\_leaf}$, which is the minimum number of samples in a leaf node. Our primary motivation was to best reduce the noise in the CDRCs due to: (i) the non-linear nature of RF and (ii) extrapolation beyond the training data. We found that the combination of $5$ and $200$ for the weighting and outcome models performed the best out of a limited parameter space, respectively. Consequently, we trained all the models with the aforementioned values.

\subparagraph{Naïve model}
The first step of the estimation process is skipped as no bias adjustment is necessary and unweighted outcome models are directly trained to learn the SFR–density relation. An outcome model is trained for each time point with environment $\bar{E}_{k}$ and star-formation rate $SFR_{k}$ as the input and target, respectively.

\subparagraph{Traditional model}
For the traditional model,

\begin{equation}
    w = \frac{f(E)}{f(E|M_{\star})},
    \label{eq:stabilised_weight_time_fixed_specific}
\end{equation}

\noindent
following the general form for time-fixed treatments (equation \ref{eq:stabilised_weight}; \citenum{robins_2000}). In the weighting models, the input is the stellar mass, and the target is the current environment. We highlight that we only input the current stellar mass and not the stellar mass history into the models, unlike in the causal model case, where we had input both the halo mass and environmental histories. The weighting models are then used to estimate the conditional densities $f(E|M_{\star})$. The marginal densities $f(E)$ are determined via kernel density estimation (KDE), and no further models are trained. Finally, the weights are constructed with both densities and incorporated into outcome models. The time-point weights are not multiplied together to form the final product weights, but rather they are applied separately to replicate the previous analyses. We train two outcome models, one with the snapshot environment and the other with the average environment, for a like-to-like comparison with the previous studies and consistency with our results, respectively. The CDRCs from the former are shown in Fig. \ref{fig:cdrc_current_traditional_model}, while the latter is plotted in Fig. \ref{fig:cdrc_joint_effect_outcome_current_model_comparison}b.

\subparagraph{Causal model (stellar mass)}
Same as the causal model's, but now with stellar mass as the time-varying confounder instead of halo mass.

\setcounter{figure}{0} % Restart figure numbering
\renewcommand\figurename{}
\renewcommand{\thefigure}{Extended Data Fig. \arabic{figure}}% Figure counter representation
\renewcommand{\theHfigure}{Extended Data Fig. \arabic{figure}}% Hyperref figure hyperlink hook

\begin{figure*}
    \centering
    \includegraphics[width=\textwidth]{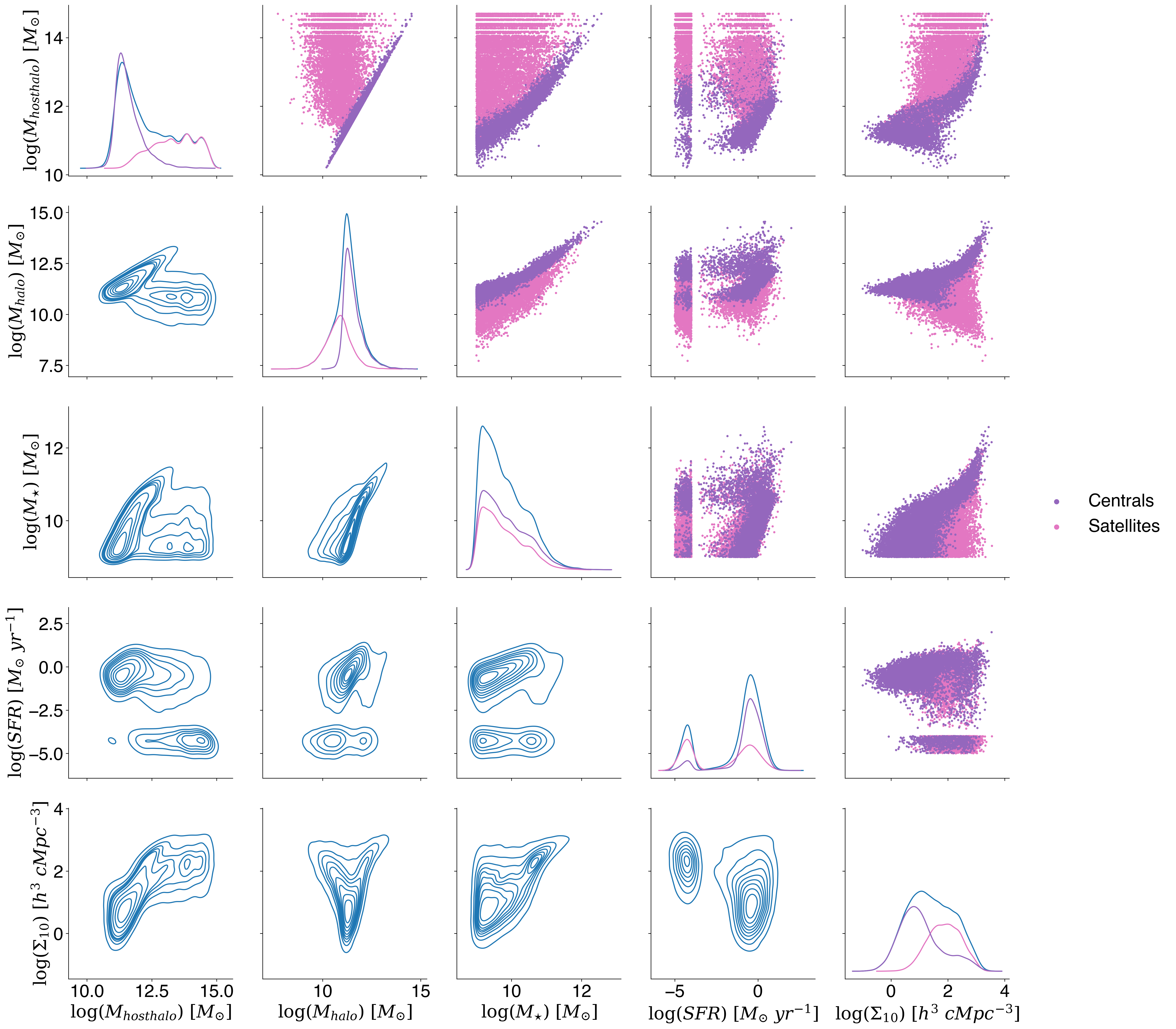}
    \caption{Distributions of fundamental halo and galaxy properties, such as host halo mass, halo mass, stellar mass, and star-formation rate (SFR), as well as the average environmental density (10th nearest neighbour density), of the galaxy sample at $z=0$. The upper triangle shows central versus satellite galaxies split.} 
    \label{fig:galaxy_properties_distributions}
\end{figure*}

\backmatter

\bmhead{Acknowledgements}
S.M. was supported by the STFC UCL Centre for Doctoral Training in Data Intensive Science (grant ST/P006736/1). O.L. acknowledges STFC Consolidated Grant ST/R000476/1 and the hospitality of All Souls College and the Physics Department, Oxford.

\bmhead{Author contribution} S.M. conceived and led the study, working on data curation, analysis, figures (with inputs from all the authors), and validation. S.M. and W.G.H. developed the causal model of galaxy formation and evolution, interpreted the results, and wrote the Article. S.M. and C.M.L. worked on the methodology, with the latter supervising the application of causal inference to astronomy. O.L. provided inputs on the text and figures and, more broadly, the `big picture'. All authors discussed, reviewed, and edited the Article.

\bmhead{Supplementary information} is provided.

\bmhead{Data availability} The data used in this study will be made available upon request.

\bmhead{Code availability} The code used in this study will be made available upon request.

\bmhead{Competing interests} The authors declare no competing interests.

\bmhead{Correspondence} and requests for materials should be addressed to S.M.


\bibliography{sn-bibliography}% common bib file

\begin{thebibliography}{100}
\expandafter\ifx\csname url\endcsname\relax
  \def\url#1{\burl{#1}}\fi
\expandafter\ifx\csname urlprefix\endcsname\relax\def\urlprefix{URL }\fi
\providecommand{\bibinfo}[2]{#2}
\providecommand{\eprint}[2][]{\url{#2}}
\providecommand{\doi}[1]{\url{https://doi.org/#1}}
\bibcommenthead

\bibitem{dressler_1980}
\bibinfo{author}{{Dressler}, A.}
\newblock \bibinfo{title}{{Galaxy morphology in rich clusters: implications for the formation and evolution of galaxies.}}
\newblock \emph{\bibinfo{journal}{\apj}} \textbf{\bibinfo{volume}{236}}, \bibinfo{pages}{351--365} (\bibinfo{year}{1980}).

\bibitem{kodama_2001}
\bibinfo{author}{{Kodama}, T.}, \bibinfo{author}{{Smail}, I.}, \bibinfo{author}{{Nakata}, F.}, \bibinfo{author}{{Okamura}, S.} \& \bibinfo{author}{{Bower}, R.~G.}
\newblock \bibinfo{title}{{The Transformation of Galaxies within the Large-Scale Structure around a z=0.41 Cluster}}.
\newblock \emph{\bibinfo{journal}{\apjl}} \textbf{\bibinfo{volume}{562}}, \bibinfo{pages}{L9--L13} (\bibinfo{year}{2001}).

\bibitem{gomez_2003}
\bibinfo{author}{{G{\'o}mez}, P.~L.} \emph{et~al.}
\newblock \bibinfo{title}{{Galaxy Star Formation as a Function of Environment in the Early Data Release of the Sloan Digital Sky Survey}}.
\newblock \emph{\bibinfo{journal}{\apj}} \textbf{\bibinfo{volume}{584}}, \bibinfo{pages}{210--227} (\bibinfo{year}{2003}).

\bibitem{hubble_1931}
\bibinfo{author}{{Hubble}, E.} \& \bibinfo{author}{{Humason}, M.~L.}
\newblock \bibinfo{title}{{The Velocity-Distance Relation among Extra-Galactic Nebulae}}.
\newblock \emph{\bibinfo{journal}{\apj}} \textbf{\bibinfo{volume}{74}}, \bibinfo{pages}{43} (\bibinfo{year}{1931}).

\bibitem{zwicky_1937}
\bibinfo{author}{{Zwicky}, F.}
\newblock \bibinfo{title}{{On the Masses of Nebulae and of Clusters of Nebulae}}.
\newblock \emph{\bibinfo{journal}{\apj}} \textbf{\bibinfo{volume}{86}}, \bibinfo{pages}{217} (\bibinfo{year}{1937}).

\bibitem{morgan_1961}
\bibinfo{author}{Morgan, W.~W.}
\newblock \bibinfo{title}{The classification of clusters of galaxies}.
\newblock \emph{\bibinfo{journal}{Proceedings of the National Academy of Sciences}} \textbf{\bibinfo{volume}{47}}, \bibinfo{pages}{905--906} (\bibinfo{year}{1961}).
\newblock \urlprefix\url{https://www.pnas.org/doi/abs/10.1073/pnas.47.7.905}.

\bibitem{abell_1965}
\bibinfo{author}{{Abell}, G.~O.}
\newblock \bibinfo{title}{{Clustering of Galaxies}}.
\newblock \emph{\bibinfo{journal}{\araa}} \textbf{\bibinfo{volume}{3}}, \bibinfo{pages}{1} (\bibinfo{year}{1965}).

\bibitem{oemler_1974}
\bibinfo{author}{{Oemler}, J., Augustus}.
\newblock \bibinfo{title}{{The Systematic Properties of Clusters of Galaxies. Photometry of 15 Clusters}}.
\newblock \emph{\bibinfo{journal}{\apj}} \textbf{\bibinfo{volume}{194}}, \bibinfo{pages}{1--20} (\bibinfo{year}{1974}).

\bibitem{davis_1976}
\bibinfo{author}{{Davis}, M.} \& \bibinfo{author}{{Geller}, M.~J.}
\newblock \bibinfo{title}{{Galaxy Correlations as a Function of Morphological Type}}.
\newblock \emph{\bibinfo{journal}{\apj}} \textbf{\bibinfo{volume}{208}}, \bibinfo{pages}{13--19} (\bibinfo{year}{1976}).

\bibitem{postman_1984}
\bibinfo{author}{{Postman}, M.} \& \bibinfo{author}{{Geller}, M.~J.}
\newblock \bibinfo{title}{{The morphology-density relation - The group connection.}}
\newblock \emph{\bibinfo{journal}{\apj}} \textbf{\bibinfo{volume}{281}}, \bibinfo{pages}{95--99} (\bibinfo{year}{1984}).

\bibitem{whitmore_1991}
\bibinfo{author}{{Whitmore}, B.~C.} \& \bibinfo{author}{{Gilmore}, D.~M.}
\newblock \bibinfo{title}{{On the Interpretation of the Morphology-Density Relation for Galaxies in Clusters}}.
\newblock \emph{\bibinfo{journal}{\apj}} \textbf{\bibinfo{volume}{367}}, \bibinfo{pages}{64} (\bibinfo{year}{1991}).

\bibitem{santiago_1992}
\bibinfo{author}{{Santiago}, B.~X.} \& \bibinfo{author}{{Strauss}, M.~A.}
\newblock \bibinfo{title}{{Large-Scale Morphological Segregation in the Center for Astrophysics Redshift Survey}}.
\newblock \emph{\bibinfo{journal}{\apj}} \textbf{\bibinfo{volume}{387}}, \bibinfo{pages}{9} (\bibinfo{year}{1992}).

\bibitem{whitmore_1993}
\bibinfo{author}{{Whitmore}, B.~C.}, \bibinfo{author}{{Gilmore}, D.~M.} \& \bibinfo{author}{{Jones}, C.}
\newblock \bibinfo{title}{{What Determines the Morphological Fractions in Clusters of Galaxies?}}
\newblock \emph{\bibinfo{journal}{\apj}} \textbf{\bibinfo{volume}{407}}, \bibinfo{pages}{489} (\bibinfo{year}{1993}).

\bibitem{hermit_1996}
\bibinfo{author}{{Hermit}, S.} \emph{et~al.}
\newblock \bibinfo{title}{{The two-point correlation function and morphological segregation in the Optical Redshift Survey.}}
\newblock \emph{\bibinfo{journal}{\mnras}} \textbf{\bibinfo{volume}{283}}, \bibinfo{pages}{709--720} (\bibinfo{year}{1996}).

\bibitem{guzzo_1997}
\bibinfo{author}{{Guzzo}, L.}, \bibinfo{author}{{Strauss}, M.~A.}, \bibinfo{author}{{Fisher}, K.~B.}, \bibinfo{author}{{Giovanelli}, R.} \& \bibinfo{author}{{Haynes}, M.~P.}
\newblock \bibinfo{title}{{Redshift-Space Distortions and the Real-Space Clustering of Different Galaxy Types}}.
\newblock \emph{\bibinfo{journal}{\apj}} \textbf{\bibinfo{volume}{489}}, \bibinfo{pages}{37--48} (\bibinfo{year}{1997}).

\bibitem{dominguez_2001}
\bibinfo{author}{{Dom{\'\i}nguez}, M.}, \bibinfo{author}{{Muriel}, H.} \& \bibinfo{author}{{Lambas}, D.~G.}
\newblock \bibinfo{title}{{Galaxy Morphological Segregation in Clusters: Local versus Global Conditions}}.
\newblock \emph{\bibinfo{journal}{\aj}} \textbf{\bibinfo{volume}{121}}, \bibinfo{pages}{1266--1274} (\bibinfo{year}{2001}).

\bibitem{giuricin_2001}
\bibinfo{author}{{Giuricin}, G.}, \bibinfo{author}{{Samurovi{\'c}}, S.}, \bibinfo{author}{{Girardi}, M.}, \bibinfo{author}{{Mezzetti}, M.} \& \bibinfo{author}{{Marinoni}, C.}
\newblock \bibinfo{title}{{The Redshift-Space Two-Point Correlation Functions of Galaxies and Groups in the Nearby Optical Galaxy Sample}}.
\newblock \emph{\bibinfo{journal}{\apj}} \textbf{\bibinfo{volume}{554}}, \bibinfo{pages}{857--872} (\bibinfo{year}{2001}).

\bibitem{treu_2003}
\bibinfo{author}{{Treu}, T.} \emph{et~al.}
\newblock \bibinfo{title}{{A Wide-Field Hubble Space Telescope Study of the Cluster Cl 0024+16 at z = 0.4. I. Morphological Distributions to 5 Mpc Radius}}.
\newblock \emph{\bibinfo{journal}{\apj}} \textbf{\bibinfo{volume}{591}}, \bibinfo{pages}{53--78} (\bibinfo{year}{2003}).

\bibitem{goto_2003}
\bibinfo{author}{{Goto}, T.} \emph{et~al.}
\newblock \bibinfo{title}{{The morphology-density relation in the Sloan Digital Sky Survey}}.
\newblock \emph{\bibinfo{journal}{\mnras}} \textbf{\bibinfo{volume}{346}}, \bibinfo{pages}{601--614} (\bibinfo{year}{2003}).

\bibitem{wilmer_1998}
\bibinfo{author}{{Willmer}, C. N.~A.}, \bibinfo{author}{{da Costa}, L.~N.} \& \bibinfo{author}{{Pellegrini}, P.~S.}
\newblock \bibinfo{title}{{Southern Sky Redshift Survey: Clustering of Local Galaxies}}.
\newblock \emph{\bibinfo{journal}{\aj}} \textbf{\bibinfo{volume}{115}}, \bibinfo{pages}{869--884} (\bibinfo{year}{1998}).

\bibitem{brown_2000}
\bibinfo{author}{{Brown}, M.~J.~I.}, \bibinfo{author}{{Webster}, R.~L.} \& \bibinfo{author}{{Boyle}, B.~J.}
\newblock \bibinfo{title}{{The clustering of colour-selected galaxies}}.
\newblock \emph{\bibinfo{journal}{\mnras}} \textbf{\bibinfo{volume}{317}}, \bibinfo{pages}{782--794} (\bibinfo{year}{2000}).

\bibitem{pimbblet_2002}
\bibinfo{author}{{Pimbblet}, K.~A.} \emph{et~al.}
\newblock \bibinfo{title}{{The Las Campanas/AAT Rich Cluster Survey - II. The environmental dependence of galaxy colours in clusters at z\raisebox{-0.5ex}\textasciitilde0.1}}.
\newblock \emph{\bibinfo{journal}{\mnras}} \textbf{\bibinfo{volume}{331}}, \bibinfo{pages}{333--350} (\bibinfo{year}{2002}).

\bibitem{zehavi_2002}
\bibinfo{author}{{Zehavi}, I.} \emph{et~al.}
\newblock \bibinfo{title}{{Galaxy Clustering in Early Sloan Digital Sky Survey Redshift Data}}.
\newblock \emph{\bibinfo{journal}{\apj}} \textbf{\bibinfo{volume}{571}}, \bibinfo{pages}{172--190} (\bibinfo{year}{2002}).

\bibitem{hogg_2004}
\bibinfo{author}{{Hogg}, D.~W.} \emph{et~al.}
\newblock \bibinfo{title}{{The Dependence on Environment of the Color-Magnitude Relation of Galaxies}}.
\newblock \emph{\bibinfo{journal}{\apjl}} \textbf{\bibinfo{volume}{601}}, \bibinfo{pages}{L29--L32} (\bibinfo{year}{2004}).

\bibitem{blanton_2005}
\bibinfo{author}{{Blanton}, M.~R.}, \bibinfo{author}{{Eisenstein}, D.}, \bibinfo{author}{{Hogg}, D.~W.}, \bibinfo{author}{{Schlegel}, D.~J.} \& \bibinfo{author}{{Brinkmann}, J.}
\newblock \bibinfo{title}{{Relationship between Environment and the Broadband Optical Properties of Galaxies in the Sloan Digital Sky Survey}}.
\newblock \emph{\bibinfo{journal}{\apj}} \textbf{\bibinfo{volume}{629}}, \bibinfo{pages}{143--157} (\bibinfo{year}{2005}).

\bibitem{martinez_2006}
\bibinfo{author}{{Mart{\'\i}nez}, H.~J.} \& \bibinfo{author}{{Muriel}, H.}
\newblock \bibinfo{title}{{Groups of galaxies: relationship between environment and galaxy properties}}.
\newblock \emph{\bibinfo{journal}{\mnras}} \textbf{\bibinfo{volume}{370}}, \bibinfo{pages}{1003--1007} (\bibinfo{year}{2006}).

\bibitem{balogh_1997}
\bibinfo{author}{{Balogh}, M.~L.}, \bibinfo{author}{{Morris}, S.~L.}, \bibinfo{author}{{Yee}, H.~K.~C.}, \bibinfo{author}{{Carlberg}, R.~G.} \& \bibinfo{author}{{Ellingson}, E.}
\newblock \bibinfo{title}{{Star Formation in Cluster Galaxies at 0.2 < Z < 0.55}}.
\newblock \emph{\bibinfo{journal}{\apjl}} \textbf{\bibinfo{volume}{488}}, \bibinfo{pages}{L75--L78} (\bibinfo{year}{1997}).

\bibitem{balogh_1998}
\bibinfo{author}{{Balogh}, M.~L.} \emph{et~al.}
\newblock \bibinfo{title}{{The Dependence of Cluster Galaxy Star Formation Rates on the Global Environment}}.
\newblock \emph{\bibinfo{journal}{\apjl}} \textbf{\bibinfo{volume}{504}}, \bibinfo{pages}{L75--L78} (\bibinfo{year}{1998}).

\bibitem{hashimoto_1998}
\bibinfo{author}{{Hashimoto}, Y.}, \bibinfo{author}{{Oemler}, J., Augustus}, \bibinfo{author}{{Lin}, H.} \& \bibinfo{author}{{Tucker}, D.~L.}
\newblock \bibinfo{title}{{The Influence of Environment on the Star Formation Rates of Galaxies}}.
\newblock \emph{\bibinfo{journal}{\apj}} \textbf{\bibinfo{volume}{499}}, \bibinfo{pages}{589--599} (\bibinfo{year}{1998}).

\bibitem{poggianti_1999}
\bibinfo{author}{{Poggianti}, B.~M.} \emph{et~al.}
\newblock \bibinfo{title}{{The Star Formation Histories of Galaxies in Distant Clusters}}.
\newblock \emph{\bibinfo{journal}{\apj}} \textbf{\bibinfo{volume}{518}}, \bibinfo{pages}{576--593} (\bibinfo{year}{1999}).

\bibitem{balogh_2000}
\bibinfo{author}{{Balogh}, M.~L.}, \bibinfo{author}{{Navarro}, J.~F.} \& \bibinfo{author}{{Morris}, S.~L.}
\newblock \bibinfo{title}{{The Origin of Star Formation Gradients in Rich Galaxy Clusters}}.
\newblock \emph{\bibinfo{journal}{\apj}} \textbf{\bibinfo{volume}{540}}, \bibinfo{pages}{113--121} (\bibinfo{year}{2000}).

\bibitem{couch_2001}
\bibinfo{author}{{Couch}, W.~J.} \emph{et~al.}
\newblock \bibinfo{title}{{A Low Global Star Formation Rate in the Rich Galaxy Cluster AC 114 at z=0.32}}.
\newblock \emph{\bibinfo{journal}{\apj}} \textbf{\bibinfo{volume}{549}}, \bibinfo{pages}{820--831} (\bibinfo{year}{2001}).

\bibitem{postman_2001}
\bibinfo{author}{{Postman}, M.}, \bibinfo{author}{{Lubin}, L.~M.} \& \bibinfo{author}{{Oke}, J.~B.}
\newblock \bibinfo{title}{{A Study of Nine High-Redshift Clusters of Galaxies. IV. Photometry and Spectra of Clusters 1324+3011 and 1604+4321}}.
\newblock \emph{\bibinfo{journal}{\aj}} \textbf{\bibinfo{volume}{122}}, \bibinfo{pages}{1125--1150} (\bibinfo{year}{2001}).

\bibitem{carter_2001}
\bibinfo{author}{{Carter}, B.~J.}, \bibinfo{author}{{Fabricant}, D.~G.}, \bibinfo{author}{{Geller}, M.~J.}, \bibinfo{author}{{Kurtz}, M.~J.} \& \bibinfo{author}{{McLean}, B.}
\newblock \bibinfo{title}{{Star Formation in a Complete Spectroscopic Survey of Galaxies}}.
\newblock \emph{\bibinfo{journal}{\apj}} \textbf{\bibinfo{volume}{559}}, \bibinfo{pages}{606--619} (\bibinfo{year}{2001}).

\bibitem{lewis_2002}
\bibinfo{author}{{Lewis}, I.} \emph{et~al.}
\newblock \bibinfo{title}{{The 2dF Galaxy Redshift Survey: the environmental dependence of galaxy star formation rates near clusters}}.
\newblock \emph{\bibinfo{journal}{\mnras}} \textbf{\bibinfo{volume}{334}}, \bibinfo{pages}{673--683} (\bibinfo{year}{2002}).

\bibitem{balogh_2004a}
\bibinfo{author}{{Balogh}, M.} \emph{et~al.}
\newblock \bibinfo{title}{{Galaxy ecology: groups and low-density environments in the SDSS and 2dFGRS}}.
\newblock \emph{\bibinfo{journal}{\mnras}} \textbf{\bibinfo{volume}{348}}, \bibinfo{pages}{1355--1372} (\bibinfo{year}{2004}).

\bibitem{tanaka_2004}
\bibinfo{author}{{Tanaka}, M.}, \bibinfo{author}{{Goto}, T.}, \bibinfo{author}{{Okamura}, S.}, \bibinfo{author}{{Shimasaku}, K.} \& \bibinfo{author}{{Brinkmann}, J.}
\newblock \bibinfo{title}{{The Environmental Dependence of Galaxy Properties in the Local Universe: Dependences on Luminosity, Local Density, and System Richness}}.
\newblock \emph{\bibinfo{journal}{\aj}} \textbf{\bibinfo{volume}{128}}, \bibinfo{pages}{2677--2695} (\bibinfo{year}{2004}).

\bibitem{rines_2005}
\bibinfo{author}{{Rines}, K.}, \bibinfo{author}{{Geller}, M.~J.}, \bibinfo{author}{{Kurtz}, M.~J.} \& \bibinfo{author}{{Diaferio}, A.}
\newblock \bibinfo{title}{{CAIRNS: The Cluster and Infall Region Nearby Survey. III. Environmental Dependence of H{\ensuremath{\alpha}} Properties of Galaxies}}.
\newblock \emph{\bibinfo{journal}{\aj}} \textbf{\bibinfo{volume}{130}}, \bibinfo{pages}{1482--1501} (\bibinfo{year}{2005}).

\bibitem{mcgaugh_1997}
\bibinfo{author}{{McGaugh}, S.~S.} \& \bibinfo{author}{{de Blok}, W.~J.~G.}
\newblock \bibinfo{title}{{Gas Mass Fractions and the Evolution of Spiral Galaxies}}.
\newblock \emph{\bibinfo{journal}{\apj}} \textbf{\bibinfo{volume}{481}}, \bibinfo{pages}{689--702} (\bibinfo{year}{1997}).

\bibitem{blanton_2003}
\bibinfo{author}{{Blanton}, M.~R.} \emph{et~al.}
\newblock \bibinfo{title}{{The Broadband Optical Properties of Galaxies with Redshifts 0.02<z<0.22}}.
\newblock \emph{\bibinfo{journal}{\apj}} \textbf{\bibinfo{volume}{594}}, \bibinfo{pages}{186--207} (\bibinfo{year}{2003}).

\bibitem{kauffmann_2003a}
\bibinfo{author}{{Kauffmann}, G.} \emph{et~al.}
\newblock \bibinfo{title}{{The dependence of star formation history and internal structure on stellar mass for {}10$^{5}$ low-redshift galaxies}}.
\newblock \emph{\bibinfo{journal}{\mnras}} \textbf{\bibinfo{volume}{341}}, \bibinfo{pages}{54--69} (\bibinfo{year}{2003}).

\bibitem{kauffmann_2003b}
\bibinfo{author}{{Kauffmann}, G.} \emph{et~al.}
\newblock \bibinfo{title}{{The host galaxies of active galactic nuclei}}.
\newblock \emph{\bibinfo{journal}{\mnras}} \textbf{\bibinfo{volume}{346}}, \bibinfo{pages}{1055--1077} (\bibinfo{year}{2003}).

\bibitem{baldry_2004a}
\bibinfo{author}{{Baldry}, I.~K.} \emph{et~al.}
\newblock \bibinfo{title}{{Quantifying the Bimodal Color-Magnitude Distribution of Galaxies}}.
\newblock \emph{\bibinfo{journal}{\apj}} \textbf{\bibinfo{volume}{600}}, \bibinfo{pages}{681--694} (\bibinfo{year}{2004}).

\bibitem{balogh_2001}
\bibinfo{author}{{Balogh}, M.~L.}, \bibinfo{author}{{Christlein}, D.}, \bibinfo{author}{{Zabludoff}, A.~I.} \& \bibinfo{author}{{Zaritsky}, D.}
\newblock \bibinfo{title}{{The Environmental Dependence of the Infrared Luminosity and Stellar Mass Functions}}.
\newblock \emph{\bibinfo{journal}{\apj}} \textbf{\bibinfo{volume}{557}}, \bibinfo{pages}{117--125} (\bibinfo{year}{2001}).

\bibitem{hogg_2003}
\bibinfo{author}{Hogg, D.~W.} \emph{et~al.}
\newblock \bibinfo{title}{The overdensities of galaxy environments as a function of luminosity and color}.
\newblock \emph{\bibinfo{journal}{The Astrophysical Journal}} \textbf{\bibinfo{volume}{585}}, \bibinfo{pages}{L5} (\bibinfo{year}{2003}).

\bibitem{mo_2004}
\bibinfo{author}{{Mo}, H.~J.}, \bibinfo{author}{{Yang}, X.}, \bibinfo{author}{{van den Bosch}, F.~C.} \& \bibinfo{author}{{Jing}, Y.~P.}
\newblock \bibinfo{title}{{The dependence of the galaxy luminosity function on large-scale environment}}.
\newblock \emph{\bibinfo{journal}{\mnras}} \textbf{\bibinfo{volume}{349}}, \bibinfo{pages}{205--212} (\bibinfo{year}{2004}).

\bibitem{croton_2005}
\bibinfo{author}{{Croton}, D.~J.} \emph{et~al.}
\newblock \bibinfo{title}{{The 2dF Galaxy Redshift Survey: luminosity functions by density environment and galaxy type}}.
\newblock \emph{\bibinfo{journal}{\mnras}} \textbf{\bibinfo{volume}{356}}, \bibinfo{pages}{1155--1167} (\bibinfo{year}{2005}).

\bibitem{hoyle_2005}
\bibinfo{author}{{Hoyle}, F.}, \bibinfo{author}{{Rojas}, R.~R.}, \bibinfo{author}{{Vogeley}, M.~S.} \& \bibinfo{author}{{Brinkmann}, J.}
\newblock \bibinfo{title}{{The Luminosity Function of Void Galaxies in the Sloan Digital Sky Survey}}.
\newblock \emph{\bibinfo{journal}{\apj}} \textbf{\bibinfo{volume}{620}}, \bibinfo{pages}{618--628} (\bibinfo{year}{2005}).

\bibitem{irwin_1995}
\bibinfo{author}{{Irwin}, J.~A.}
\newblock \bibinfo{title}{{Galaxies and Their Environments}}.
\newblock \emph{\bibinfo{journal}{\pasp}} \textbf{\bibinfo{volume}{107}}, \bibinfo{pages}{715} (\bibinfo{year}{1995}).

\bibitem{kauffmann_2004}
\bibinfo{author}{{Kauffmann}, G.} \emph{et~al.}
\newblock \bibinfo{title}{{The environmental dependence of the relations between stellar mass, structure, star formation and nuclear activity in galaxies}}.
\newblock \emph{\bibinfo{journal}{\mnras}} \textbf{\bibinfo{volume}{353}}, \bibinfo{pages}{713--731} (\bibinfo{year}{2004}).

\bibitem{balogh_2004b}
\bibinfo{author}{{Balogh}, M.~L.} \emph{et~al.}
\newblock \bibinfo{title}{{The Bimodal Galaxy Color Distribution: Dependence on Luminosity and Environment}}.
\newblock \emph{\bibinfo{journal}{\apjl}} \textbf{\bibinfo{volume}{615}}, \bibinfo{pages}{L101--L104} (\bibinfo{year}{2004}).

\bibitem{baldry_2004b}
\bibinfo{author}{{Baldry}, I.~K.}, \bibinfo{author}{{Balogh}, M.~L.}, \bibinfo{author}{{Bower}, R.}, \bibinfo{author}{{Glazebrook}, K.} \& \bibinfo{author}{{Nichol}, R.~C.}
\newblock \bibinfo{editor}{{Allen}, R.~E.}, \bibinfo{editor}{{Nanopoulos}, D.~V.} \& \bibinfo{editor}{{Pope}, C.~N.} (eds) \emph{\bibinfo{title}{{Color bimodality: Implications for galaxy evolution}}}.
\newblock (eds \bibinfo{editor}{{Allen}, R.~E.}, \bibinfo{editor}{{Nanopoulos}, D.~V.} \& \bibinfo{editor}{{Pope}, C.~N.}) \emph{\bibinfo{booktitle}{The New Cosmology: Conference on Strings and Cosmology}}, Vol. \bibinfo{volume}{743} of \emph{\bibinfo{series}{American Institute of Physics Conference Series}}, \bibinfo{pages}{106--119} (\bibinfo{year}{2004}).
\newblock \eprint{astro-ph/0410603}.

\bibitem{baldry_2006}
\bibinfo{author}{{Baldry}, I.~K.} \emph{et~al.}
\newblock \bibinfo{title}{{Galaxy bimodality versus stellar mass and environment}}.
\newblock \emph{\bibinfo{journal}{\mnras}} \textbf{\bibinfo{volume}{373}}, \bibinfo{pages}{469--483} (\bibinfo{year}{2006}).

\bibitem{weinmann_2006}
\bibinfo{author}{{Weinmann}, S.~M.}, \bibinfo{author}{{van den Bosch}, F.~C.}, \bibinfo{author}{{Yang}, X.} \& \bibinfo{author}{{Mo}, H.~J.}
\newblock \bibinfo{title}{{Properties of galaxy groups in the Sloan Digital Sky Survey - I. The dependence of colour, star formation and morphology on halo mass}}.
\newblock \emph{\bibinfo{journal}{\mnras}} \textbf{\bibinfo{volume}{366}}, \bibinfo{pages}{2--28} (\bibinfo{year}{2006}).

\bibitem{bamford_2009}
\bibinfo{author}{{Bamford}, S.~P.} \emph{et~al.}
\newblock \bibinfo{title}{{Galaxy Zoo: the dependence of morphology and colour on environment*}}.
\newblock \emph{\bibinfo{journal}{\mnras}} \textbf{\bibinfo{volume}{393}}, \bibinfo{pages}{1324--1352} (\bibinfo{year}{2009}).

\bibitem{skibba_2009}
\bibinfo{author}{{Skibba}, R.~A.} \emph{et~al.}
\newblock \bibinfo{title}{{Galaxy Zoo: disentangling the environmental dependence of morphology and colour}}.
\newblock \emph{\bibinfo{journal}{\mnras}} \textbf{\bibinfo{volume}{399}}, \bibinfo{pages}{966--982} (\bibinfo{year}{2009}).

\bibitem{lucia_2012}
\bibinfo{author}{{De Lucia}, G.}, \bibinfo{author}{{Weinmann}, S.}, \bibinfo{author}{{Poggianti}, B.~M.}, \bibinfo{author}{{Arag{\'o}n-Salamanca}, A.} \& \bibinfo{author}{{Zaritsky}, D.}
\newblock \bibinfo{title}{{The environmental history of group and cluster galaxies in a {\ensuremath{\Lambda}} cold dark matter universe}}.
\newblock \emph{\bibinfo{journal}{\mnras}} \textbf{\bibinfo{volume}{423}}, \bibinfo{pages}{1277--1292} (\bibinfo{year}{2012}).

\bibitem{vanderweele_2016}
\bibinfo{author}{VanderWeele, T.~J.}, \bibinfo{author}{Jackson, J.~W.} \& \bibinfo{author}{Li, S.}
\newblock \bibinfo{title}{Causal inference and longitudinal data: a case study of religion and mental health}.
\newblock \emph{\bibinfo{journal}{Social psychiatry and psychiatric epidemiology}} \textbf{\bibinfo{volume}{51}}, \bibinfo{pages}{1457--1466} (\bibinfo{year}{2016}).

\bibitem{lee_2017}
\bibinfo{author}{Lee, C.~M.} \& \bibinfo{author}{Spekkens, R.~W.}
\newblock \bibinfo{title}{Causal inference via algebraic geometry: feasibility tests for functional causal structures with two binary observed variables}.
\newblock \emph{\bibinfo{journal}{Journal of Causal Inference}} \textbf{\bibinfo{volume}{5}} (\bibinfo{year}{2017}).

\bibitem{lee_2020}
\bibinfo{author}{Gilligan-Lee, C.}
\newblock \bibinfo{title}{Causing trouble}.
\newblock \emph{\bibinfo{journal}{New Scientist}} \textbf{\bibinfo{volume}{246}}, \bibinfo{pages}{32--35} (\bibinfo{year}{2020}).

\bibitem{dhir_2020}
\bibinfo{author}{Dhir, A.} \& \bibinfo{author}{Lee, C.~M.}
\newblock \bibinfo{title}{Integrating overlapping datasets using bivariate causal discovery} (\bibinfo{year}{2020}).

\bibitem{lee_2022}
\bibinfo{author}{Gilligan-Lee, C.~M.}, \bibinfo{author}{Hart, C.}, \bibinfo{author}{Richens, J.} \& \bibinfo{author}{Johri, S.}
\newblock \bibinfo{title}{Leveraging directed causal discovery to detect latent common causes in cause-effect pairs}.
\newblock \emph{\bibinfo{journal}{IEEE Transactions on Neural Networks and Learning Systems}}  (\bibinfo{year}{2022}).

\bibitem{jeunen_2022}
\bibinfo{author}{Jeunen, O.}, \bibinfo{author}{Gilligan-Lee, C.}, \bibinfo{author}{Mehrotra, R.} \& \bibinfo{author}{Lalmas, M.}
\newblock \bibinfo{title}{Disentangling causal effects from sets of interventions in the presence of unobserved confounders}.
\newblock \emph{\bibinfo{journal}{Advances in Neural Information Processing Systems}} \textbf{\bibinfo{volume}{35}}, \bibinfo{pages}{27850--27861} (\bibinfo{year}{2022}).

\bibitem{zeitler_2023}
\bibinfo{author}{Zeitler, J.}, \bibinfo{author}{Vlontzos, A.} \& \bibinfo{author}{Gilligan-Lee, C.~M.}
\newblock \bibinfo{title}{Non-parametric identifiability and sensitivity analysis of synthetic control models} (\bibinfo{year}{2023}).

\bibitem{goffrier_2023}
\bibinfo{author}{Van~Goffrier, G.}, \bibinfo{author}{Maystre, L.} \& \bibinfo{author}{Gilligan-Lee, C.~M.}
\newblock \bibinfo{title}{Estimating long-term causal effects from short-term experiments and long-term observational data with unobserved confounding} (\bibinfo{year}{2023}).

\bibitem{pearl_2010}
\bibinfo{author}{Pearl, J.}
\newblock \bibinfo{title}{The foundations of causal inference}.
\newblock \emph{\bibinfo{journal}{Sociological Methodology}} \textbf{\bibinfo{volume}{40}}, \bibinfo{pages}{75--149} (\bibinfo{year}{2010}).

\bibitem{angrist_1991}
\bibinfo{author}{Angrist, J.~D.} \& \bibinfo{author}{Krueger, A.~B.}
\newblock \bibinfo{title}{Does compulsory school attendance affect schooling and earnings?}
\newblock \emph{\bibinfo{journal}{The Quarterly Journal of Economics}} \textbf{\bibinfo{volume}{106}}, \bibinfo{pages}{979--1014} (\bibinfo{year}{1991}).

\bibitem{card_1993}
\bibinfo{author}{Card, D.} \& \bibinfo{author}{Krueger, A.}
\newblock \bibinfo{title}{Minimum wages and employment: A case study of the fast food industry in new jersey and pennsylvania}.
\newblock \emph{\bibinfo{journal}{American Economic Review}} \textbf{\bibinfo{volume}{84}} (\bibinfo{year}{1993}).

\bibitem{cengiz_2019}
\bibinfo{author}{Cengiz, D.}, \bibinfo{author}{Dube, A.}, \bibinfo{author}{Lindner, A.} \& \bibinfo{author}{Zipperer, B.}
\newblock \bibinfo{title}{The effect of minimum wages on low-wage jobs}.
\newblock \emph{\bibinfo{journal}{The Quarterly Journal of Economics}} \textbf{\bibinfo{volume}{134}}, \bibinfo{pages}{1405--1454} (\bibinfo{year}{2019}).

\bibitem{kam_2008}
\bibinfo{author}{Kam, C.~D.} \& \bibinfo{author}{Palmer, C.~L.}
\newblock \bibinfo{title}{Reconsidering the effects of education on political participation}.
\newblock \emph{\bibinfo{journal}{The Journal of Politics}} \textbf{\bibinfo{volume}{70}}, \bibinfo{pages}{612--631} (\bibinfo{year}{2008}).

\bibitem{angrist_1999}
\bibinfo{author}{Angrist, J.~D.} \& \bibinfo{author}{Lavy, V.}
\newblock \bibinfo{title}{Using maimonides' rule to estimate the effect of class size on scholastic achievement}.
\newblock \emph{\bibinfo{journal}{The Quarterly journal of economics}} \textbf{\bibinfo{volume}{114}}, \bibinfo{pages}{533--575} (\bibinfo{year}{1999}).

\bibitem{carlsson_2015}
\bibinfo{author}{Carlsson, M.}, \bibinfo{author}{Dahl, G.~B.}, \bibinfo{author}{{\"O}ckert, B.} \& \bibinfo{author}{Rooth, D.-O.}
\newblock \bibinfo{title}{The effect of schooling on cognitive skills}.
\newblock \emph{\bibinfo{journal}{Review of Economics and Statistics}} \textbf{\bibinfo{volume}{97}}, \bibinfo{pages}{533--547} (\bibinfo{year}{2015}).

\bibitem{ghosh_2018}
\bibinfo{author}{Ghosh, A.}, \bibinfo{author}{Simon, K.} \& \bibinfo{author}{Sommers, B.}
\newblock \bibinfo{title}{The effect of health insurance on prescription drug use among low-income adults:evidence from recent medicaid expansions}.
\newblock \emph{\bibinfo{journal}{Journal of Health Economics}} \textbf{\bibinfo{volume}{63}} (\bibinfo{year}{2018}).

\bibitem{doll_1950}
\bibinfo{author}{Doll, R.} \& \bibinfo{author}{Hill, A.~B.}
\newblock \bibinfo{title}{Smoking and carcinoma of the lung}.
\newblock \emph{\bibinfo{journal}{British medical journal}} \textbf{\bibinfo{volume}{2}}, \bibinfo{pages}{739} (\bibinfo{year}{1950}).

\bibitem{chay_2003}
\bibinfo{author}{Chay, K.~Y.} \& \bibinfo{author}{Greenstone, M.}
\newblock \bibinfo{title}{The impact of air pollution on infant mortality: evidence from geographic variation in pollution shocks induced by a recession}.
\newblock \emph{\bibinfo{journal}{The quarterly journal of economics}} \textbf{\bibinfo{volume}{118}}, \bibinfo{pages}{1121--1167} (\bibinfo{year}{2003}).

\bibitem{clark_2013}
\bibinfo{author}{Clark, D.} \& \bibinfo{author}{Royer, H.}
\newblock \bibinfo{title}{The effect of education on adult mortality and health: Evidence from britain}.
\newblock \emph{\bibinfo{journal}{American Economic Review}} \textbf{\bibinfo{volume}{103}}, \bibinfo{pages}{2087--2120} (\bibinfo{year}{2013}).

\bibitem{desouza_2022}
\bibinfo{author}{Desouza, P.~N.} \emph{et~al.}
\newblock \bibinfo{title}{Robust relationship between ambient air pollution and infant mortality in india}.
\newblock \emph{\bibinfo{journal}{Science of The Total Environment}} \textbf{\bibinfo{volume}{815}}, \bibinfo{pages}{152755} (\bibinfo{year}{2022}).

\bibitem{allen_2017}
\bibinfo{author}{Allen, J.-M.~A.}, \bibinfo{author}{Barrett, J.}, \bibinfo{author}{Horsman, D.~C.}, \bibinfo{author}{Lee, C.~M.} \& \bibinfo{author}{Spekkens, R.~W.}
\newblock \bibinfo{title}{Quantum common causes and quantum causal models}.
\newblock \emph{\bibinfo{journal}{Physical Review X}} \textbf{\bibinfo{volume}{7}}, \bibinfo{pages}{031021} (\bibinfo{year}{2017}).

\bibitem{lee_2018}
\bibinfo{author}{Lee, C.~M.} \& \bibinfo{author}{Hoban, M.~J.}
\newblock \bibinfo{title}{Towards device-independent information processing on general quantum networks}.
\newblock \emph{\bibinfo{journal}{Physical review letters}} \textbf{\bibinfo{volume}{120}}, \bibinfo{pages}{020504} (\bibinfo{year}{2018}).

\bibitem{lee_2019}
\bibinfo{author}{Lee, C.~M.}
\newblock \bibinfo{title}{Device-independent certification of non-classical joint measurements via causal models}.
\newblock \emph{\bibinfo{journal}{npj Quantum Information}} \textbf{\bibinfo{volume}{5}}, \bibinfo{pages}{34} (\bibinfo{year}{2019}).

\bibitem{scholkopf_2015}
\bibinfo{author}{{Sch{\"o}lkopf}, B.} \emph{et~al.}
\newblock \bibinfo{title}{{Removing systematic errors for exoplanet search via latent causes}}.
\newblock \emph{\bibinfo{journal}{arXiv e-prints}} \bibinfo{pages}{arXiv:1505.03036} (\bibinfo{year}{2015}).

\bibitem{wang_2016}
\bibinfo{author}{{Wang}, D.}, \bibinfo{author}{{Hogg}, D.~W.}, \bibinfo{author}{{Foreman-Mackey}, D.} \& \bibinfo{author}{{Sch{\"o}lkopf}, B.}
\newblock \bibinfo{title}{{A Causal, Data-driven Approach to Modeling the Kepler Data}}.
\newblock \emph{\bibinfo{journal}{\pasp}} \textbf{\bibinfo{volume}{128}}, \bibinfo{pages}{094503} (\bibinfo{year}{2016}).

\bibitem{pasquato_2019}
\bibinfo{author}{{Pasquato}, M.} \& \bibinfo{author}{{Matsiuk}, N.}
\newblock \bibinfo{title}{{Quasi-experimental Approach to Open Cluster Dynamics}}.
\newblock \emph{\bibinfo{journal}{Research Notes of the American Astronomical Society}} \textbf{\bibinfo{volume}{3}}, \bibinfo{pages}{179} (\bibinfo{year}{2019}).

\bibitem{pang_2021}
\bibinfo{author}{{Pang}, X.} \emph{et~al.}
\newblock \bibinfo{title}{{Disruption of Hierarchical Clustering in the Vela OB2 Complex and the Cluster Pair Collinder 135 and UBC 7 with Gaia EDR3: Evidence of Supernova Quenching}}.
\newblock \emph{\bibinfo{journal}{\apj}} \textbf{\bibinfo{volume}{923}}, \bibinfo{pages}{20} (\bibinfo{year}{2021}).

\bibitem{pasquato_2023}
\bibinfo{author}{{Pasquato}, M.}, \bibinfo{author}{{Jin}, Z.}, \bibinfo{author}{{Lemos}, P.}, \bibinfo{author}{{Davis}, B.~L.} \& \bibinfo{author}{{Macci{\`o}}, A.~V.}
\newblock \bibinfo{title}{{Causa prima: cosmology meets causal discovery for the first time}}.
\newblock \emph{\bibinfo{journal}{arXiv e-prints}} \bibinfo{pages}{arXiv:2311.15160} (\bibinfo{year}{2023}).

\bibitem{jin_2024}
\bibinfo{author}{{Jin}, Z.} \emph{et~al.}
\newblock \bibinfo{title}{{A Data-driven Discovery of the Causal Connection between Galaxy and Black Hole Evolution}}.
\newblock \emph{\bibinfo{journal}{arXiv e-prints}} \bibinfo{pages}{arXiv:2410.00965} (\bibinfo{year}{2024}).

\bibitem{mucesh_2021}
\bibinfo{author}{{Mucesh}, S.} \emph{et~al.}
\newblock \bibinfo{title}{{A machine learning approach to galaxy properties: joint redshift-stellar mass probability distributions with Random Forest}}.
\newblock \emph{\bibinfo{journal}{\mnras}} \textbf{\bibinfo{volume}{502}}, \bibinfo{pages}{2770--2786} (\bibinfo{year}{2021}).

\bibitem{baron_2019}
\bibinfo{author}{{Baron}, D.}
\newblock \bibinfo{title}{{Machine Learning in Astronomy: a practical overview}}.
\newblock \emph{\bibinfo{journal}{arXiv e-prints}} \bibinfo{pages}{arXiv:1904.07248} (\bibinfo{year}{2019}).

\bibitem{fluke_2020}
\bibinfo{author}{{Fluke}, C.~J.} \& \bibinfo{author}{{Jacobs}, C.}
\newblock \bibinfo{title}{{Surveying the reach and maturity of machine learning and artificial intelligence in astronomy}}.
\newblock \emph{\bibinfo{journal}{WIREs Data Mining and Knowledge Discovery}} \textbf{\bibinfo{volume}{10}}, \bibinfo{pages}{e1349} (\bibinfo{year}{2020}).

\bibitem{teimoorinia_2016}
\bibinfo{author}{{Teimoorinia}, H.}, \bibinfo{author}{{Bluck}, A. F.~L.} \& \bibinfo{author}{{Ellison}, S.~L.}
\newblock \bibinfo{title}{{An artificial neural network approach for ranking quenching parameters in central galaxies}}.
\newblock \emph{\bibinfo{journal}{\mnras}} \textbf{\bibinfo{volume}{457}}, \bibinfo{pages}{2086--2106} (\bibinfo{year}{2016}).

\bibitem{bluck_2019}
\bibinfo{author}{{Bluck}, A. F.~L.} \emph{et~al.}
\newblock \bibinfo{title}{{What shapes a galaxy? - unraveling the role of mass, environment, and star formation in forming galactic structure}}.
\newblock \emph{\bibinfo{journal}{\mnras}} \textbf{\bibinfo{volume}{485}}, \bibinfo{pages}{666--696} (\bibinfo{year}{2019}).

\bibitem{bluck_2020a}
\bibinfo{author}{{Bluck}, A. F.~L.} \emph{et~al.}
\newblock \bibinfo{title}{{Are galactic star formation and quenching governed by local, global, or environmental phenomena?}}
\newblock \emph{\bibinfo{journal}{\mnras}} \textbf{\bibinfo{volume}{492}}, \bibinfo{pages}{96--139} (\bibinfo{year}{2020}).

\bibitem{bluck_2020b}
\bibinfo{author}{{Bluck}, A. F.~L.} \emph{et~al.}
\newblock \bibinfo{title}{{How do central and satellite galaxies quench? - Insights from spatially resolved spectroscopy in the MaNGA survey}}.
\newblock \emph{\bibinfo{journal}{\mnras}} \textbf{\bibinfo{volume}{499}}, \bibinfo{pages}{230--268} (\bibinfo{year}{2020}).

\bibitem{bluck_2022}
\bibinfo{author}{{Bluck}, A. F.~L.} \emph{et~al.}
\newblock \bibinfo{title}{{The quenching of galaxies, bulges, and disks since cosmic noon. A machine learning approach for identifying causality in astronomical data}}.
\newblock \emph{\bibinfo{journal}{\aap}} \textbf{\bibinfo{volume}{659}}, \bibinfo{pages}{A160} (\bibinfo{year}{2022}).

\bibitem{brownson_2022}
\bibinfo{author}{{Brownson}, S.}, \bibinfo{author}{{Bluck}, A. F.~L.}, \bibinfo{author}{{Maiolino}, R.} \& \bibinfo{author}{{Jones}, G.~C.}
\newblock \bibinfo{title}{{What drives galaxy quenching? A deep connection between galaxy kinematics and quenching in the local Universe}}.
\newblock \emph{\bibinfo{journal}{\mnras}} \textbf{\bibinfo{volume}{511}}, \bibinfo{pages}{1913--1941} (\bibinfo{year}{2022}).

\bibitem{piotrowska_2022}
\bibinfo{author}{{Piotrowska}, J.~M.}, \bibinfo{author}{{Bluck}, A. F.~L.}, \bibinfo{author}{{Maiolino}, R.} \& \bibinfo{author}{{Peng}, Y.}
\newblock \bibinfo{title}{{On the quenching of star formation in observed and simulated central galaxies: evidence for the role of integrated AGN feedback}}.
\newblock \emph{\bibinfo{journal}{\mnras}} \textbf{\bibinfo{volume}{512}}, \bibinfo{pages}{1052--1090} (\bibinfo{year}{2022}).

\bibitem{mcgibbon_2022}
\bibinfo{author}{{McGibbon}, R.~J.} \& \bibinfo{author}{{Khochfar}, S.}
\newblock \bibinfo{title}{{Multi-epoch machine learning 1: Unravelling nature versus nurture for galaxy formation}}.
\newblock \emph{\bibinfo{journal}{\mnras}} \textbf{\bibinfo{volume}{513}}, \bibinfo{pages}{5423--5437} (\bibinfo{year}{2022}).

\bibitem{kaddour_2022}
\bibinfo{author}{{Kaddour}, J.}, \bibinfo{author}{{Lynch}, A.}, \bibinfo{author}{{Liu}, Q.}, \bibinfo{author}{{Kusner}, M.~J.} \& \bibinfo{author}{{Silva}, R.}
\newblock \bibinfo{title}{{Causal Machine Learning: A Survey and Open Problems}}.
\newblock \emph{\bibinfo{journal}{arXiv e-prints}} \bibinfo{pages}{arXiv:2206.15475} (\bibinfo{year}{2022}).

\bibitem{richens_2020}
\bibinfo{author}{Richens, J.~G.}, \bibinfo{author}{Lee, C.~M.} \& \bibinfo{author}{Johri, S.}
\newblock \bibinfo{title}{Improving the accuracy of medical diagnosis with causal machine learning}.
\newblock \emph{\bibinfo{journal}{Nature communications}} \textbf{\bibinfo{volume}{11}}, \bibinfo{pages}{1--9} (\bibinfo{year}{2020}).

\bibitem{perov_2020}
\bibinfo{author}{Perov, Y.} \emph{et~al.}
\newblock \bibinfo{title}{Multiverse: causal reasoning using importance sampling in probabilistic programming} (\bibinfo{year}{2020}).

\bibitem{reynaud_2022}
\bibinfo{author}{Reynaud, H.} \emph{et~al.}
\newblock \bibinfo{title}{D'artagnan: Counterfactual video generation}.
\newblock \emph{\bibinfo{journal}{arXiv preprint arXiv:2206.01651}}  (\bibinfo{year}{2022}).

\bibitem{vlontzos_2023}
\bibinfo{author}{Vlontzos, A.}, \bibinfo{author}{Kainz, B.} \& \bibinfo{author}{Gilligan-Lee, C.~M.}
\newblock \bibinfo{title}{Estimating categorical counterfactuals via deep twin networks}.
\newblock \emph{\bibinfo{journal}{Nature Machine Intelligence}} \textbf{\bibinfo{volume}{5}}, \bibinfo{pages}{159--168} (\bibinfo{year}{2023}).

\bibitem{sanchez_2022}
\bibinfo{author}{{Sanchez}, P.} \emph{et~al.}
\newblock \bibinfo{title}{{Causal machine learning for healthcare and precision medicine}}.
\newblock \emph{\bibinfo{journal}{Royal Society Open Science}} \textbf{\bibinfo{volume}{9}}, \bibinfo{pages}{220638} (\bibinfo{year}{2022}).

\bibitem{pillepich_2018a}
\bibinfo{author}{{Pillepich}, A.} \emph{et~al.}
\newblock \bibinfo{title}{{First results from the IllustrisTNG simulations: the stellar mass content of groups and clusters of galaxies}}.
\newblock \emph{\bibinfo{journal}{\mnras}} \textbf{\bibinfo{volume}{475}}, \bibinfo{pages}{648--675} (\bibinfo{year}{2018}).

\bibitem{springel_2018}
\bibinfo{author}{{Springel}, V.} \emph{et~al.}
\newblock \bibinfo{title}{{First results from the IllustrisTNG simulations: matter and galaxy clustering}}.
\newblock \emph{\bibinfo{journal}{\mnras}} \textbf{\bibinfo{volume}{475}}, \bibinfo{pages}{676--698} (\bibinfo{year}{2018}).

\bibitem{nelson_2018}
\bibinfo{author}{{Nelson}, D.} \emph{et~al.}
\newblock \bibinfo{title}{{First results from the IllustrisTNG simulations: the galaxy colour bimodality}}.
\newblock \emph{\bibinfo{journal}{\mnras}} \textbf{\bibinfo{volume}{475}}, \bibinfo{pages}{624--647} (\bibinfo{year}{2018}).

\bibitem{naiman_2018}
\bibinfo{author}{{Naiman}, J.~P.} \emph{et~al.}
\newblock \bibinfo{title}{{First results from the IllustrisTNG simulations: a tale of two elements - chemical evolution of magnesium and europium}}.
\newblock \emph{\bibinfo{journal}{\mnras}} \textbf{\bibinfo{volume}{477}}, \bibinfo{pages}{1206--1224} (\bibinfo{year}{2018}).

\bibitem{marinacci_2018}
\bibinfo{author}{{Marinacci}, F.} \emph{et~al.}
\newblock \bibinfo{title}{{First results from the IllustrisTNG simulations: radio haloes and magnetic fields}}.
\newblock \emph{\bibinfo{journal}{\mnras}} \textbf{\bibinfo{volume}{480}}, \bibinfo{pages}{5113--5139} (\bibinfo{year}{2018}).

\bibitem{nelson_2019}
\bibinfo{author}{{Nelson}, D.} \emph{et~al.}
\newblock \bibinfo{title}{{The IllustrisTNG simulations: public data release}}.
\newblock \emph{\bibinfo{journal}{Computational Astrophysics and Cosmology}} \textbf{\bibinfo{volume}{6}}, \bibinfo{pages}{2} (\bibinfo{year}{2019}).

\bibitem{chalmers_1981}
\bibinfo{author}{Chalmers, T.~C.} \emph{et~al.}
\newblock \bibinfo{title}{A method for assessing the quality of a randomized control trial}.
\newblock \emph{\bibinfo{journal}{Controlled clinical trials}} \textbf{\bibinfo{volume}{2}}, \bibinfo{pages}{31--49} (\bibinfo{year}{1981}).

\bibitem{white_1978}
\bibinfo{author}{{White}, S.~D.~M.} \& \bibinfo{author}{{Rees}, M.~J.}
\newblock \bibinfo{title}{{Core condensation in heavy halos: a two-stage theory for galaxy formation and clustering.}}
\newblock \emph{\bibinfo{journal}{\mnras}} \textbf{\bibinfo{volume}{183}}, \bibinfo{pages}{341--358} (\bibinfo{year}{1978}).

\bibitem{efstathiou_1983}
\bibinfo{author}{{Efstathiou}, G.} \& \bibinfo{author}{{Silk}, J.}
\newblock \bibinfo{title}{{The Formation of Galaxies}}.
\newblock \emph{\bibinfo{journal}{\fcp}} \textbf{\bibinfo{volume}{9}}, \bibinfo{pages}{1--138} (\bibinfo{year}{1983}).

\bibitem{blumenthal_1984}
\bibinfo{author}{{Blumenthal}, G.~R.}, \bibinfo{author}{{Faber}, S.~M.}, \bibinfo{author}{{Primack}, J.~R.} \& \bibinfo{author}{{Rees}, M.~J.}
\newblock \bibinfo{title}{{Formation of galaxies and large-scale structure with cold dark matter.}}
\newblock \emph{\bibinfo{journal}{\nat}} \textbf{\bibinfo{volume}{311}}, \bibinfo{pages}{517--525} (\bibinfo{year}{1984}).

\bibitem{white_1991}
\bibinfo{author}{{White}, S. D.~M.} \& \bibinfo{author}{{Frenk}, C.~S.}
\newblock \bibinfo{title}{{Galaxy Formation through Hierarchical Clustering}}.
\newblock \emph{\bibinfo{journal}{\apj}} \textbf{\bibinfo{volume}{379}}, \bibinfo{pages}{52} (\bibinfo{year}{1991}).

\bibitem{cole_1991}
\bibinfo{author}{{Cole}, S.}
\newblock \bibinfo{title}{{Modeling Galaxy Formation in Evolving Dark Matter Halos}}.
\newblock \emph{\bibinfo{journal}{\apj}} \textbf{\bibinfo{volume}{367}}, \bibinfo{pages}{45} (\bibinfo{year}{1991}).

\bibitem{kauffmann_1993}
\bibinfo{author}{{Kauffmann}, G.}, \bibinfo{author}{{White}, S.~D.~M.} \& \bibinfo{author}{{Guiderdoni}, B.}
\newblock \bibinfo{title}{{The formation and evolution of galaxies within merging dark matter haloes.}}
\newblock \emph{\bibinfo{journal}{\mnras}} \textbf{\bibinfo{volume}{264}}, \bibinfo{pages}{201--218} (\bibinfo{year}{1993}).

\bibitem{cole_1994}
\bibinfo{author}{{Cole}, S.}, \bibinfo{author}{{Aragon-Salamanca}, A.}, \bibinfo{author}{{Frenk}, C.~S.}, \bibinfo{author}{{Navarro}, J.~F.} \& \bibinfo{author}{{Zepf}, S.~E.}
\newblock \bibinfo{title}{{A recipe for galaxy formation.}}
\newblock \emph{\bibinfo{journal}{\mnras}} \textbf{\bibinfo{volume}{271}}, \bibinfo{pages}{781--806} (\bibinfo{year}{1994}).

\bibitem{kauffmann_1999}
\bibinfo{author}{{Kauffmann}, G.}, \bibinfo{author}{{Colberg}, J.~M.}, \bibinfo{author}{{Diaferio}, A.} \& \bibinfo{author}{{White}, S. D.~M.}
\newblock \bibinfo{title}{{Clustering of galaxies in a hierarchical universe - I. Methods and results at z=0}}.
\newblock \emph{\bibinfo{journal}{\mnras}} \textbf{\bibinfo{volume}{303}}, \bibinfo{pages}{188--206} (\bibinfo{year}{1999}).

\bibitem{somerville_1999}
\bibinfo{author}{{Somerville}, R.~S.} \& \bibinfo{author}{{Primack}, J.~R.}
\newblock \bibinfo{title}{{Semi-analytic modelling of galaxy formation: the local Universe}}.
\newblock \emph{\bibinfo{journal}{\mnras}} \textbf{\bibinfo{volume}{310}}, \bibinfo{pages}{1087--1110} (\bibinfo{year}{1999}).

\bibitem{springel_2001}
\bibinfo{author}{{Springel}, V.}, \bibinfo{author}{{White}, S. D.~M.}, \bibinfo{author}{{Tormen}, G.} \& \bibinfo{author}{{Kauffmann}, G.}
\newblock \bibinfo{title}{{Populating a cluster of galaxies - I. Results at [formmu2]z=0}}.
\newblock \emph{\bibinfo{journal}{\mnras}} \textbf{\bibinfo{volume}{328}}, \bibinfo{pages}{726--750} (\bibinfo{year}{2001}).

\bibitem{hatton_2003}
\bibinfo{author}{{Hatton}, S.} \emph{et~al.}
\newblock \bibinfo{title}{{GALICS- I. A hybrid N-body/semi-analytic model of hierarchical galaxy formation}}.
\newblock \emph{\bibinfo{journal}{\mnras}} \textbf{\bibinfo{volume}{343}}, \bibinfo{pages}{75--106} (\bibinfo{year}{2003}).

\bibitem{springel_2005}
\bibinfo{author}{{Springel}, V.} \emph{et~al.}
\newblock \bibinfo{title}{{Simulations of the formation, evolution and clustering of galaxies and quasars}}.
\newblock \emph{\bibinfo{journal}{\nat}} \textbf{\bibinfo{volume}{435}}, \bibinfo{pages}{629--636} (\bibinfo{year}{2005}).

\bibitem{kang_2005}
\bibinfo{author}{{Kang}, X.}, \bibinfo{author}{{Jing}, Y.~P.}, \bibinfo{author}{{Mo}, H.~J.} \& \bibinfo{author}{{B{\"o}rner}, G.}
\newblock \bibinfo{title}{{Semianalytical Model of Galaxy Formation with High-Resolution N-Body Simulations}}.
\newblock \emph{\bibinfo{journal}{\apj}} \textbf{\bibinfo{volume}{631}}, \bibinfo{pages}{21--40} (\bibinfo{year}{2005}).

\bibitem{lu_2011}
\bibinfo{author}{{Lu}, Y.}, \bibinfo{author}{{Mo}, H.~J.}, \bibinfo{author}{{Weinberg}, M.~D.} \& \bibinfo{author}{{Katz}, N.}
\newblock \bibinfo{title}{{A Bayesian approach to the semi-analytic model of galaxy formation: methodology}}.
\newblock \emph{\bibinfo{journal}{\mnras}} \textbf{\bibinfo{volume}{416}}, \bibinfo{pages}{1949--1964} (\bibinfo{year}{2011}).

\bibitem{benson_2012}
\bibinfo{author}{{Benson}, A.~J.}
\newblock \bibinfo{title}{{G ALACTICUS: A semi-analytic model of galaxy formation}}.
\newblock \emph{\bibinfo{journal}{\na}} \textbf{\bibinfo{volume}{17}}, \bibinfo{pages}{175--197} (\bibinfo{year}{2012}).

\bibitem{herniques_2015}
\bibinfo{author}{{Henriques}, B. M.~B.} \emph{et~al.}
\newblock \bibinfo{title}{{Galaxy formation in the Planck cosmology - I. Matching the observed evolution of star formation rates, colours and stellar masses}}.
\newblock \emph{\bibinfo{journal}{\mnras}} \textbf{\bibinfo{volume}{451}}, \bibinfo{pages}{2663--2680} (\bibinfo{year}{2015}).

\bibitem{baugh_2005}
\bibinfo{author}{{Baugh}, C.~M.}
\newblock \bibinfo{title}{{A primer on hierarchical galaxy formation: the semi-analytical approach}}.
\newblock \emph{\bibinfo{journal}{Reports on Progress in Physics}} \textbf{\bibinfo{volume}{69}}, \bibinfo{pages}{3101--3156} (\bibinfo{year}{2006}).

\bibitem{benson_2010}
\bibinfo{author}{{Benson}, A.~J.}
\newblock \bibinfo{title}{{Galaxy formation theory}}.
\newblock \emph{\bibinfo{journal}{\physrep}} \textbf{\bibinfo{volume}{495}}, \bibinfo{pages}{33--86} (\bibinfo{year}{2010}).

\bibitem{wright_1921}
\bibinfo{author}{Wright, S.}
\newblock \bibinfo{title}{Correlation and causation}.
\newblock \emph{\bibinfo{journal}{Journal of agricultural research}} \textbf{\bibinfo{volume}{20}}, \bibinfo{pages}{557--585} (\bibinfo{year}{1921}).

\bibitem{reichenbach_1956}
\bibinfo{author}{Reichenbach, H.}
\newblock \emph{\bibinfo{title}{The Direction of Time}}  (\bibinfo{publisher}{Mineola, N.Y.: Dover Publications}, \bibinfo{year}{1956}).

\bibitem{pearl_2009a}
\bibinfo{author}{Pearl, J.}
\newblock \emph{\bibinfo{title}{Causality}}  (\bibinfo{publisher}{Cambridge university press}, \bibinfo{year}{2009}).

\bibitem{hagedoorn_2021}
\bibinfo{author}{Hagedoorn, P.} \& \bibinfo{author}{Helbich, M.}
\newblock \bibinfo{title}{Longitudinal exposure assessments of neighbourhood effects in health research: What can be learned from people's residential histories?}
\newblock \emph{\bibinfo{journal}{Health \& Place}} \textbf{\bibinfo{volume}{68}}, \bibinfo{pages}{102543} (\bibinfo{year}{2021}).

\bibitem{naimi_2017}
\bibinfo{author}{Naimi, A.~I.}, \bibinfo{author}{Cole, S.~R.} \& \bibinfo{author}{Kennedy, E.~H.}
\newblock \bibinfo{title}{An introduction to g methods}.
\newblock \emph{\bibinfo{journal}{International journal of epidemiology}} \textbf{\bibinfo{volume}{46}}, \bibinfo{pages}{756--762} (\bibinfo{year}{2017}).

\bibitem{robins_2000}
\bibinfo{author}{Robins, J.~M.}, \bibinfo{author}{Hernan, M.~A.} \& \bibinfo{author}{Brumback, B.}
\newblock \bibinfo{title}{Marginal structural models and causal inference in epidemiology} (\bibinfo{year}{2000}).

\bibitem{fewell_2004}
\bibinfo{author}{Fewell, Z.} \emph{et~al.}
\newblock \bibinfo{title}{Controlling for time-dependent confounding using marginal structural models}.
\newblock \emph{\bibinfo{journal}{The Stata Journal}} \textbf{\bibinfo{volume}{4}}, \bibinfo{pages}{402--420} (\bibinfo{year}{2004}).

\bibitem{mortimer_2005}
\bibinfo{author}{Mortimer, K.~M.}, \bibinfo{author}{Neugebauer, R.}, \bibinfo{author}{Van Der~Laan, M.} \& \bibinfo{author}{Tager, I.~B.}
\newblock \bibinfo{title}{An application of model-fitting procedures for marginal structural models}.
\newblock \emph{\bibinfo{journal}{American Journal of Epidemiology}} \textbf{\bibinfo{volume}{162}}, \bibinfo{pages}{382--388} (\bibinfo{year}{2005}).

\bibitem{mansournia_2012}
\bibinfo{author}{Mansournia, M.~A.} \emph{et~al.}
\newblock \bibinfo{title}{Effect of physical activity on functional performance and knee pain in patients with osteoarthritis: analysis with marginal structural models}.
\newblock \emph{\bibinfo{journal}{Epidemiology}} \bibinfo{pages}{631--640} (\bibinfo{year}{2012}).

\bibitem{nandi_2012}
\bibinfo{author}{Nandi, A.}, \bibinfo{author}{Glymour, M.}, \bibinfo{author}{Kawachi, I.} \& \bibinfo{author}{VanderWeele, T.}
\newblock \bibinfo{title}{Using marginal structural models to estimate the direct effect of fadverse childhood social conditions on onset of heart disease, diabetes, and stroke}.
\newblock \emph{\bibinfo{journal}{Epidemiology (Cambridge, Mass.)}} \textbf{\bibinfo{volume}{23}}, \bibinfo{pages}{223--32} (\bibinfo{year}{2012}).

\bibitem{li_2016}
\bibinfo{author}{Li, S.}, \bibinfo{author}{Okereke, O.~I.}, \bibinfo{author}{Chang, S.-C.}, \bibinfo{author}{Kawachi, I.} \& \bibinfo{author}{VanderWeele, T.~J.}
\newblock \bibinfo{title}{Religious service attendance and lower depression among women—a prospective cohort study}.
\newblock \emph{\bibinfo{journal}{Annals of Behavioral Medicine}} \textbf{\bibinfo{volume}{50}}, \bibinfo{pages}{876--884} (\bibinfo{year}{2016}).

\bibitem{cerda_2010}
\bibinfo{author}{Cerd{\'a}, M.}, \bibinfo{author}{Diez-Roux, A.~V.}, \bibinfo{author}{Tchetgen, E.~T.}, \bibinfo{author}{Gordon-Larsen, P.} \& \bibinfo{author}{Kiefe, C.}
\newblock \bibinfo{title}{The relationship between neighborhood poverty and alcohol use: estimation by marginal structural models}.
\newblock \emph{\bibinfo{journal}{Epidemiology (Cambridge, Mass.)}} \textbf{\bibinfo{volume}{21}}, \bibinfo{pages}{482} (\bibinfo{year}{2010}).

\bibitem{breiman_2001}
\bibinfo{author}{Breiman, L.}
\newblock \bibinfo{title}{Random forests}.
\newblock \emph{\bibinfo{journal}{Machine learning}} \textbf{\bibinfo{volume}{45}}, \bibinfo{pages}{5--32} (\bibinfo{year}{2001}).

\bibitem{holland_1986}
\bibinfo{author}{Holland, P.~W.}
\newblock \bibinfo{title}{Statistics and causal inference}.
\newblock \emph{\bibinfo{journal}{Journal of the American statistical Association}} \textbf{\bibinfo{volume}{81}}, \bibinfo{pages}{945--960} (\bibinfo{year}{1986}).

\bibitem{patel_2009}
\bibinfo{author}{{Patel}, S.~G.}, \bibinfo{author}{{Holden}, B.~P.}, \bibinfo{author}{{Kelson}, D.~D.}, \bibinfo{author}{{Illingworth}, G.~D.} \& \bibinfo{author}{{Franx}, M.}
\newblock \bibinfo{title}{{The Dependence of Star Formation Rates on Stellar Mass and Environment at z \raisebox{-0.5ex}\textasciitilde 0.8}}.
\newblock \emph{\bibinfo{journal}{\apjl}} \textbf{\bibinfo{volume}{705}}, \bibinfo{pages}{L67--L70} (\bibinfo{year}{2009}).

\bibitem{muzzin_2012}
\bibinfo{author}{{Muzzin}, A.} \emph{et~al.}
\newblock \bibinfo{title}{{The Gemini Cluster Astrophysics Spectroscopic Survey (GCLASS): The Role of Environment and Self-regulation in Galaxy Evolution at z \raisebox{-0.5ex}\textasciitilde 1}}.
\newblock \emph{\bibinfo{journal}{\apj}} \textbf{\bibinfo{volume}{746}}, \bibinfo{pages}{188} (\bibinfo{year}{2012}).

\bibitem{quadri_2012}
\bibinfo{author}{{Quadri}, R.~F.}, \bibinfo{author}{{Williams}, R.~J.}, \bibinfo{author}{{Franx}, M.} \& \bibinfo{author}{{Hildebrandt}, H.}
\newblock \bibinfo{title}{{Tracing the Star-formation-Density Relation to z \raisebox{-0.5ex}\textasciitilde 2}}.
\newblock \emph{\bibinfo{journal}{\apj}} \textbf{\bibinfo{volume}{744}}, \bibinfo{pages}{88} (\bibinfo{year}{2012}).

\bibitem{chartab_2020}
\bibinfo{author}{{Chartab}, N.} \emph{et~al.}
\newblock \bibinfo{title}{{Large-scale Structures in the CANDELS Fields: The Role of the Environment in Star Formation Activity}}.
\newblock \emph{\bibinfo{journal}{\apj}} \textbf{\bibinfo{volume}{890}}, \bibinfo{pages}{7} (\bibinfo{year}{2020}).

\bibitem{feruglio_2010}
\bibinfo{author}{{Feruglio}, C.} \emph{et~al.}
\newblock \bibinfo{title}{{Obscured Star Formation and Environment in the COSMOS Field}}.
\newblock \emph{\bibinfo{journal}{\apj}} \textbf{\bibinfo{volume}{721}}, \bibinfo{pages}{607--614} (\bibinfo{year}{2010}).

\bibitem{grutzbauch_2011}
\bibinfo{author}{{Gr{\"u}tzbauch}, R.} \emph{et~al.}
\newblock \bibinfo{title}{{How does galaxy environment matter? The relationship between galaxy environments, colour and stellar mass at 0.4 < z < 1 in the Palomar/DEEP2 survey}}.
\newblock \emph{\bibinfo{journal}{\mnras}} \textbf{\bibinfo{volume}{411}}, \bibinfo{pages}{929--946} (\bibinfo{year}{2011}).

\bibitem{scoville_2013}
\bibinfo{author}{{Scoville}, N.} \emph{et~al.}
\newblock \bibinfo{title}{{Evolution of Galaxies and Their Environments at z = 0.1-3 in COSMOS}}.
\newblock \emph{\bibinfo{journal}{\apjs}} \textbf{\bibinfo{volume}{206}}, \bibinfo{pages}{3} (\bibinfo{year}{2013}).

\bibitem{ziparo_2014}
\bibinfo{author}{{Ziparo}, F.} \emph{et~al.}
\newblock \bibinfo{title}{{Reversal or no reversal: the evolution of the star formation rate-density relation up to z {\ensuremath{\sim}} 1.6}}.
\newblock \emph{\bibinfo{journal}{\mnras}} \textbf{\bibinfo{volume}{437}}, \bibinfo{pages}{458--474} (\bibinfo{year}{2014}).

\bibitem{darvish_2016}
\bibinfo{author}{{Darvish}, B.} \emph{et~al.}
\newblock \bibinfo{title}{{The Effects of the Local Environment and Stellar Mass on Galaxy Quenching to z {\ensuremath{\sim}} 3}}.
\newblock \emph{\bibinfo{journal}{\apj}} \textbf{\bibinfo{volume}{825}}, \bibinfo{pages}{113} (\bibinfo{year}{2016}).

\bibitem{elbaz_2007}
\bibinfo{author}{{Elbaz}, D.} \emph{et~al.}
\newblock \bibinfo{title}{{The reversal of the star formation-density relation in the distant universe}}.
\newblock \emph{\bibinfo{journal}{\aap}} \textbf{\bibinfo{volume}{468}}, \bibinfo{pages}{33--48} (\bibinfo{year}{2007}).

\bibitem{cooper_2008}
\bibinfo{author}{{Cooper}, M.~C.} \emph{et~al.}
\newblock \bibinfo{title}{{The DEEP2 Galaxy Redshift Survey: the role of galaxy environment in the cosmic star formation history}}.
\newblock \emph{\bibinfo{journal}{\mnras}} \textbf{\bibinfo{volume}{383}}, \bibinfo{pages}{1058--1078} (\bibinfo{year}{2008}).

\bibitem{tran_2010}
\bibinfo{author}{{Tran}, K.-V.~H.} \emph{et~al.}
\newblock \bibinfo{title}{{Reversal of Fortune: Confirmation of an Increasing Star Formation-Density Relation in a Cluster at z = 1.62}}.
\newblock \emph{\bibinfo{journal}{\apjl}} \textbf{\bibinfo{volume}{719}}, \bibinfo{pages}{L126--L129} (\bibinfo{year}{2010}).

\bibitem{popesso_2011}
\bibinfo{author}{{Popesso}, P.} \emph{et~al.}
\newblock \bibinfo{title}{{The effect of environment on star forming galaxies at redshift. I. First insight from PACS}}.
\newblock \emph{\bibinfo{journal}{\aap}} \textbf{\bibinfo{volume}{532}}, \bibinfo{pages}{A145} (\bibinfo{year}{2011}).

\bibitem{santos_2015}
\bibinfo{author}{{Santos}, J.~S.} \emph{et~al.}
\newblock \bibinfo{title}{{The reversal of the SF-density relation in a massive, X-ray-selected galaxy cluster at z = 1.58: results from Herschel.}}
\newblock \emph{\bibinfo{journal}{\mnras}} \textbf{\bibinfo{volume}{447}}, \bibinfo{pages}{L65--L69} (\bibinfo{year}{2015}).

\bibitem{lemaux_2022}
\bibinfo{author}{{Lemaux}, B.~C.} \emph{et~al.}
\newblock \bibinfo{title}{{The VIMOS Ultra Deep Survey: The reversal of the star-formation rate {\ensuremath{-}} density relation at 2 < z < 5}}.
\newblock \emph{\bibinfo{journal}{\aap}} \textbf{\bibinfo{volume}{662}}, \bibinfo{pages}{A33} (\bibinfo{year}{2022}).

\bibitem{shi_2024}
\bibinfo{author}{{Shi}, K.}, \bibinfo{author}{{Malavasi}, N.}, \bibinfo{author}{{Toshikawa}, J.} \& \bibinfo{author}{{Zheng}, X.}
\newblock \bibinfo{title}{{Nature versus Nurture: Revisiting the Environmental Impact on Star Formation Activities of Galaxies}}.
\newblock \emph{\bibinfo{journal}{\apj}} \textbf{\bibinfo{volume}{961}}, \bibinfo{pages}{39} (\bibinfo{year}{2024}).

\bibitem{muldrew_2012}
\bibinfo{author}{{Muldrew}, S.~I.} \emph{et~al.}
\newblock \bibinfo{title}{{Measures of galaxy environment - I. What is 'environment'?}}
\newblock \emph{\bibinfo{journal}{\mnras}} \textbf{\bibinfo{volume}{419}}, \bibinfo{pages}{2670--2682} (\bibinfo{year}{2012}).

\bibitem{tonnesen_2014}
\bibinfo{author}{{Tonnesen}, S.} \& \bibinfo{author}{{Cen}, R.}
\newblock \bibinfo{title}{{On the Reversal of Star formation Rate-Density Relation at z = 1: Insights from Simulations}}.
\newblock \emph{\bibinfo{journal}{\apj}} \textbf{\bibinfo{volume}{788}}, \bibinfo{pages}{133} (\bibinfo{year}{2014}).

\bibitem{hwang_2019}
\bibinfo{author}{{Hwang}, H.~S.}, \bibinfo{author}{{Shin}, J.} \& \bibinfo{author}{{Song}, H.}
\newblock \bibinfo{title}{{Evolution of star formation rate-density relation over cosmic time in a simulated universe: the observed reversal reproduced}}.
\newblock \emph{\bibinfo{journal}{\mnras}} \textbf{\bibinfo{volume}{489}}, \bibinfo{pages}{339--348} (\bibinfo{year}{2019}).

\bibitem{fevre_2000}
\bibinfo{author}{{Le F{\`e}vre}, O.} \emph{et~al.}
\newblock \bibinfo{title}{{Hubble Space Telescope imaging of the CFRS and LDSS redshift surveys - IV. Influence of mergers in the evolution of faint field galaxies from z\raisebox{-0.5ex}\textasciitilde1}}.
\newblock \emph{\bibinfo{journal}{\mnras}} \textbf{\bibinfo{volume}{311}}, \bibinfo{pages}{565--575} (\bibinfo{year}{2000}).

\bibitem{kampczyk_2007}
\bibinfo{author}{{Kampczyk}, P.} \emph{et~al.}
\newblock \bibinfo{title}{{Simulating the Cosmos: The Fraction of Merging Galaxies at High Redshift}}.
\newblock \emph{\bibinfo{journal}{\apjs}} \textbf{\bibinfo{volume}{172}}, \bibinfo{pages}{329--340} (\bibinfo{year}{2007}).

\bibitem{kartaltepe_2007}
\bibinfo{author}{{Kartaltepe}, J.~S.} \emph{et~al.}
\newblock \bibinfo{title}{{Evolution of the Frequency of Luminous (>=L$^{*}$$_{V}$) Close Galaxy Pairs at z < 1.2 in the COSMOS Field}}.
\newblock \emph{\bibinfo{journal}{\apjs}} \textbf{\bibinfo{volume}{172}}, \bibinfo{pages}{320--328} (\bibinfo{year}{2007}).

\bibitem{lotz_2011}
\bibinfo{author}{{Lotz}, J.~M.} \emph{et~al.}
\newblock \bibinfo{title}{{The Major and Minor Galaxy Merger Rates at z < 1.5}}.
\newblock \emph{\bibinfo{journal}{\apj}} \textbf{\bibinfo{volume}{742}}, \bibinfo{pages}{103} (\bibinfo{year}{2011}).

\bibitem{smethurst_2017}
\bibinfo{author}{{Smethurst}, R.~J.} \emph{et~al.}
\newblock \bibinfo{title}{{Galaxy Zoo: the interplay of quenching mechanisms in the group environment $\star$}}.
\newblock \emph{\bibinfo{journal}{\mnras}} \textbf{\bibinfo{volume}{469}}, \bibinfo{pages}{3670--3687} (\bibinfo{year}{2017}).

\bibitem{cowie_1996}
\bibinfo{author}{{Cowie}, L.~L.}, \bibinfo{author}{{Songaila}, A.}, \bibinfo{author}{{Hu}, E.~M.} \& \bibinfo{author}{{Cohen}, J.~G.}
\newblock \bibinfo{title}{{New Insight on Galaxy Formation and Evolution From Keck Spectroscopy of the Hawaii Deep Fields}}.
\newblock \emph{\bibinfo{journal}{\aj}} \textbf{\bibinfo{volume}{112}}, \bibinfo{pages}{839} (\bibinfo{year}{1996}).

\bibitem{heavens_2004}
\bibinfo{author}{{Heavens}, A.}, \bibinfo{author}{{Panter}, B.}, \bibinfo{author}{{Jimenez}, R.} \& \bibinfo{author}{{Dunlop}, J.}
\newblock \bibinfo{title}{{The star-formation history of the Universe from the stellar populations of nearby galaxies}}.
\newblock \emph{\bibinfo{journal}{\nat}} \textbf{\bibinfo{volume}{428}}, \bibinfo{pages}{625--627} (\bibinfo{year}{2004}).

\bibitem{kodama_2004}
\bibinfo{author}{{Kodama}, T.} \emph{et~al.}
\newblock \bibinfo{title}{{Down-sizing in galaxy formation at z\raisebox{-0.5ex}\textasciitilde 1 in the Subaru/XMM-Newton Deep Survey (SXDS)}}.
\newblock \emph{\bibinfo{journal}{\mnras}} \textbf{\bibinfo{volume}{350}}, \bibinfo{pages}{1005--1014} (\bibinfo{year}{2004}).

\bibitem{jimenez_2005}
\bibinfo{author}{{Jimenez}, R.}, \bibinfo{author}{{Panter}, B.}, \bibinfo{author}{{Heavens}, A.~F.} \& \bibinfo{author}{{Verde}, L.}
\newblock \bibinfo{title}{{Baryonic conversion tree: the global assembly of stars and dark matter in galaxies from the Sloan Digital Sky Survey}}.
\newblock \emph{\bibinfo{journal}{\mnras}} \textbf{\bibinfo{volume}{356}}, \bibinfo{pages}{495--501} (\bibinfo{year}{2005}).

\bibitem{juneau_2005}
\bibinfo{author}{{Juneau}, S.} \emph{et~al.}
\newblock \bibinfo{title}{{Cosmic Star Formation History and Its Dependence on Galaxy Stellar Mass}}.
\newblock \emph{\bibinfo{journal}{\apjl}} \textbf{\bibinfo{volume}{619}}, \bibinfo{pages}{L135--L138} (\bibinfo{year}{2005}).

\bibitem{thomas_2005}
\bibinfo{author}{{Thomas}, D.}, \bibinfo{author}{{Maraston}, C.}, \bibinfo{author}{{Bender}, R.} \& \bibinfo{author}{{Mendes de Oliveira}, C.}
\newblock \bibinfo{title}{{The Epochs of Early-Type Galaxy Formation as a Function of Environment}}.
\newblock \emph{\bibinfo{journal}{\apj}} \textbf{\bibinfo{volume}{621}}, \bibinfo{pages}{673--694} (\bibinfo{year}{2005}).

\bibitem{bauer_2005}
\bibinfo{author}{{Bauer}, A.~E.}, \bibinfo{author}{{Drory}, N.}, \bibinfo{author}{{Hill}, G.~J.} \& \bibinfo{author}{{Feulner}, G.}
\newblock \bibinfo{title}{{Specific Star Formation Rates to Redshift 1.5}}.
\newblock \emph{\bibinfo{journal}{\apjl}} \textbf{\bibinfo{volume}{621}}, \bibinfo{pages}{L89--L92} (\bibinfo{year}{2005}).

\bibitem{bell_2005}
\bibinfo{author}{{Bell}, E.~F.} \emph{et~al.}
\newblock \bibinfo{title}{{Toward an Understanding of the Rapid Decline of the Cosmic Star Formation Rate}}.
\newblock \emph{\bibinfo{journal}{\apj}} \textbf{\bibinfo{volume}{625}}, \bibinfo{pages}{23--36} (\bibinfo{year}{2005}).

\bibitem{nelan_2005}
\bibinfo{author}{{Nelan}, J.~E.} \emph{et~al.}
\newblock \bibinfo{title}{{NOAO Fundamental Plane Survey. II. Age and Metallicity along the Red Sequence from Line-Strength Data}}.
\newblock \emph{\bibinfo{journal}{\apj}} \textbf{\bibinfo{volume}{632}}, \bibinfo{pages}{137--156} (\bibinfo{year}{2005}).

\bibitem{feulner_2005}
\bibinfo{author}{{Feulner}, G.} \emph{et~al.}
\newblock \bibinfo{title}{{Specific Star Formation Rates to Redshift 5 from the FORS Deep Field and the GOODS-S Field}}.
\newblock \emph{\bibinfo{journal}{\apjl}} \textbf{\bibinfo{volume}{633}}, \bibinfo{pages}{L9--L12} (\bibinfo{year}{2005}).

\bibitem{bundy_2006}
\bibinfo{author}{{Bundy}, K.} \emph{et~al.}
\newblock \bibinfo{title}{{The Mass Assembly History of Field Galaxies: Detection of an Evolving Mass Limit for Star-Forming Galaxies}}.
\newblock \emph{\bibinfo{journal}{\apj}} \textbf{\bibinfo{volume}{651}}, \bibinfo{pages}{120--141} (\bibinfo{year}{2006}).

\bibitem{drory_2008}
\bibinfo{author}{{Drory}, N.} \& \bibinfo{author}{{Alvarez}, M.}
\newblock \bibinfo{title}{{The Contribution of Star Formation and Merging to Stellar Mass Buildup in Galaxies}}.
\newblock \emph{\bibinfo{journal}{\apj}} \textbf{\bibinfo{volume}{680}}, \bibinfo{pages}{41--53} (\bibinfo{year}{2008}).

\bibitem{vergani_2008}
\bibinfo{author}{{Vergani}, D.} \emph{et~al.}
\newblock \bibinfo{title}{{The VIMOS VLT Deep Survey. Tracing the galaxy stellar mass assembly history over the last 8 Gyr}}.
\newblock \emph{\bibinfo{journal}{\aap}} \textbf{\bibinfo{volume}{487}}, \bibinfo{pages}{89--101} (\bibinfo{year}{2008}).

\bibitem{mortlock_2011}
\bibinfo{author}{{Mortlock}, A.} \emph{et~al.}
\newblock \bibinfo{title}{{A deep probe of the galaxy stellar mass functions at z{\ensuremath{\sim}} 1-3 with the GOODS NICMOS Survey}}.
\newblock \emph{\bibinfo{journal}{\mnras}} \textbf{\bibinfo{volume}{413}}, \bibinfo{pages}{2845--2859} (\bibinfo{year}{2011}).

\bibitem{fontanot_2009}
\bibinfo{author}{{Fontanot}, F.}, \bibinfo{author}{{De Lucia}, G.}, \bibinfo{author}{{Monaco}, P.}, \bibinfo{author}{{Somerville}, R.~S.} \& \bibinfo{author}{{Santini}, P.}
\newblock \bibinfo{title}{{The many manifestations of downsizing: hierarchical galaxy formation models confront observations}}.
\newblock \emph{\bibinfo{journal}{\mnras}} \textbf{\bibinfo{volume}{397}}, \bibinfo{pages}{1776--1790} (\bibinfo{year}{2009}).

\bibitem{conroy_2013}
\bibinfo{author}{{Conroy}, C.}
\newblock \bibinfo{title}{{Modeling the Panchromatic Spectral Energy Distributions of Galaxies}}.
\newblock \emph{\bibinfo{journal}{\araa}} \textbf{\bibinfo{volume}{51}}, \bibinfo{pages}{393--455} (\bibinfo{year}{2013}).

\bibitem{sarpa_2022}
\bibinfo{author}{{Sarpa}, E.}, \bibinfo{author}{{Longobardi}, A.}, \bibinfo{author}{{Kraljic}, K.}, \bibinfo{author}{{Veropalumbo}, A.} \& \bibinfo{author}{{Schimd}, C.}
\newblock \bibinfo{title}{{Tracing the environmental history of observed galaxies via extended fast action minimization method}}.
\newblock \emph{\bibinfo{journal}{\mnras}} \textbf{\bibinfo{volume}{516}}, \bibinfo{pages}{231--244} (\bibinfo{year}{2022}).

\bibitem{euclid_2011}
\bibinfo{author}{{Laureijs}, R.} \emph{et~al.}
\newblock \bibinfo{title}{{Euclid Definition Study Report}}.
\newblock \emph{\bibinfo{journal}{arXiv e-prints}} \bibinfo{pages}{arXiv:1110.3193} (\bibinfo{year}{2011}).

\bibitem{lsst_2009}
\bibinfo{author}{{LSST Science Collaboration}} \emph{et~al.}
\newblock \bibinfo{title}{{LSST Science Book, Version 2.0}}.
\newblock \emph{\bibinfo{journal}{arXiv e-prints}} \bibinfo{pages}{arXiv:0912.0201} (\bibinfo{year}{2009}).

\bibitem{roman_2015}
\bibinfo{author}{{Spergel}, D.} \emph{et~al.}
\newblock \bibinfo{title}{{Wide-Field InfrarRed Survey Telescope-Astrophysics Focused Telescope Assets WFIRST-AFTA 2015 Report}}.
\newblock \emph{\bibinfo{journal}{arXiv e-prints}} \bibinfo{pages}{arXiv:1503.03757} (\bibinfo{year}{2015}).

\bibitem{springel_2010}
\bibinfo{author}{{Springel}, V.}
\newblock \bibinfo{title}{{E pur si muove: Galilean-invariant cosmological hydrodynamical simulations on a moving mesh}}.
\newblock \emph{\bibinfo{journal}{\mnras}} \textbf{\bibinfo{volume}{401}}, \bibinfo{pages}{791--851} (\bibinfo{year}{2010}).

\bibitem{planck_2016}
\bibinfo{author}{{Planck Collaboration}} \emph{et~al.}
\newblock \bibinfo{title}{{Planck 2015 results. XIII. Cosmological parameters}}.
\newblock \emph{\bibinfo{journal}{\aap}} \textbf{\bibinfo{volume}{594}}, \bibinfo{pages}{A13} (\bibinfo{year}{2016}).

\bibitem{zeldovich_1970}
\bibinfo{author}{{Zel'dovich}, Y.~B.}
\newblock \bibinfo{title}{{Gravitational instability: An approximate theory for large density perturbations.}}
\newblock \emph{\bibinfo{journal}{\aap}} \textbf{\bibinfo{volume}{5}}, \bibinfo{pages}{84--89} (\bibinfo{year}{1970}).

\bibitem{springel_2015}
\bibinfo{author}{{Springel}, V.}
\newblock \bibinfo{title}{{N-GenIC: Cosmological structure initial conditions}}.
\newblock \bibinfo{howpublished}{Astrophysics Source Code Library, record ascl:1502.003} (\bibinfo{year}{2015}).
\newblock \eprint{1502.003}.

\bibitem{davis_1985}
\bibinfo{author}{{Davis}, M.}, \bibinfo{author}{{Efstathiou}, G.}, \bibinfo{author}{{Frenk}, C.~S.} \& \bibinfo{author}{{White}, S.~D.~M.}
\newblock \bibinfo{title}{{The evolution of large-scale structure in a universe dominated by cold dark matter}}.
\newblock \emph{\bibinfo{journal}{\apj}} \textbf{\bibinfo{volume}{292}}, \bibinfo{pages}{371--394} (\bibinfo{year}{1985}).

\bibitem{dolag_2009}
\bibinfo{author}{{Dolag}, K.}, \bibinfo{author}{{Borgani}, S.}, \bibinfo{author}{{Murante}, G.} \& \bibinfo{author}{{Springel}, V.}
\newblock \bibinfo{title}{{Substructures in hydrodynamical cluster simulations}}.
\newblock \emph{\bibinfo{journal}{\mnras}} \textbf{\bibinfo{volume}{399}}, \bibinfo{pages}{497--514} (\bibinfo{year}{2009}).

\bibitem{rodriguez_2015}
\bibinfo{author}{{Rodriguez-Gomez}, V.} \emph{et~al.}
\newblock \bibinfo{title}{{The merger rate of galaxies in the Illustris simulation: a comparison with observations and semi-empirical models}}.
\newblock \emph{\bibinfo{journal}{\mnras}} \textbf{\bibinfo{volume}{449}}, \bibinfo{pages}{49--64} (\bibinfo{year}{2015}).

\bibitem{donnari_2019}
\bibinfo{author}{{Donnari}, M.} \emph{et~al.}
\newblock \bibinfo{title}{{The star formation activity of IllustrisTNG galaxies: main sequence, UVJ diagram, quenched fractions, and systematics}}.
\newblock \emph{\bibinfo{journal}{\mnras}} \textbf{\bibinfo{volume}{485}}, \bibinfo{pages}{4817--4840} (\bibinfo{year}{2019}).

\bibitem{cassata_2007}
\bibinfo{author}{{Cassata}, P.} \emph{et~al.}
\newblock \bibinfo{title}{{The Cosmic Evolution Survey (COSMOS): The Morphological Content and Environmental Dependence of the Galaxy Color-Magnitude Relation at z \raisebox{-0.5ex}\textasciitilde 0.7}}.
\newblock \emph{\bibinfo{journal}{\apjs}} \textbf{\bibinfo{volume}{172}}, \bibinfo{pages}{270--283} (\bibinfo{year}{2007}).

\bibitem{sobral_2011}
\bibinfo{author}{{Sobral}, D.} \emph{et~al.}
\newblock \bibinfo{title}{{The dependence of star formation activity on environment and stellar mass at z{\ensuremath{\sim}} 1 from the HiZELS-H{\ensuremath{\alpha}} survey}}.
\newblock \emph{\bibinfo{journal}{\mnras}} \textbf{\bibinfo{volume}{411}}, \bibinfo{pages}{675--692} (\bibinfo{year}{2011}).

\bibitem{haas_2012}
\bibinfo{author}{{Haas}, M.~R.}, \bibinfo{author}{{Schaye}, J.} \& \bibinfo{author}{{Jeeson-Daniel}, A.}
\newblock \bibinfo{title}{{Disentangling galaxy environment and host halo mass}}.
\newblock \emph{\bibinfo{journal}{\mnras}} \textbf{\bibinfo{volume}{419}}, \bibinfo{pages}{2133--2146} (\bibinfo{year}{2012}).

\bibitem{chesnaye_2022}
\bibinfo{author}{Chesnaye, N.~C.} \emph{et~al.}
\newblock \bibinfo{title}{An introduction to inverse probability of treatment weighting in observational research}.
\newblock \emph{\bibinfo{journal}{Clinical Kidney Journal}} \textbf{\bibinfo{volume}{15}}, \bibinfo{pages}{14--20} (\bibinfo{year}{2022}).

\bibitem{rosenbaum_1983}
\bibinfo{author}{Rosenbaum, P.~R.} \& \bibinfo{author}{Rubin, D.~B.}
\newblock \bibinfo{title}{The central role of the propensity score in observational studies for causal effects}.
\newblock \emph{\bibinfo{journal}{Biometrika}} \textbf{\bibinfo{volume}{70}}, \bibinfo{pages}{41--55} (\bibinfo{year}{1983}).

\bibitem{cannas_2019}
\bibinfo{author}{Cannas, M.} \& \bibinfo{author}{Arpino, B.}
\newblock \bibinfo{title}{A comparison of machine learning algorithms and covariate balance measures for propensity score matching and weighting}.
\newblock \emph{\bibinfo{journal}{Biometrical Journal}} \textbf{\bibinfo{volume}{61}}, \bibinfo{pages}{1049--1072} (\bibinfo{year}{2019}).

\bibitem{scikit-learn}
\bibinfo{author}{Pedregosa, F.} \emph{et~al.}
\newblock \bibinfo{title}{Scikit-learn: Machine learning in {P}ython}.
\newblock \emph{\bibinfo{journal}{Journal of Machine Learning Research}} \textbf{\bibinfo{volume}{12}}, \bibinfo{pages}{2825--2830} (\bibinfo{year}{2011}).

\bibitem{pearl_2009b}
\bibinfo{author}{Pearl, J.}
\newblock \bibinfo{title}{{Causal inference in statistics: An overview}}.
\newblock \emph{\bibinfo{journal}{Statistics Surveys}} \textbf{\bibinfo{volume}{3}}, \bibinfo{pages}{96 -- 146} (\bibinfo{year}{2009}).
\newblock \urlprefix\url{https://doi.org/10.1214/09-SS057}.

\bibitem{neyman_1923}
\bibinfo{author}{Splawa-Neyman, J.}, \bibinfo{author}{Dabrowska, D.~M.} \& \bibinfo{author}{Speed, T.~P.}
\newblock \bibinfo{title}{{On the Application of Probability Theory to Agricultural Experiments. Essay on Principles. Section 9}}.
\newblock \emph{\bibinfo{journal}{Statistical Science}} \textbf{\bibinfo{volume}{5}}, \bibinfo{pages}{465 -- 472} (\bibinfo{year}{1990}).
\newblock \urlprefix\url{https://doi.org/10.1214/ss/1177012031}.

\bibitem{rubin_1974}
\bibinfo{author}{Rubin, D.~B.}
\newblock \bibinfo{title}{Estimating causal effects of treatments in randomized and nonrandomized studies.}
\newblock \emph{\bibinfo{journal}{Journal of educational Psychology}} \textbf{\bibinfo{volume}{66}}, \bibinfo{pages}{688} (\bibinfo{year}{1974}).

\bibitem{yao_2020}
\bibinfo{author}{{Yao}, L.} \emph{et~al.}
\newblock \bibinfo{title}{{A Survey on Causal Inference}}.
\newblock \emph{\bibinfo{journal}{arXiv e-prints}} \bibinfo{pages}{arXiv:2002.02770} (\bibinfo{year}{2020}).

\bibitem{cox_1958}
\bibinfo{author}{Cox, D.}
\newblock \emph{\bibinfo{title}{Planning of Experiments}} Wiley Series in Probability and Statistics - Applied Probability and Statistics Section (\bibinfo{publisher}{Wiley}, \bibinfo{year}{1958}).
\newblock \urlprefix\url{https://books.google.co.uk/books?id=sZFQAAAAMAAJ}.

\bibitem{rubin_1980}
\bibinfo{author}{Rubin, D.~B.}
\newblock \bibinfo{title}{Randomization analysis of experimental data: The fisher randomization test comment}.
\newblock \emph{\bibinfo{journal}{Journal of the American statistical association}} \textbf{\bibinfo{volume}{75}}, \bibinfo{pages}{591--593} (\bibinfo{year}{1980}).

\bibitem{guth_1982}
\bibinfo{author}{{Guth}, A.~H.} \& \bibinfo{author}{{Pi}, S.~Y.}
\newblock \bibinfo{title}{{Fluctuations in the New Inflationary Universe}}.
\newblock \emph{\bibinfo{journal}{\prl}} \textbf{\bibinfo{volume}{49}}, \bibinfo{pages}{1110--1113} (\bibinfo{year}{1982}).

\bibitem{hawking_1982}
\bibinfo{author}{{Hawking}, S.~W.}
\newblock \bibinfo{title}{{The development of irregularities in a single bubble inflationary universe}}.
\newblock \emph{\bibinfo{journal}{Physics Letters B}} \textbf{\bibinfo{volume}{115}}, \bibinfo{pages}{295--297} (\bibinfo{year}{1982}).

\bibitem{linde_1982}
\bibinfo{author}{{Linde}, A.~D.}
\newblock \bibinfo{title}{{A new inflationary universe scenario: A possible solution of the horizon, flatness, homogeneity, isotropy and primordial monopole problems}}.
\newblock \emph{\bibinfo{journal}{Physics Letters B}} \textbf{\bibinfo{volume}{108}}, \bibinfo{pages}{389--393} (\bibinfo{year}{1982}).

\bibitem{starobinsky_1982}
\bibinfo{author}{{Starobinsky}, A.~A.}
\newblock \bibinfo{title}{{Dynamics of phase transition in the new inflationary universe scenario and generation of perturbations}}.
\newblock \emph{\bibinfo{journal}{Physics Letters B}} \textbf{\bibinfo{volume}{117}}, \bibinfo{pages}{175--178} (\bibinfo{year}{1982}).

\bibitem{bardeen_1983}
\bibinfo{author}{{Bardeen}, J.~M.}, \bibinfo{author}{{Steinhardt}, P.~J.} \& \bibinfo{author}{{Turner}, M.~S.}
\newblock \bibinfo{title}{{Spontaneous creation of almost scale-free density perturbations in an inflationary universe}}.
\newblock \emph{\bibinfo{journal}{\prd}} \textbf{\bibinfo{volume}{28}}, \bibinfo{pages}{679--693} (\bibinfo{year}{1983}).

\bibitem{mo_2010}
\bibinfo{author}{Mo, H.}, \bibinfo{author}{van~den Bosch, F.} \& \bibinfo{author}{White, S.}
\newblock \emph{\bibinfo{title}{Galaxy Formation and Evolution}} Galaxy Formation and Evolution (\bibinfo{publisher}{Cambridge University Press}, \bibinfo{year}{2010}).
\newblock \urlprefix\url{https://books.google.co.uk/books?id=Zj7fDU3Z4wsC}.

\bibitem{silk_1977}
\bibinfo{author}{{Silk}, J.}
\newblock \bibinfo{title}{{On the fragmentation of cosmic gas clouds. I. The formation of galaxies and the first generation of stars.}}
\newblock \emph{\bibinfo{journal}{\apj}} \textbf{\bibinfo{volume}{211}}, \bibinfo{pages}{638--648} (\bibinfo{year}{1977}).

\bibitem{rees_1977}
\bibinfo{author}{{Rees}, M.~J.} \& \bibinfo{author}{{Ostriker}, J.~P.}
\newblock \bibinfo{title}{{Cooling, dynamics and fragmentation of massive gas clouds: clues to the masses and radii of galaxies and clusters.}}
\newblock \emph{\bibinfo{journal}{\mnras}} \textbf{\bibinfo{volume}{179}}, \bibinfo{pages}{541--559} (\bibinfo{year}{1977}).

\bibitem{binney_1977}
\bibinfo{author}{{Binney}, J.}
\newblock \bibinfo{title}{{The physics of dissipational galaxy formation.}}
\newblock \emph{\bibinfo{journal}{\apj}} \textbf{\bibinfo{volume}{215}}, \bibinfo{pages}{483--491} (\bibinfo{year}{1977}).

\bibitem{lacey_1991}
\bibinfo{author}{{Lacey}, C.} \& \bibinfo{author}{{Silk}, J.}
\newblock \bibinfo{title}{{Tidally Triggered Galaxy Formation. I. Evolution of the Galaxy Luminosity Function}}.
\newblock \emph{\bibinfo{journal}{\apj}} \textbf{\bibinfo{volume}{381}}, \bibinfo{pages}{14} (\bibinfo{year}{1991}).

\bibitem{fall_1980}
\bibinfo{author}{{Fall}, S.~M.} \& \bibinfo{author}{{Efstathiou}, G.}
\newblock \bibinfo{title}{{Formation and rotation of disc galaxies with haloes.}}
\newblock \emph{\bibinfo{journal}{\mnras}} \textbf{\bibinfo{volume}{193}}, \bibinfo{pages}{189--206} (\bibinfo{year}{1980}).

\bibitem{mo_1988}
\bibinfo{author}{{Mo}, H.~J.}, \bibinfo{author}{{Mao}, S.} \& \bibinfo{author}{{White}, S. D.~M.}
\newblock \bibinfo{title}{{The formation of galactic discs}}.
\newblock \emph{\bibinfo{journal}{\mnras}} \textbf{\bibinfo{volume}{295}}, \bibinfo{pages}{319--336} (\bibinfo{year}{1998}).

\bibitem{bonnell_1997}
\bibinfo{author}{{Bonnell}, I.~A.}, \bibinfo{author}{{Bate}, M.~R.}, \bibinfo{author}{{Clarke}, C.~J.} \& \bibinfo{author}{{Pringle}, J.~E.}
\newblock \bibinfo{title}{{Accretion and the stellar mass spectrum in small clusters}}.
\newblock \emph{\bibinfo{journal}{\mnras}} \textbf{\bibinfo{volume}{285}}, \bibinfo{pages}{201--208} (\bibinfo{year}{1997}).

\bibitem{krumholz_2005}
\bibinfo{author}{{Krumholz}, M.~R.}, \bibinfo{author}{{McKee}, C.~F.} \& \bibinfo{author}{{Klein}, R.~I.}
\newblock \bibinfo{title}{{Stars Form By Gravitational Collapse, Not Competitive Accretion}}.
\newblock \emph{\bibinfo{journal}{arXiv e-prints}} \bibinfo{pages}{astro--ph/0510412} (\bibinfo{year}{2005}).

\bibitem{schmidt_1959}
\bibinfo{author}{{Schmidt}, M.}
\newblock \bibinfo{title}{{The Rate of Star Formation.}}
\newblock \emph{\bibinfo{journal}{\apj}} \textbf{\bibinfo{volume}{129}}, \bibinfo{pages}{243} (\bibinfo{year}{1959}).

\bibitem{kennicutt_1998}
\bibinfo{author}{{Kennicutt}, J., Robert~C.}
\newblock \bibinfo{title}{{Star Formation in Galaxies Along the Hubble Sequence}}.
\newblock \emph{\bibinfo{journal}{\araa}} \textbf{\bibinfo{volume}{36}}, \bibinfo{pages}{189--232} (\bibinfo{year}{1998}).

\bibitem{larson_1974}
\bibinfo{author}{{Larson}, R.~B.}
\newblock \bibinfo{title}{{Effects of supernovae on the early evolution of galaxies}}.
\newblock \emph{\bibinfo{journal}{\mnras}} \textbf{\bibinfo{volume}{169}}, \bibinfo{pages}{229--246} (\bibinfo{year}{1974}).

\bibitem{dekel_1986}
\bibinfo{author}{{Dekel}, A.} \& \bibinfo{author}{{Silk}, J.}
\newblock \bibinfo{title}{{The Origin of Dwarf Galaxies, Cold Dark Matter, and Biased Galaxy Formation}}.
\newblock \emph{\bibinfo{journal}{\apj}} \textbf{\bibinfo{volume}{303}}, \bibinfo{pages}{39} (\bibinfo{year}{1986}).

\bibitem{hopkins_2014}
\bibinfo{author}{{Hopkins}, P.~F.} \emph{et~al.}
\newblock \bibinfo{title}{{Galaxies on FIRE (Feedback In Realistic Environments): stellar feedback explains cosmologically inefficient star formation}}.
\newblock \emph{\bibinfo{journal}{\mnras}} \textbf{\bibinfo{volume}{445}}, \bibinfo{pages}{581--603} (\bibinfo{year}{2014}).

\bibitem{heckmann_1990}
\bibinfo{author}{{Heckman}, T.~M.}, \bibinfo{author}{{Armus}, L.} \& \bibinfo{author}{{Miley}, G.~K.}
\newblock \bibinfo{title}{{On the Nature and Implications of Starburst-driven Galactic Superwinds}}.
\newblock \emph{\bibinfo{journal}{\apjs}} \textbf{\bibinfo{volume}{74}}, \bibinfo{pages}{833} (\bibinfo{year}{1990}).

\bibitem{martin_1999}
\bibinfo{author}{{Martin}, C.~L.}
\newblock \bibinfo{title}{{Properties of Galactic Outflows: Measurements of the Feedback from Star Formation}}.
\newblock \emph{\bibinfo{journal}{\apj}} \textbf{\bibinfo{volume}{513}}, \bibinfo{pages}{156--160} (\bibinfo{year}{1999}).

\bibitem{scannapieco_2008}
\bibinfo{author}{{Scannapieco}, C.}, \bibinfo{author}{{Tissera}, P.~B.}, \bibinfo{author}{{White}, S. D.~M.} \& \bibinfo{author}{{Springel}, V.}
\newblock \bibinfo{title}{{Effects of supernova feedback on the formation of galaxy discs}}.
\newblock \emph{\bibinfo{journal}{\mnras}} \textbf{\bibinfo{volume}{389}}, \bibinfo{pages}{1137--1149} (\bibinfo{year}{2008}).

\bibitem{croton_2006}
\bibinfo{author}{{Croton}, D.~J.} \emph{et~al.}
\newblock \bibinfo{title}{{The many lives of active galactic nuclei: cooling flows, black holes and the luminosities and colours of galaxies}}.
\newblock \emph{\bibinfo{journal}{\mnras}} \textbf{\bibinfo{volume}{365}}, \bibinfo{pages}{11--28} (\bibinfo{year}{2006}).

\bibitem{fabian_2012}
\bibinfo{author}{{Fabian}, A.~C.}
\newblock \bibinfo{title}{{Observational Evidence of Active Galactic Nuclei Feedback}}.
\newblock \emph{\bibinfo{journal}{\araa}} \textbf{\bibinfo{volume}{50}}, \bibinfo{pages}{455--489} (\bibinfo{year}{2012}).

\bibitem{heckman_2014}
\bibinfo{author}{{Heckman}, T.~M.} \& \bibinfo{author}{{Best}, P.~N.}
\newblock \bibinfo{title}{{The Coevolution of Galaxies and Supermassive Black Holes: Insights from Surveys of the Contemporary Universe}}.
\newblock \emph{\bibinfo{journal}{\araa}} \textbf{\bibinfo{volume}{52}}, \bibinfo{pages}{589--660} (\bibinfo{year}{2014}).

\bibitem{soltan_1982}
\bibinfo{author}{{Soltan}, A.}
\newblock \bibinfo{title}{{Masses of quasars.}}
\newblock \emph{\bibinfo{journal}{\mnras}} \textbf{\bibinfo{volume}{200}}, \bibinfo{pages}{115--122} (\bibinfo{year}{1982}).

\bibitem{silk_1998}
\bibinfo{author}{{Silk}, J.} \& \bibinfo{author}{{Rees}, M.~J.}
\newblock \bibinfo{title}{{Quasars and galaxy formation}}.
\newblock \emph{\bibinfo{journal}{\aap}} \textbf{\bibinfo{volume}{331}}, \bibinfo{pages}{L1--L4} (\bibinfo{year}{1998}).

\bibitem{boselli_2006}
\bibinfo{author}{{Boselli}, A.} \& \bibinfo{author}{{Gavazzi}, G.}
\newblock \bibinfo{title}{{Environmental Effects on Late-Type Galaxies in Nearby Clusters}}.
\newblock \emph{\bibinfo{journal}{\pasp}} \textbf{\bibinfo{volume}{118}}, \bibinfo{pages}{517--559} (\bibinfo{year}{2006}).

\bibitem{toomre_1972}
\bibinfo{author}{{Toomre}, A.} \& \bibinfo{author}{{Toomre}, J.}
\newblock \bibinfo{title}{{Galactic Bridges and Tails}}.
\newblock \emph{\bibinfo{journal}{\apj}} \textbf{\bibinfo{volume}{178}}, \bibinfo{pages}{623--666} (\bibinfo{year}{1972}).

\bibitem{toomre_1977}
\bibinfo{author}{{Toomre}, A.}
\newblock \bibinfo{editor}{{Tinsley}, B.~M.} \& \bibinfo{editor}{{Larson}, D.~C., Richard B.~Gehret} (eds) \emph{\bibinfo{title}{{Mergers and Some Consequences}}}.
\newblock (eds \bibinfo{editor}{{Tinsley}, B.~M.} \& \bibinfo{editor}{{Larson}, D.~C., Richard B.~Gehret}) \emph{\bibinfo{booktitle}{Evolution of Galaxies and Stellar Populations}}, \bibinfo{pages}{401} (\bibinfo{year}{1977}).

\bibitem{hernquist_1992}
\bibinfo{author}{{Hernquist}, L.}
\newblock \bibinfo{title}{{Structure of Merger Remnants. I. Bulgeless Progenitors}}.
\newblock \emph{\bibinfo{journal}{\apj}} \textbf{\bibinfo{volume}{400}}, \bibinfo{pages}{460} (\bibinfo{year}{1992}).

\bibitem{hernquist_1993}
\bibinfo{author}{{Hernquist}, L.}
\newblock \bibinfo{title}{{Structure of Merger Remnants. II. Progenitors with Rotating Bulges}}.
\newblock \emph{\bibinfo{journal}{\apj}} \textbf{\bibinfo{volume}{409}}, \bibinfo{pages}{548} (\bibinfo{year}{1993}).

\bibitem{barnes_1998}
\bibinfo{author}{{Barnes}, J.~E.}
\newblock \bibinfo{title}{{Encounters of Disk/Halo Galaxies}}.
\newblock \emph{\bibinfo{journal}{\apj}} \textbf{\bibinfo{volume}{331}}, \bibinfo{pages}{699} (\bibinfo{year}{1988}).

\bibitem{barnes_2002}
\bibinfo{author}{{Barnes}, J.~E.}
\newblock \bibinfo{title}{{Formation of gas discs in merging galaxies}}.
\newblock \emph{\bibinfo{journal}{\mnras}} \textbf{\bibinfo{volume}{333}}, \bibinfo{pages}{481--494} (\bibinfo{year}{2002}).

\bibitem{cox_2006}
\bibinfo{author}{{Cox}, T.~J.} \emph{et~al.}
\newblock \bibinfo{title}{{The Kinematic Structure of Merger Remnants}}.
\newblock \emph{\bibinfo{journal}{\apj}} \textbf{\bibinfo{volume}{650}}, \bibinfo{pages}{791--811} (\bibinfo{year}{2006}).

\bibitem{hopkins_2009a}
\bibinfo{author}{{Hopkins}, P.~F.}, \bibinfo{author}{{Cox}, T.~J.}, \bibinfo{author}{{Younger}, J.~D.} \& \bibinfo{author}{{Hernquist}, L.}
\newblock \bibinfo{title}{{How do Disks Survive Mergers?}}
\newblock \emph{\bibinfo{journal}{\apj}} \textbf{\bibinfo{volume}{691}}, \bibinfo{pages}{1168--1201} (\bibinfo{year}{2009}).

\bibitem{sparre_2016}
\bibinfo{author}{{Sparre}, M.} \& \bibinfo{author}{{Springel}, V.}
\newblock \bibinfo{title}{{Zooming in on major mergers: dense, starbursting gas in cosmological simulations}}.
\newblock \emph{\bibinfo{journal}{\mnras}} \textbf{\bibinfo{volume}{462}}, \bibinfo{pages}{2418--2430} (\bibinfo{year}{2016}).

\bibitem{pontzen_2017}
\bibinfo{author}{{Pontzen}, A.} \emph{et~al.}
\newblock \bibinfo{title}{{How to quench a galaxy}}.
\newblock \emph{\bibinfo{journal}{\mnras}} \textbf{\bibinfo{volume}{465}}, \bibinfo{pages}{547--558} (\bibinfo{year}{2017}).

\bibitem{mihos_1994}
\bibinfo{author}{{Mihos}, J.~C.} \& \bibinfo{author}{{Hernquist}, L.}
\newblock \bibinfo{title}{{Ultraluminous Starbursts in Major Mergers}}.
\newblock \emph{\bibinfo{journal}{\apjl}} \textbf{\bibinfo{volume}{431}}, \bibinfo{pages}{L9} (\bibinfo{year}{1994}).

\bibitem{mihos_1996}
\bibinfo{author}{{Mihos}, J.~C.} \& \bibinfo{author}{{Hernquist}, L.}
\newblock \bibinfo{title}{{Gasdynamics and Starbursts in Major Mergers}}.
\newblock \emph{\bibinfo{journal}{\apj}} \textbf{\bibinfo{volume}{464}}, \bibinfo{pages}{641} (\bibinfo{year}{1996}).

\bibitem{hopkins_2006}
\bibinfo{author}{{Hopkins}, P.~F.} \emph{et~al.}
\newblock \bibinfo{title}{{A Unified, Merger-driven Model of the Origin of Starbursts, Quasars, the Cosmic X-Ray Background, Supermassive Black Holes, and Galaxy Spheroids}}.
\newblock \emph{\bibinfo{journal}{\apjs}} \textbf{\bibinfo{volume}{163}}, \bibinfo{pages}{1--49} (\bibinfo{year}{2006}).

\bibitem{hopkins_2008a}
\bibinfo{author}{{Hopkins}, P.~F.}, \bibinfo{author}{{Hernquist}, L.}, \bibinfo{author}{{Cox}, T.~J.} \& \bibinfo{author}{{Kere{\v{s}}}, D.}
\newblock \bibinfo{title}{{A Cosmological Framework for the Co-Evolution of Quasars, Supermassive Black Holes, and Elliptical Galaxies. I. Galaxy Mergers and Quasar Activity}}.
\newblock \emph{\bibinfo{journal}{\apjs}} \textbf{\bibinfo{volume}{175}}, \bibinfo{pages}{356--389} (\bibinfo{year}{2008}).

\bibitem{hopkins_2008b}
\bibinfo{author}{{Hopkins}, P.~F.}, \bibinfo{author}{{Cox}, T.~J.}, \bibinfo{author}{{Kere{\v{s}}}, D.} \& \bibinfo{author}{{Hernquist}, L.}
\newblock \bibinfo{title}{{A Cosmological Framework for the Co-Evolution of Quasars, Supermassive Black Holes, and Elliptical Galaxies. II. Formation of Red Ellipticals}}.
\newblock \emph{\bibinfo{journal}{\apjs}} \textbf{\bibinfo{volume}{175}}, \bibinfo{pages}{390--422} (\bibinfo{year}{2008}).

\bibitem{snyder_2011}
\bibinfo{author}{{Snyder}, G.~F.}, \bibinfo{author}{{Cox}, T.~J.}, \bibinfo{author}{{Hayward}, C.~C.}, \bibinfo{author}{{Hernquist}, L.} \& \bibinfo{author}{{Jonsson}, P.}
\newblock \bibinfo{title}{{K+A Galaxies as the Aftermath of Gas-rich Mergers: Simulating the Evolution of Galaxies as Seen by Spectroscopic Surveys}}.
\newblock \emph{\bibinfo{journal}{\apj}} \textbf{\bibinfo{volume}{741}}, \bibinfo{pages}{77} (\bibinfo{year}{2011}).

\bibitem{hayward_2014}
\bibinfo{author}{{Hayward}, C.~C.}, \bibinfo{author}{{Torrey}, P.}, \bibinfo{author}{{Springel}, V.}, \bibinfo{author}{{Hernquist}, L.} \& \bibinfo{author}{{Vogelsberger}, M.}
\newblock \bibinfo{title}{{Galaxy mergers on a moving mesh: a comparison with smoothed particle hydrodynamics}}.
\newblock \emph{\bibinfo{journal}{\mnras}} \textbf{\bibinfo{volume}{442}}, \bibinfo{pages}{1992--2016} (\bibinfo{year}{2014}).

\bibitem{sanders_1998}
\bibinfo{author}{{Sanders}, D.~B.} \emph{et~al.}
\newblock \bibinfo{title}{{Ultraluminous Infrared Galaxies and the Origin of Quasars}}.
\newblock \emph{\bibinfo{journal}{\apj}} \textbf{\bibinfo{volume}{325}}, \bibinfo{pages}{74} (\bibinfo{year}{1988}).

\bibitem{dimatteo_2005}
\bibinfo{author}{{Di Matteo}, T.}, \bibinfo{author}{{Springel}, V.} \& \bibinfo{author}{{Hernquist}, L.}
\newblock \bibinfo{title}{{Energy input from quasars regulates the growth and activity of black holes and their host galaxies}}.
\newblock \emph{\bibinfo{journal}{\nat}} \textbf{\bibinfo{volume}{433}}, \bibinfo{pages}{604--607} (\bibinfo{year}{2005}).

\bibitem{hopkins_2009b}
\bibinfo{author}{{Hopkins}, P.~F.} \& \bibinfo{author}{{Hernquist}, L.}
\newblock \bibinfo{title}{{A Characteristic Division Between the Fueling of Quasars and Seyferts: Five Simple Tests}}.
\newblock \emph{\bibinfo{journal}{\apj}} \textbf{\bibinfo{volume}{694}}, \bibinfo{pages}{599--609} (\bibinfo{year}{2009}).

\bibitem{treister_2012}
\bibinfo{author}{{Treister}, E.}, \bibinfo{author}{{Schawinski}, K.}, \bibinfo{author}{{Urry}, C.~M.} \& \bibinfo{author}{{Simmons}, B.~D.}
\newblock \bibinfo{title}{{Major Galaxy Mergers Only Trigger the Most Luminous Active Galactic Nuclei}}.
\newblock \emph{\bibinfo{journal}{\apjl}} \textbf{\bibinfo{volume}{758}}, \bibinfo{pages}{L39} (\bibinfo{year}{2012}).

\bibitem{chandrasekhar_1943a}
\bibinfo{author}{{Chandrasekhar}, S.}
\newblock \bibinfo{title}{{Dynamical Friction. I. General Considerations: the Coefficient of Dynamical Friction.}}
\newblock \emph{\bibinfo{journal}{\apj}} \textbf{\bibinfo{volume}{97}}, \bibinfo{pages}{255} (\bibinfo{year}{1943}).

\bibitem{chandrasekhar_1943b}
\bibinfo{author}{{Chandrasekhar}, S.}
\newblock \bibinfo{title}{{Dynamical Friction. II. The Rate of Escape of Stars from Clusters and the Evidence for the Operation of Dynamical Friction.}}
\newblock \emph{\bibinfo{journal}{\apj}} \textbf{\bibinfo{volume}{97}}, \bibinfo{pages}{263} (\bibinfo{year}{1943}).

\bibitem{chandrasekhar_1943c}
\bibinfo{author}{{Chandrasekhar}, S.}
\newblock \bibinfo{title}{{Dynamical Friction. III. a More Exact Theory of the Rate of Escape of Stars from Clusters.}}
\newblock \emph{\bibinfo{journal}{\apj}} \textbf{\bibinfo{volume}{98}}, \bibinfo{pages}{54} (\bibinfo{year}{1943}).

\bibitem{gunn_1972}
\bibinfo{author}{{Gunn}, J.~E.} \& \bibinfo{author}{{Gott}, I., J.~Richard}.
\newblock \bibinfo{title}{{On the Infall of Matter Into Clusters of Galaxies and Some Effects on Their Evolution}}.
\newblock \emph{\bibinfo{journal}{\apj}} \textbf{\bibinfo{volume}{176}}, \bibinfo{pages}{1} (\bibinfo{year}{1972}).

\bibitem{dressler_1983}
\bibinfo{author}{{Dressler}, A.} \& \bibinfo{author}{{Gunn}, J.~E.}
\newblock \bibinfo{title}{{Spectroscopy of galaxies in distant clusters. II. The population of the 3C 295 cluster.}}
\newblock \emph{\bibinfo{journal}{\apj}} \textbf{\bibinfo{volume}{270}}, \bibinfo{pages}{7--19} (\bibinfo{year}{1983}).

\bibitem{gavazzi_1995}
\bibinfo{author}{{Gavazzi}, G.} \emph{et~al.}
\newblock \bibinfo{title}{{The radio and optical structure of three peculiar galaxies in A 1367.}}
\newblock \emph{\bibinfo{journal}{\aap}} \textbf{\bibinfo{volume}{304}}, \bibinfo{pages}{325} (\bibinfo{year}{1995}).

\bibitem{vollmer_2001}
\bibinfo{author}{{Vollmer}, B.}, \bibinfo{author}{{Cayatte}, V.}, \bibinfo{author}{{Balkowski}, C.} \& \bibinfo{author}{{Duschl}, W.~J.}
\newblock \bibinfo{title}{{Ram Pressure Stripping and Galaxy Orbits: The Case of the Virgo Cluster}}.
\newblock \emph{\bibinfo{journal}{\apj}} \textbf{\bibinfo{volume}{561}}, \bibinfo{pages}{708--726} (\bibinfo{year}{2001}).

\bibitem{abadi_1999}
\bibinfo{author}{{Abadi}, M.~G.}, \bibinfo{author}{{Moore}, B.} \& \bibinfo{author}{{Bower}, R.~G.}
\newblock \bibinfo{title}{{Ram pressure stripping of spiral galaxies in clusters}}.
\newblock \emph{\bibinfo{journal}{\mnras}} \textbf{\bibinfo{volume}{308}}, \bibinfo{pages}{947--954} (\bibinfo{year}{1999}).

\bibitem{moore_1999}
\bibinfo{author}{{Moore}, B.}, \bibinfo{author}{{Lake}, G.}, \bibinfo{author}{{Quinn}, T.} \& \bibinfo{author}{{Stadel}, J.}
\newblock \bibinfo{title}{{On the survival and destruction of spiral galaxies in clusters}}.
\newblock \emph{\bibinfo{journal}{\mnras}} \textbf{\bibinfo{volume}{304}}, \bibinfo{pages}{465--474} (\bibinfo{year}{1999}).

\bibitem{kampakoglou_2007}
\bibinfo{author}{{Kampakoglou}, M.} \& \bibinfo{author}{{Benson}, A.~J.}
\newblock \bibinfo{title}{{Tidal mass loss from collisionless systems}}.
\newblock \emph{\bibinfo{journal}{\mnras}} \textbf{\bibinfo{volume}{374}}, \bibinfo{pages}{775--786} (\bibinfo{year}{2007}).

\bibitem{ghigna_1998}
\bibinfo{author}{{Ghigna}, S.} \emph{et~al.}
\newblock \bibinfo{title}{{Dark matter haloes within clusters}}.
\newblock \emph{\bibinfo{journal}{\mnras}} \textbf{\bibinfo{volume}{300}}, \bibinfo{pages}{146--162} (\bibinfo{year}{1998}).

\bibitem{farouki_1981}
\bibinfo{author}{{Farouki}, R.} \& \bibinfo{author}{{Shapiro}, S.~L.}
\newblock \bibinfo{title}{{Computer simulations of environmental influences on galaxy evolution in dense clusters. II - Rapid tidal encounters}}.
\newblock \emph{\bibinfo{journal}{\apj}} \textbf{\bibinfo{volume}{243}}, \bibinfo{pages}{32--41} (\bibinfo{year}{1981}).

\bibitem{moore_1996}
\bibinfo{author}{{Moore}, B.}, \bibinfo{author}{{Katz}, N.}, \bibinfo{author}{{Lake}, G.}, \bibinfo{author}{{Dressler}, A.} \& \bibinfo{author}{{Oemler}, A.}
\newblock \bibinfo{title}{{Galaxy harassment and the evolution of clusters of galaxies}}.
\newblock \emph{\bibinfo{journal}{\nat}} \textbf{\bibinfo{volume}{379}}, \bibinfo{pages}{613--616} (\bibinfo{year}{1996}).

\bibitem{moore_1998}
\bibinfo{author}{{Moore}, B.}, \bibinfo{author}{{Lake}, G.} \& \bibinfo{author}{{Katz}, N.}
\newblock \bibinfo{title}{{Morphological Transformation from Galaxy Harassment}}.
\newblock \emph{\bibinfo{journal}{\apj}} \textbf{\bibinfo{volume}{495}}, \bibinfo{pages}{139--151} (\bibinfo{year}{1998}).

\bibitem{gnedin_2003}
\bibinfo{author}{{Gnedin}, O.~Y.}
\newblock \bibinfo{title}{{Tidal Effects in Clusters of Galaxies}}.
\newblock \emph{\bibinfo{journal}{\apj}} \textbf{\bibinfo{volume}{582}}, \bibinfo{pages}{141--161} (\bibinfo{year}{2003}).

\bibitem{mastropietro_2005}
\bibinfo{author}{{Mastropietro}, C.} \emph{et~al.}
\newblock \bibinfo{title}{{Morphological evolution of discs in clusters}}.
\newblock \emph{\bibinfo{journal}{\mnras}} \textbf{\bibinfo{volume}{364}}, \bibinfo{pages}{607--619} (\bibinfo{year}{2005}).

\bibitem{aguerri_2009}
\bibinfo{author}{{Aguerri}, J.~A.~L.} \& \bibinfo{author}{{Gonz{\'a}lez-Garc{\'\i}a}, A.~C.}
\newblock \bibinfo{title}{{On the origin of dwarf elliptical galaxies: the fundamental plane}}.
\newblock \emph{\bibinfo{journal}{\aap}} \textbf{\bibinfo{volume}{494}}, \bibinfo{pages}{891--904} (\bibinfo{year}{2009}).

\bibitem{smith_2010}
\bibinfo{author}{{Smith}, R.}, \bibinfo{author}{{Davies}, J.~I.} \& \bibinfo{author}{{Nelson}, A.~H.}
\newblock \bibinfo{title}{{How effective is harassment on infalling late-type dwarfs?}}
\newblock \emph{\bibinfo{journal}{\mnras}} \textbf{\bibinfo{volume}{405}}, \bibinfo{pages}{1723--1735} (\bibinfo{year}{2010}).

\bibitem{smith_2015}
\bibinfo{author}{{Smith}, R.} \emph{et~al.}
\newblock \bibinfo{title}{{The sensitivity of harassment to orbit: mass loss from early-type dwarfs in galaxy clusters}}.
\newblock \emph{\bibinfo{journal}{\mnras}} \textbf{\bibinfo{volume}{454}}, \bibinfo{pages}{2502--2516} (\bibinfo{year}{2015}).

\bibitem{bialas_2015}
\bibinfo{author}{{Bialas}, D.}, \bibinfo{author}{{Lisker}, T.}, \bibinfo{author}{{Olczak}, C.}, \bibinfo{author}{{Spurzem}, R.} \& \bibinfo{author}{{Kotulla}, R.}
\newblock \bibinfo{title}{{On the occurrence of galaxy harassment}}.
\newblock \emph{\bibinfo{journal}{\aap}} \textbf{\bibinfo{volume}{576}}, \bibinfo{pages}{A103} (\bibinfo{year}{2015}).

\bibitem{larson_1980}
\bibinfo{author}{{Larson}, R.~B.}, \bibinfo{author}{{Tinsley}, B.~M.} \& \bibinfo{author}{{Caldwell}, C.~N.}
\newblock \bibinfo{title}{{The evolution of disk galaxies and the origin of S0 galaxies}}.
\newblock \emph{\bibinfo{journal}{\apj}} \textbf{\bibinfo{volume}{237}}, \bibinfo{pages}{692--707} (\bibinfo{year}{1980}).

\bibitem{benson_2000}
\bibinfo{author}{{Benson}, A.~J.}, \bibinfo{author}{{Bower}, R.~G.}, \bibinfo{author}{{Frenk}, C.~S.} \& \bibinfo{author}{{White}, S.~D.~M.}
\newblock \bibinfo{title}{{Diffuse X-ray emission from late-type galaxy haloes}}.
\newblock \emph{\bibinfo{journal}{\mnras}} \textbf{\bibinfo{volume}{314}}, \bibinfo{pages}{557--565} (\bibinfo{year}{2000}).

\bibitem{cowie_1977}
\bibinfo{author}{{Cowie}, L.~L.} \& \bibinfo{author}{{Songaila}, A.}
\newblock \bibinfo{title}{{Thermal evaporation of gas within galaxies by a hot intergalactic medium}}.
\newblock \emph{\bibinfo{journal}{\nat}} \textbf{\bibinfo{volume}{266}}, \bibinfo{pages}{501--503} (\bibinfo{year}{1977}).

\bibitem{nulsen_1982}
\bibinfo{author}{{Nulsen}, P.~E.~J.}
\newblock \bibinfo{title}{{Transport processes and the stripping of cluster galaxies.}}
\newblock \emph{\bibinfo{journal}{\mnras}} \textbf{\bibinfo{volume}{198}}, \bibinfo{pages}{1007--1016} (\bibinfo{year}{1982}).

\bibitem{athanassoula_1992}
\bibinfo{author}{{Athanassoula}, E.}
\newblock \bibinfo{title}{{The existence and shapes of dust lanes in galactic bars.}}
\newblock \emph{\bibinfo{journal}{\mnras}} \textbf{\bibinfo{volume}{259}}, \bibinfo{pages}{345--364} (\bibinfo{year}{1992}).

\bibitem{zurita_2004}
\bibinfo{author}{{Zurita}, A.}, \bibinfo{author}{{Rela{\~n}o}, M.}, \bibinfo{author}{{Beckman}, J.~E.} \& \bibinfo{author}{{Knapen}, J.~H.}
\newblock \bibinfo{title}{{Ionized gas kinematics and massive star formation in NGC 1530}}.
\newblock \emph{\bibinfo{journal}{\aap}} \textbf{\bibinfo{volume}{413}}, \bibinfo{pages}{73--89} (\bibinfo{year}{2004}).

\bibitem{sheth_2005}
\bibinfo{author}{{Sheth}, K.}, \bibinfo{author}{{Vogel}, S.~N.}, \bibinfo{author}{{Regan}, M.~W.}, \bibinfo{author}{{Thornley}, M.~D.} \& \bibinfo{author}{{Teuben}, P.~J.}
\newblock \bibinfo{title}{{Secular Evolution via Bar-driven Gas Inflow: Results from BIMA SONG}}.
\newblock \emph{\bibinfo{journal}{\apj}} \textbf{\bibinfo{volume}{632}}, \bibinfo{pages}{217--226} (\bibinfo{year}{2005}).

\bibitem{kormendy_2004}
\bibinfo{author}{{Kormendy}, J.} \& \bibinfo{author}{{Kennicutt}, J., Robert~C.}
\newblock \bibinfo{title}{{Secular Evolution and the Formation of Pseudobulges in Disk Galaxies}}.
\newblock \emph{\bibinfo{journal}{\araa}} \textbf{\bibinfo{volume}{42}}, \bibinfo{pages}{603--683} (\bibinfo{year}{2004}).

\bibitem{fang_2013}
\bibinfo{author}{{Fang}, J.~J.}, \bibinfo{author}{{Faber}, S.~M.}, \bibinfo{author}{{Koo}, D.~C.} \& \bibinfo{author}{{Dekel}, A.}
\newblock \bibinfo{title}{{A Link between Star Formation Quenching and Inner Stellar Mass Density in Sloan Digital Sky Survey Central Galaxies}}.
\newblock \emph{\bibinfo{journal}{\apj}} \textbf{\bibinfo{volume}{776}}, \bibinfo{pages}{63} (\bibinfo{year}{2013}).

\bibitem{kormendy_1995}
\bibinfo{author}{{Kormendy}, J.} \& \bibinfo{author}{{Richstone}, D.}
\newblock \bibinfo{title}{{Inward Bound---The Search For Supermassive Black Holes In Galactic Nuclei}}.
\newblock \emph{\bibinfo{journal}{\araa}} \textbf{\bibinfo{volume}{33}}, \bibinfo{pages}{581} (\bibinfo{year}{1995}).

\bibitem{magorrian_1998}
\bibinfo{author}{{Magorrian}, J.} \emph{et~al.}
\newblock \bibinfo{title}{{The Demography of Massive Dark Objects in Galaxy Centers}}.
\newblock \emph{\bibinfo{journal}{\aj}} \textbf{\bibinfo{volume}{115}}, \bibinfo{pages}{2285--2305} (\bibinfo{year}{1998}).

\bibitem{ferrarese_2000}
\bibinfo{author}{{Ferrarese}, L.} \& \bibinfo{author}{{Merritt}, D.}
\newblock \bibinfo{title}{{A Fundamental Relation between Supermassive Black Holes and Their Host Galaxies}}.
\newblock \emph{\bibinfo{journal}{\apjl}} \textbf{\bibinfo{volume}{539}}, \bibinfo{pages}{L9--L12} (\bibinfo{year}{2000}).

\bibitem{gebhardt_2000}
\bibinfo{author}{{Gebhardt}, K.} \emph{et~al.}
\newblock \bibinfo{title}{{A Relationship between Nuclear Black Hole Mass and Galaxy Velocity Dispersion}}.
\newblock \emph{\bibinfo{journal}{\apjl}} \textbf{\bibinfo{volume}{539}}, \bibinfo{pages}{L13--L16} (\bibinfo{year}{2000}).

\bibitem{haring_2004}
\bibinfo{author}{{H{\"a}ring}, N.} \& \bibinfo{author}{{Rix}, H.-W.}
\newblock \bibinfo{title}{{On the Black Hole Mass-Bulge Mass Relation}}.
\newblock \emph{\bibinfo{journal}{\apjl}} \textbf{\bibinfo{volume}{604}}, \bibinfo{pages}{L89--L92} (\bibinfo{year}{2004}).

\bibitem{kormendy_2013}
\bibinfo{author}{{Kormendy}, J.} \& \bibinfo{author}{{Ho}, L.~C.}
\newblock \bibinfo{title}{{Coevolution (Or Not) of Supermassive Black Holes and Host Galaxies}}.
\newblock \emph{\bibinfo{journal}{\araa}} \textbf{\bibinfo{volume}{51}}, \bibinfo{pages}{511--653} (\bibinfo{year}{2013}).

\bibitem{hirano_2004}
\bibinfo{author}{Hirano, K.} \& \bibinfo{author}{Imbens, G.~W.}
\newblock \bibinfo{title}{The propensity score with continuous treatments}.
\newblock \emph{\bibinfo{journal}{Applied Bayesian modeling and causal inference from incomplete-data perspectives}} \textbf{\bibinfo{volume}{226164}}, \bibinfo{pages}{73--84} (\bibinfo{year}{2004}).

\bibitem{agostino_1998}
\bibinfo{author}{D'Agostino~Jr, R.~B.}
\newblock \bibinfo{title}{Propensity score methods for bias reduction in the comparison of a treatment to a non-randomized control group}.
\newblock \emph{\bibinfo{journal}{Statistics in medicine}} \textbf{\bibinfo{volume}{17}}, \bibinfo{pages}{2265--2281} (\bibinfo{year}{1998}).

\bibitem{austin_2011}
\bibinfo{author}{Austin, P.~C.}
\newblock \bibinfo{title}{An introduction to propensity score methods for reducing the effects of confounding in observational studies}.
\newblock \emph{\bibinfo{journal}{Multivariate behavioral research}} \textbf{\bibinfo{volume}{46}}, \bibinfo{pages}{399--424} (\bibinfo{year}{2011}).

\bibitem{ellison_2008}
\bibinfo{author}{{Ellison}, S.~L.}, \bibinfo{author}{{Patton}, D.~R.}, \bibinfo{author}{{Simard}, L.} \& \bibinfo{author}{{McConnachie}, A.~W.}
\newblock \bibinfo{title}{{Galaxy Pairs in the Sloan Digital Sky Survey. I. Star Formation, Active Galactic Nucleus Fraction, and the Mass-Metallicity Relation}}.
\newblock \emph{\bibinfo{journal}{\aj}} \textbf{\bibinfo{volume}{135}}, \bibinfo{pages}{1877--1899} (\bibinfo{year}{2008}).

\bibitem{garduno_2021}
\bibinfo{author}{{Gardu{\~n}o}, L.~E.} \emph{et~al.}
\newblock \bibinfo{title}{{Galaxy And Mass Assembly (GAMA): the interplay between galaxy mass, SFR, and heavy element abundance in paired galaxy sets}}.
\newblock \emph{\bibinfo{journal}{\mnras}} \textbf{\bibinfo{volume}{501}}, \bibinfo{pages}{2969--2982} (\bibinfo{year}{2021}).

\bibitem{sotillo_2021}
\bibinfo{author}{{Sotillo-Ramos}, D.} \emph{et~al.}
\newblock \bibinfo{title}{{Galaxy and mass assembly (GAMA): The environmental impact on SFR and metallicity in galaxy groups}}.
\newblock \emph{\bibinfo{journal}{\mnras}} \textbf{\bibinfo{volume}{508}}, \bibinfo{pages}{1817--1830} (\bibinfo{year}{2021}).

\bibitem{williamson_2017}
\bibinfo{author}{Williamson, T.} \& \bibinfo{author}{Ravani, P.}
\newblock \bibinfo{title}{Marginal structural models in clinical research: when and how to use them?}
\newblock \emph{\bibinfo{journal}{Nephrology Dialysis Transplantation}} \textbf{\bibinfo{volume}{32}}, \bibinfo{pages}{ii84--ii90} (\bibinfo{year}{2017}).

\bibitem{thoemmes_2016}
\bibinfo{author}{Thoemmes, F.} \& \bibinfo{author}{Ong, A.~D.}
\newblock \bibinfo{title}{A primer on inverse probability of treatment weighting and marginal structural models}.
\newblock \emph{\bibinfo{journal}{Emerging Adulthood}} \textbf{\bibinfo{volume}{4}}, \bibinfo{pages}{40--59} (\bibinfo{year}{2016}).

\bibitem{kloek_1978}
\bibinfo{author}{Kloek, T.} \& \bibinfo{author}{Van~Dijk, H.~K.}
\newblock \bibinfo{title}{Bayesian estimates of equation system parameters: an application of integration by monte carlo}.
\newblock \emph{\bibinfo{journal}{Econometrica: Journal of the Econometric Society}} \bibinfo{pages}{1--19} (\bibinfo{year}{1978}).

\bibitem{horvitz_1952}
\bibinfo{author}{Horvitz, D.~G.} \& \bibinfo{author}{Thompson, D.~J.}
\newblock \bibinfo{title}{A generalization of sampling without replacement from a finite universe}.
\newblock \emph{\bibinfo{journal}{Journal of the American statistical Association}} \textbf{\bibinfo{volume}{47}}, \bibinfo{pages}{663--685} (\bibinfo{year}{1952}).

\bibitem{hernan_2023}
\bibinfo{author}{Hernan, M.} \& \bibinfo{author}{Robins, J.}
\newblock \emph{\bibinfo{title}{Causal Inference}} Chapman \& Hall/CRC Monographs on Statistics \& Applied Probab (\bibinfo{publisher}{Taylor \& Francis}, \bibinfo{year}{2023}).
\newblock \urlprefix\url{https://books.google.co.uk/books?id=\_KnHIAAACAAJ}.

\bibitem{vanderweele_2008}
\bibinfo{author}{VanderWeele, T.~J.}
\newblock \bibinfo{title}{Ignorability and stability assumptions in neighborhood effects research}.
\newblock \emph{\bibinfo{journal}{Statistics in Medicine}} \textbf{\bibinfo{volume}{27}}, \bibinfo{pages}{1934--1943} (\bibinfo{year}{2008}).
\newblock \urlprefix\url{https://onlinelibrary.wiley.com/doi/abs/10.1002/sim.3139}.

\bibitem{zhu_2015}
\bibinfo{author}{Zhu, Y.}, \bibinfo{author}{Coffman, D.~L.} \& \bibinfo{author}{Ghosh, D.}
\newblock \bibinfo{title}{A boosting algorithm for estimating generalized propensity scores with continuous treatments}.
\newblock \emph{\bibinfo{journal}{Journal of causal inference}} \textbf{\bibinfo{volume}{3}}, \bibinfo{pages}{25--40} (\bibinfo{year}{2015}).

\bibitem{kendall_1938}
\bibinfo{author}{Kendall, M.~G.}
\newblock \bibinfo{title}{{A NEW MEASURE OF RANK CORRELATION}}.
\newblock \emph{\bibinfo{journal}{Biometrika}} \textbf{\bibinfo{volume}{30}}, \bibinfo{pages}{81--93} (\bibinfo{year}{1938}).
\newblock \urlprefix\url{https://doi.org/10.1093/biomet/30.1-2.81}.

\bibitem{hernan_2006}
\bibinfo{author}{Hern{\'a}n, M.~A.} \& \bibinfo{author}{Robins, J.~M.}
\newblock \bibinfo{title}{Estimating causal effects from epidemiological data}.
\newblock \emph{\bibinfo{journal}{Journal of Epidemiology \& Community Health}} \textbf{\bibinfo{volume}{60}}, \bibinfo{pages}{578--586} (\bibinfo{year}{2006}).

\bibitem{cole_2008}
\bibinfo{author}{Cole, S.~R.} \& \bibinfo{author}{Hernán, M.~A.}
\newblock \bibinfo{title}{{Constructing Inverse Probability Weights for Marginal Structural Models}}.
\newblock \emph{\bibinfo{journal}{American Journal of Epidemiology}} \textbf{\bibinfo{volume}{168}}, \bibinfo{pages}{656--664} (\bibinfo{year}{2008}).
\newblock \urlprefix\url{https://doi.org/10.1093/aje/kwn164}.

\end{thebibliography}

\begin{appendices}

\onecolumn
\setcounter{figure}{0} % Restart figure numbering
\renewcommand\figurename{Fig.}
\renewcommand{\thefigure}{A\arabic{figure}}% Figure counter representation
\renewcommand{\theHfigure}{A\arabic{figure}}% Hyperref figure hyperlink hook

\section{Causal inference}
\label{sec:causal_inference}
Causal inference is concerned with inferring cause and effect \citep[see][for an overview]{pearl_2009b}. The goal is to identify and quantify the causal effect of one thing on another, e.g., a vaccine on a disease. A correlation between the vaccine and outcome (disease cured or not) hints at its effectiveness but does not guarantee it as ``correlation does not imply causation''. The observed correlation can be due to a common cause \citep{reichenbach_1956} that causes both the vaccine and outcome. For example, age is a potential common cause as it typically influences whether an individual can receive a vaccine and their chance of recovery from a disease. In the extreme case, there may not be a causal connection between the vaccine and outcome, and the correlation may be entirely due to age, which would signify that the vaccine is ineffective. However, if the vaccine has a causal effect on the outcome, the correlation will be partly due to age and the vaccine. Regardless of the situation, the measured effectiveness of the vaccine without considering age will be biased. Herein lies the fundamental difference between statistical and causal inference: the former ascertains a relationship between two quantities (assuming one exists), while the latter can establish the causal nature of the relationship. In this section, we give a brief overview of causal inference, starting with its key tenets: causal models and causal graphs. We note that since ``correlation'' technically only refers to the degree to which a pair of variables are linearly related, we will use the broader term ``association'' from here to refer to statistical dependence because it describes any relationship between variables, linear or not.

\subsection{Causal models and graphs}
The principal component for any causal inference task is a causal model or structural causal model (SCM), i.e., a model that describes causal relationships between variables. Formally, an SCM specifies a set of exogenous, or latent, variables $U=\{u_1,\dots,u_n\}$ distributed as $P(U)$, a set of endogenous, or observable, variables $V=\{v_1,\dots, v_m\}$, a directed acyclic graph (DAG), called the \emph{causal structure} of the model, whose nodes are the variables $U\cup V$, and a collection of functions $F=\{f_1,\dots, f_n\}$, such that $v_i = f_i(PA_i, u_i), \text{ for } i=1,\dots, n,$ where $PA$ denotes the parent observed nodes of an observed variable \citep{pearl_2009a}. The collection of functions and distribution over latent variables induces a distribution over observable variables: $P(V=v) := \sum_{\{u_i \mid f_i(PA_i, u_i)\,=\,v_i\}} P(u_i)$. We can thus assign uncertainty over observable variables despite the fact the underlying dynamics are deterministic. 

\begin{figure}
    \centering
    \includegraphics[width=0.3\textwidth]{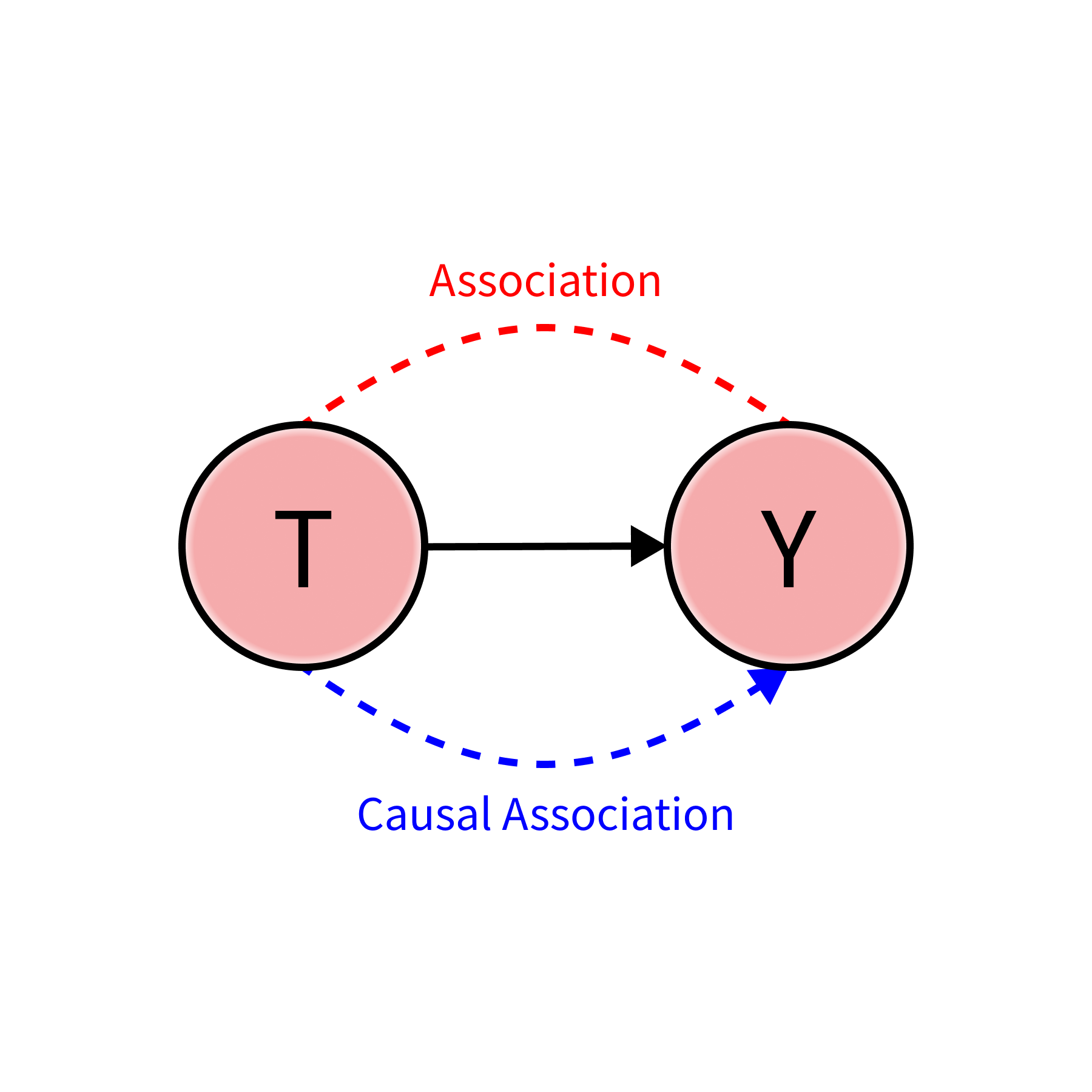}
    \caption{Causal graph representing the causal relationship between two variables, $T$ and $Y$. The direct edge (black arrow) from $T$ to $Y$ implies that $T$ directly causes $Y$. The causal association (i.e., the association due to causation) `flows' asymmetrically from $T$ to $Y$ (represented by the blue dashed line arrow), while association `flows' symmetrically (represented by the red dashed line). The causal graph is an example of a directed acyclic graph (DAG). A DAG is a graph that is: (i) directed (i.e., has edges that imply a direction) and (ii) acyclic (i.e., a variable does not cause itself either directly or through another variable).} 
    \label{fig:treatment_outcome_dag}
\end{figure}

Structural equations ($F$) fully capture and mathematically describe a causal model. However, a graphical representation of a causal model in the form of causal graphs (also called causal diagrams) is more intuitive for understanding causal relationships. A causal graph is a probabilistic graphical model and consists of a collection of nodes and edges that connect the nodes \citep{wright_1921}. The nodes represent variables, while the edges communicate the causes of the variables. Fig. \ref{fig:treatment_outcome_dag} shows the fundamental causal graph between two variables, $T$ and $Y$. The direct edge (black arrow) from $T$ to $Y$ implies that $T$ directly causes $Y$. The causal graph is an example of a DAG because: (i) it is directed (i.e., has edges that imply a direction) and (ii) acyclic (i.e., a variable does not cause itself either directly or through another variable).

DAGs make it easy to deduce if two variables share a causal or non-causal relationship. More importantly, they allow one to effortlessly conclude if association is causation with a few basic rules. For example, causal association (i.e., the association due to causation) can be imagined as `flowing' asymmetrically along directed paths (a sequence of adjacent nodes with direct edges all in the same direction), while association `flowing' symmetrically along directed paths. In the DAG in Fig. \ref{fig:treatment_outcome_dag}, the causal association flows in one direction from $T$ to $Y$ along the direct path (as shown by the blue dashed line arrow). However, the association flows in both directions along the same path (as shown by the red dashed line). Hence, association alone does not provide any information on the direction of causality. It does not distinguish between the following possible causal relationships between the two variables: 

\begin{enumerate}
    \item $T$ causes $Y$ (direct causation)
    \item $Y$ causes $T$ (reverse causation)
    \item $T$ and $Y$ share a common cause (common causation)
    \item $T$ and $Y$ cause each other (cyclic causation)
\end{enumerate}

Furthermore, it is also possible that $T$ and $Y$ are not related at all, and the association is coincidental. Consequently, association does not imply causation. Nevertheless, the DAG in this instance conveys that all association is causal as there is only a solitary direct path between $T$ and $Y$. If the DAG represents the true causal model of the vaccine–disease example, where $T$ is the vaccine and $Y$ is the outcome, the observed correlation does imply causation. In the following section, we describe the mathematical framework we adopt for reasoning and quantifying causality.

\subsection{Causal framework}
The Rubin causal model \citep{holland_1986}, also known as the Neyman–Rubin causal model \citep{neyman_1923, rubin_1974}, is a mathematical framework of causal inference based on the idea of potential outcomes \citep[see][for a recent review]{yao_2020}. The framework is inspired by how humans reason about causality. We compare an outcome $Y$ given an action $T$ with the outcome under no action. If there is a difference in the two outcomes, we reason that the action has had a causal effect on the outcome. The individual causal effect (ICE) on a unit $i$,

\begin{equation}
    \tau_{i} = Y_{i}(1) - Y_{i}(0),
    \label{eq:individual_causal_effect}
\end{equation}

\noindent
where $Y_{i}(1)$ and $Y_{i}(0)$ are the two potential outcomes under action and no action, respectively. It is impossible to know both potential outcomes given that the two potential realities, the one in which the action takes place and the other in which it does not, cannot be observed simultaneously. The potential outcome that is observed is ``factual'', whereas the unobserved potential outcome is ``counterfactual''. This dilemma is the ``fundamental problem of causal inference'' \citep{holland_1986}. The impossible nature of the task is the reason why causality is such a challenging subject to tackle. Nevertheless, it is possible to estimate rather than compute causal effects. Generally, it is difficult to accurately estimate unit-level causal effects, but it is feasible to reliably estimate an average of the causal effect within a population—the average causal effect (ACE; \citenum{holland_1986}),

\begin{equation}
    \tau = \mathbb{E}[{\tau_{i}}] = \mathbb{E}[{Y_{i}(1)} - Y_{i}(0)] = \mathbb{E}[{Y_{i}(1)}] - \mathbb{E}[Y_{i}(0)].
    \label{eq:average_causal_effect_causal}
\end{equation}

\noindent
We make the reader aware of the terminology we use throughout the Article: the action $T$ is the quantity we are interested in measuring the causal effect of, and the outcome $Y$ is the quantity we want to measure the causal effect on. Furthermore, the action is sometimes referred to as an intervention, an exposure, or a treatment, depending on the scientific nature of the study. We refer to the action as the treatment.

\subsection{Causal assumptions}
\label{sec:causal_assumptions}
Causal inference necessitates the following assumptions:

\begin{enumerate}
    \item \textit{Exchangeability} – the potential outcomes are independent of the treatment.
    \begin{equation}
        Y(t) \perp\!\!\!\perp T \ 
    \end{equation}
    
    \item \textit{Positivity} – the probability of receiving treatment is greater than zero but less than one.
    \begin{equation}
        0 < P(T) < 1
    \end{equation}
    
    \item \textit{Consistency} – the treatment is well-defined such that the observed outcome is equal to the potential outcome under treatment.
    \begin{equation}
        Y = Y(t)
    \end{equation}
    
    \item \textit{No interference} – the potential outcome of a unit only depends on its treatment and not on the treatment of other units \citep{cox_1958}. 
        \begin{equation}
            Y_{i} = Y_i(t_{i})
        \end{equation}
\end{enumerate}

The exchangeability assumption states that potential outcomes must be independent of the treatment. In other words, it must be possible to exchange treatment groups without changing their potential outcomes. To understand why the assumption is essential, consider the aforementioned vaccine–disease example, with age as a common cause. If age influences who receives the vaccine, then the treatment and control groups are not exchangeable because their age distributions are dissimilar. And if age also impacts one's ability to recover from the disease, then the causal effect estimated using the groups is biased as it is an admixture of the causal effects of the vaccine and age. Exchangeability ensures that the causal effect is bias-free because if the treatment groups are similar in all of their characteristics except for the treatment, then any outstanding causal effect must be the result of the treatment only.

Positivity states that there must be a non-zero probability of receiving any treatment. This assumption is important because its violation leads to undefined causal effects. For example, consider the situation where everyone or no one receives the vaccine. In such a scenario, the causal effect of the vaccine would be mathematically impossible to estimate because the counterfactual would always be missing. Intuitively, causal effects are only meaningful if the outcome under ``treatment'' is contrasted to the outcome under ``no treatment'' within the potential outcomes framework of causal inference.

Consistency states that the observed outcome must equal the potential outcome under treatment. When this assumption is not met, the causal effect is inconclusive. Following the vaccine–disease example, there must be only one version of the vaccine if the goal is to estimate its efficacy. If multiple versions exist and they are labelled as the treatment, then the causal effect will be a mixture of the individual causal effects of the different vaccines. Furthermore, if the temperature of the vaccine affects the outcome, then all individuals must receive the vaccine at the same temperature. Simply put, the treatment must be well-defined.

Lastly, the no interference assumption states that the potential outcome of a unit must only depend on its treatment. Violation of this assumption makes the causal effect of treatment ill-defined in the strict sense because the treatment is now an admixture of multiple units' treatment and must be redefined. This situation is common in many real-world cases and is referred to as spillover effects. For example, the likelihood of contracting COVID-19 depends not only on one's immunity to the disease but also on the immunity within the population.

The consistency and no interference assumptions are sometimes grouped into the so-called stable unit treatment value assumption (SUTVA; \citenum{rubin_1980}) because their violation results in ill-defined causal effects. If all of the above assumptions are met, the ACE is identifiable and is the statistical quantity,

\begin{equation}
    \tau = \mathbb{E}[Y(1)] - \mathbb{E}[Y(0)] = \mathbb{E}[Y|T=1] - \mathbb{E}[Y|T=0].
    \label{eq:average_causal_effect_statistical}
\end{equation}

\noindent
We explain these causal assumptions in detail in the context of our study in Appendix \ref{sec:validation}.

\subsection{Biases and adjustments}
The gold standard for causal inference is a randomised control trial (RCT; \citenum{chalmers_1981}). A well-conducted RCT outputs a true measure of the ACE because the causal assumptions are met by construction. However, it is not always possible to perform RCTs because they can be unethical, infeasible, or outright impossible. More often than not, only observational data is available that is prone to many biases, unlike experimental data. The biases violate the causal assumptions and distort the true causal effect. Here, causal graphs truly come into their own as they make it easy to identify such biases and adjust for them such that the causal assumptions hold, resulting in valid estimates of the causal effect. There are many different types of biases, but the primary two are confounding bias and selection (or collider) bias.

\subsubsection{Confounding bias}
\label{sec:confounding_bias}
\begin{figure}
    \centering
    \includegraphics[width=0.6\textwidth]{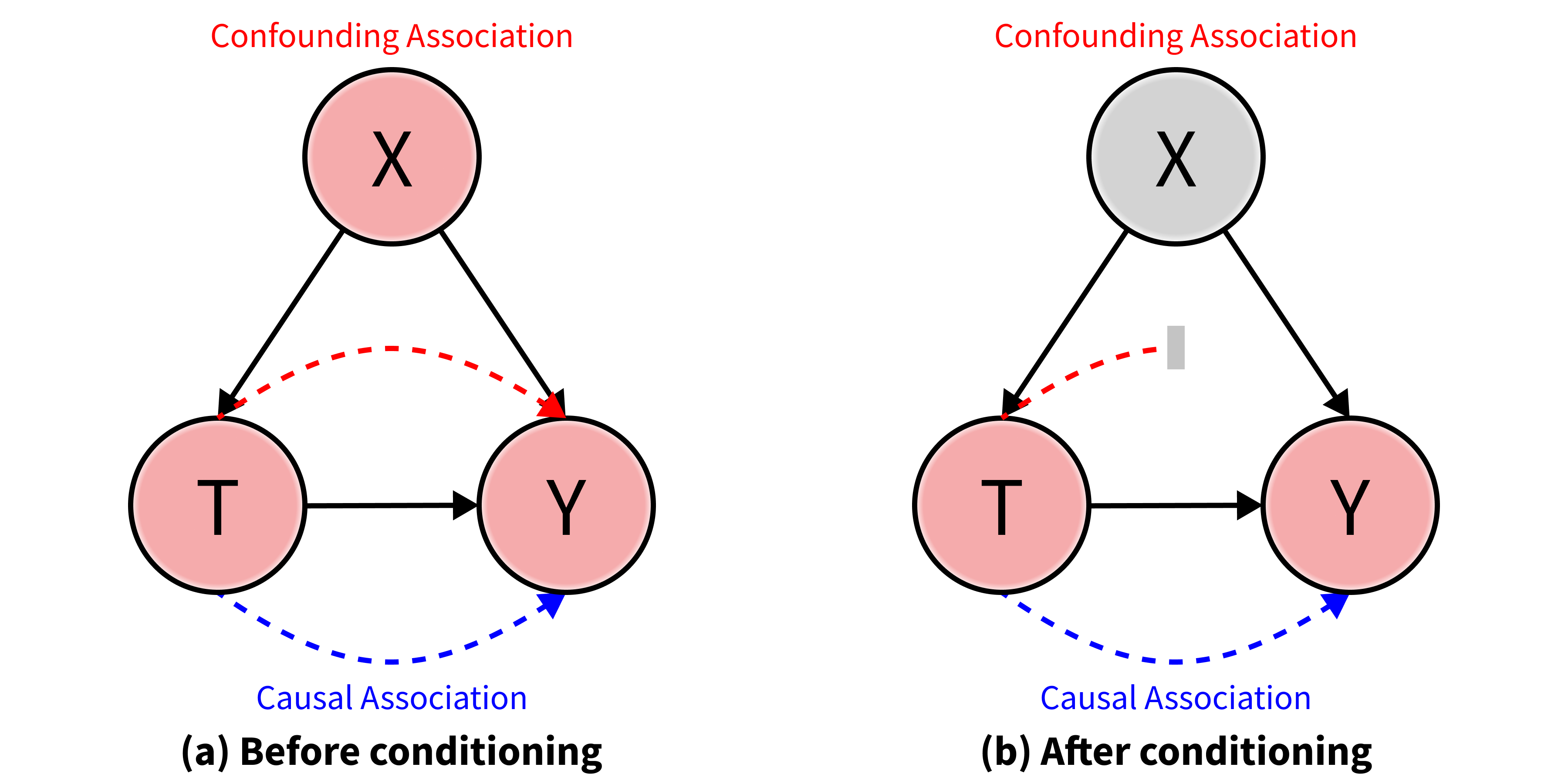}
    \caption{Illustration of confounding bias. DAGs representing the causal relationships between treatment $T$, outcome $Y$, and their common cause or confounder $X$. (a) There are two paths for association to flow between $T$ and $Y$: (i) the direct path between $T$ and $Y$ and (ii) the backdoor path linking $T$ and $Y$ via $X$. The causal association (depicted with the blue dashed line arrow) flows through the former, while the non-causal confounding association (depicted with the red dashed line arrow) flows through the latter. The admixture of the causal and non-causal associations means association is not causation. (b) The act of conditioning on $X$ (visualised with the greyed-out node) blocks the non-causal confounding association from flowing via the backdoor path.}
    \label{fig:confounding_bias}
\end{figure}

Confounding bias arises in the presence of a common cause or confounder $X$ that causes both the treatment and the outcome, as illustrated in Fig. \ref{fig:confounding_bias}. Unlike the DAG in Fig. \ref{fig:treatment_outcome_dag}, there are two paths for association to flow between $T$ and $Y$: (i) the direct path between $T$ and $Y$ and (ii) the backdoor path linking $T$ and $Y$ via $X$. The causal association flows through the former, and the non-causal confounding association flows through the latter. The amalgam of causal and non-causal associations means association is not causation, and the causal effect is biased. Specifically, the causal effect is an admixture of the causal effects of the treatment and confounder. Intuitively, if age influences the treatment and outcome in the aforementioned vaccine–disease example, then it is difficult to separate the causal effect that age has on the outcome from the causal effect of the treatment. In terms of the causal assumptions, the presence of confounders violates exchangeability because the treatment is not independent.

In experimental data, confounding is not an issue as RCTs remove its effect via randomisation of the treatment. In DAGs, treatment randomisation translates to removing the direct edge from $X$ to $T$, making $T$ independent, so confounding association cannot flow via the backdoor path as it does not exist. As a result, association is causation because exchangeability holds, and the causal effect does not suffer from confounding bias. In contrast, confounding is a major issue in observational data because, by its nature, the treatment is not randomised beforehand via experimentation. The goal with observational data is not to solve the issue directly but rather to negate it by adjusting the data such that the estimates of causal effects are bias-free. The first step is to modify the causal assumptions to be appropriate for observational data. For example, conditional exchangeability must hold for observational data, as the prevalence of confounders will always violate exchangeability. Conditional exchangeability states that potential outcomes are independent of the treatment given confounders.

\begin{equation}
    Y(t) \perp\!\!\!\perp T|X.
\end{equation}

\noindent
Visually, conditioning on a confounder blocks the non-causal confounding association from flowing from $T$ to $Y$ via the backdoor path, as shown in Fig. \ref{fig:confounding_bias}b, leaving only the causal association. Also, as exchangeability and confounding are intertwined concepts, the conditional exchangeability assumption is sometimes referred to as unconfoundedness. An alteration of the positivity assumption is also necessary to account for confounding. Following on from the original definition, the conditional probability of receiving treatment given confounders must be greater than zero and less than one.

\begin{equation}
    0 < P(T|X) < 1.
\end{equation}

\noindent
Given the prior consistency and no interference assumptions, in addition to conditional exchageability and positivity, the ACE

\begin{equation}
    \tau = \mathbb{E}[Y(1)] - \mathbb{E}[Y(0)] = \mathbb{E_{X}}[\mathbb{E}[Y|T=1, X] -\mathbb{E}[Y|T=0,X]].
    \label{eq:adjustment_formula}
\end{equation}

\noindent
This is known as the adjustment formula because adjustments are made post-data generation to infer true, unbiased causal effects.

\begin{figure}
    \centering
    \includegraphics[width=0.6\textwidth]{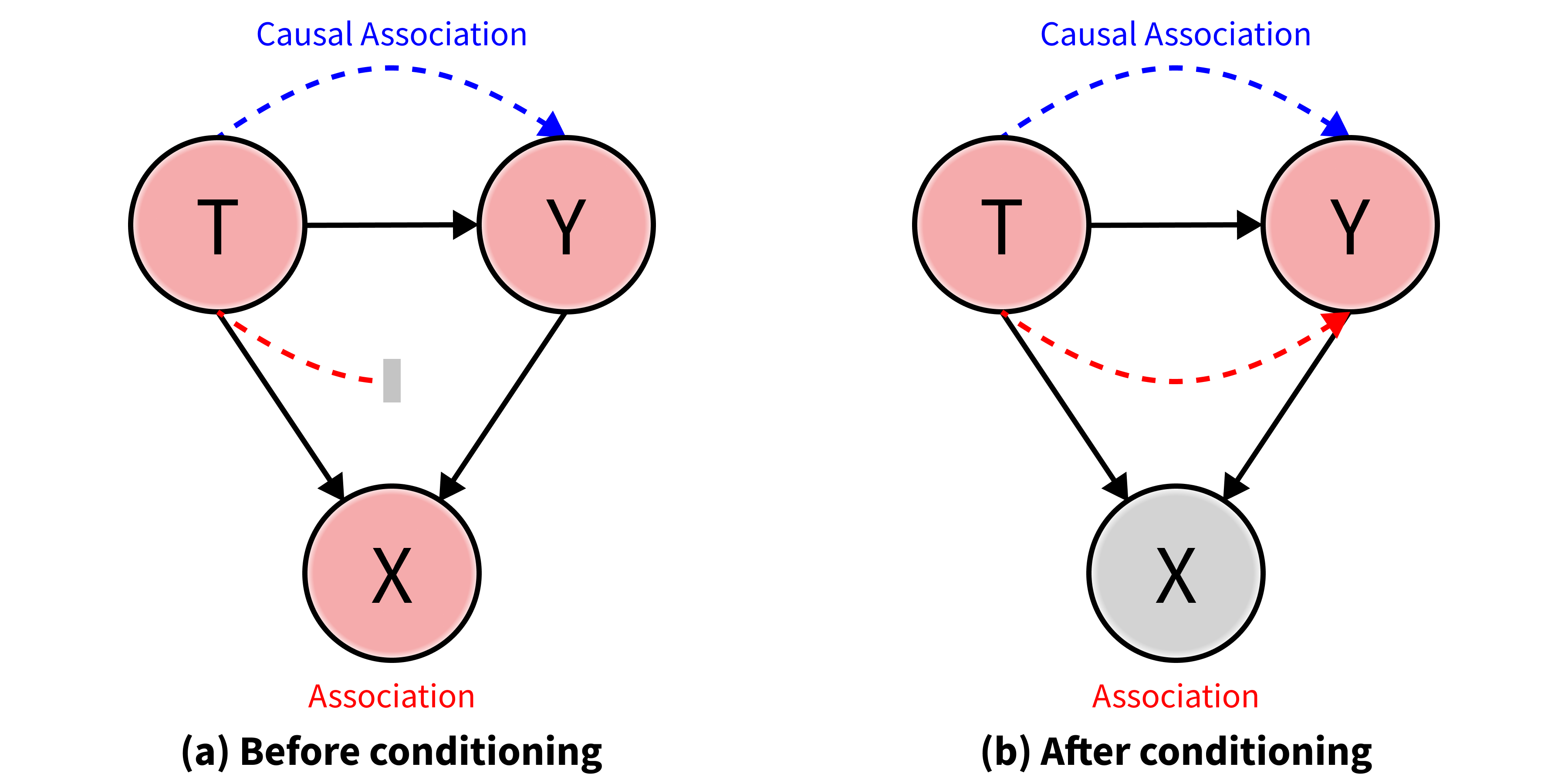}
    \caption{Illustration of selection (or collider) bias. DAGs representing the causal relationships between treatment $T$, outcome $Y$, and their common effect or collider $X$. (a) Similar to Fig. \ref{fig:confounding_bias}, there is a direct path and a backdoor path between $T$ and $Y$ for association to flow. As before, the causal association (depicted with the blue dashed line arrow) flows through the former. However, the non-causal association (depicted with the red dashed line arrow) cannot flow through the latter as it is now blocked because of the collider. (b) The act of conditioning on $X$ (visualised with the greyed-out node) unblocks the previously blocked backdoor path, allowing once again the non-causal association to flow. As a result, association is not causation as it is once again an admixture of the causal and non-causal associations.}
   \label{fig:selection_bias}
\end{figure}

\subsubsection{Selection bias}
\label{sec:selection_bias}
While confounding bias persists when there is a lack of adjustment of a common cause, selection bias occurs precisely due to adjustment of a common effect $X$, as illustrated in Fig. \ref{fig:selection_bias}. A common effect is a variable that is caused by both the treatment and the outcome. As previously, there is a direct path and a backdoor path between $T$ and $Y$ for association to flow. The causal association flows through the former as before, but the non-causal association does not flow through the latter as it is now a blocked path. The flow of association from $T$ and $Y$ `collides' on $X$, as shown in Fig. \ref{fig:selection_bias}a. Hence, $X$ is also referred to as a collider. In this scenario, association is causation, and the causal effect is not biased. By incorrectly conditioning on $X$, the backdoor path is unblocked, allowing the non-causal association to flow as shown in Fig. \ref{fig:selection_bias}b, which ultimately induces selection bias.

Continuing the vaccine–disease example: assume the vaccine has side effects and can cause hospitalisations in rare cases. The disease can also cause hospitalisations by deteriorating the health of individuals. Fig. \ref{fig:selection_bias} represents this exact situation, where hospitalisation $X$ is the common effect of the vaccine $T$ and disease $Y$. Conditioning on $X$ by selecting only the hospitalised patients induces a non-causal association between the vaccine and disease. Specifically, a positive association between the vaccine and disease would be observed as the hospitalised population is more likely to be vaccinated or have the disease than the general population. The conclusion that one would draw from the selected data is that the vaccine causes the disease, which would be detrimental as it would dissuade people from receiving the vaccine. In terms of the causal assumptions, conditioning or selecting on the common effect also violates exchangeability.

\renewcommand{\thefigure}{B\arabic{figure}}% Figure counter representation
\renewcommand{\theHfigure}{B\arabic{figure}}% Hyperref figure hyperlink hook

\section{Constructing the causal model of galaxy formation and evolution}
\label{sec:galaxy_formation_and_evolution}
In this section, we construct a causal model of galaxy formation and evolution. We assume the cold dark matter (CDM) paradigm, in which galaxies form and evolve in dark matter haloes \citep{white_1978, efstathiou_1983, blumenthal_1984}. To build our causal model, we review established theories of galaxy formation and evolution, and in particular ideas from semi-analytic modelling (SAM; \citenum{white_1991, cole_1991, kauffmann_1993, cole_1994, kauffmann_1999, somerville_1999, springel_2001, hatton_2003, springel_2005, kang_2005, lu_2011, benson_2012, herniques_2015}; also see \citenum{baugh_2005, benson_2010}, for reviews), and express them as causal graphs. We carefully consider all the relevant physical processes and assemble the causal model step-by-step with mini causal models before connecting all the pieces. 

\begin{figure}
    \centering
    \includegraphics[width=\textwidth]{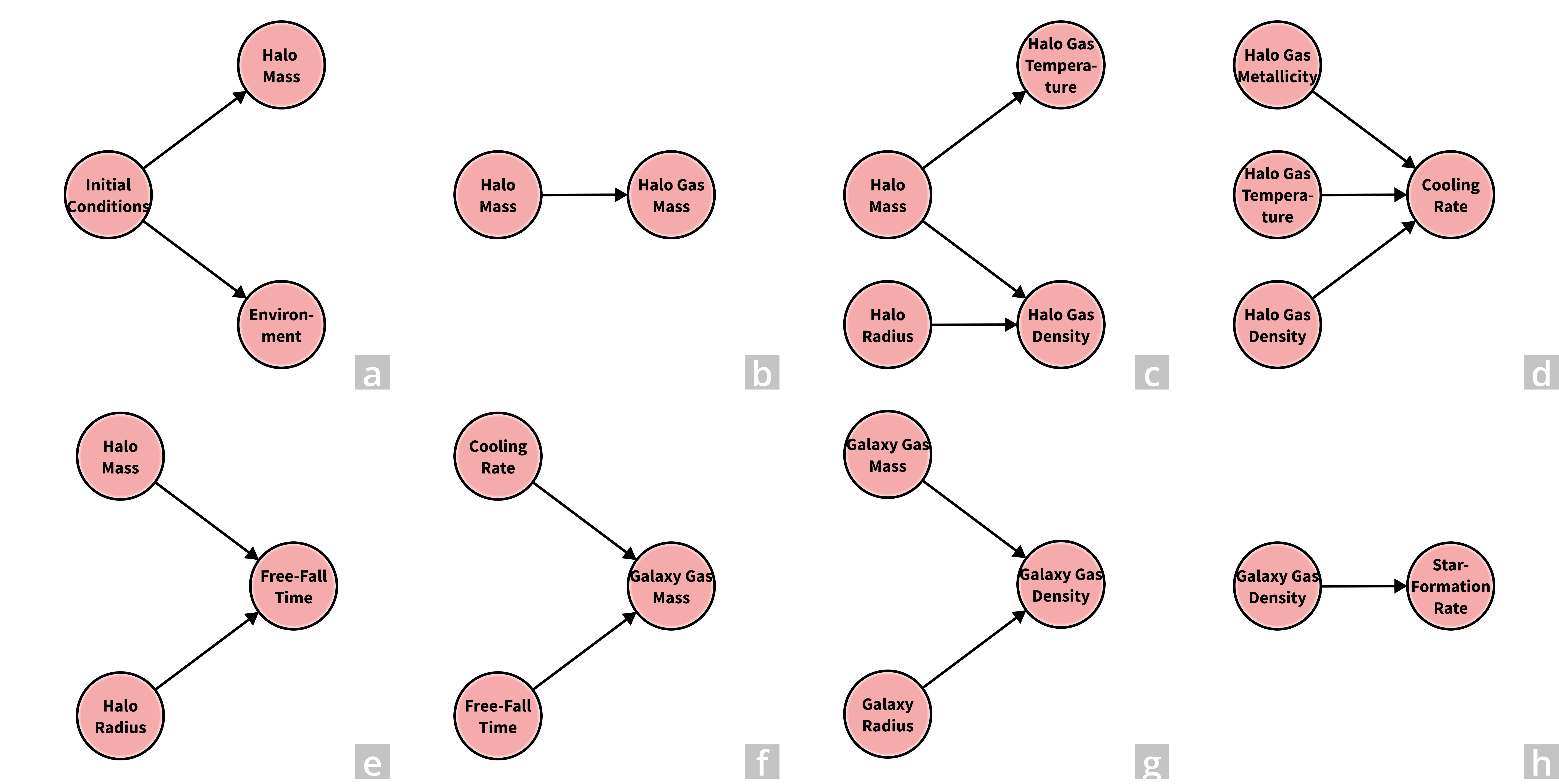}
    \caption{Mini causal models of different stages of galaxy formation and standard physical processes occurring in galaxies. The nodes are variables, and the edges communicate the causes. For visual clarity, not all the possible causal connections of a variable with others are drawn.} 
    \label{fig:mini_causal_models_galaxy_formation}
\end{figure}

We adopt a straightforward naming convention in the causal model: any variables associated with the halo and galaxy are preceded by them, respectively. Furthermore, halo refers to the dark matter halo that hosts a galaxy, and host halo refers to the parent dark matter halo that hosts other haloes. As such, halo refers to both distinct haloes and subhaloes. In the following section, we describe the galaxy formation process. Fig. \ref{fig:mini_causal_models_galaxy_formation} shows the mini causal models of the different stages of galaxy formation and standard physical processes occurring in galaxies.

\subsection{Galaxy formation}
\label{sec:galaxy_formation}
In the very early universe, quantum fluctuations of the scalar field drive inflation and generate density perturbations in the initial matter density field, sowing the seeds for galaxy formation \citep{guth_1982, hawking_1982, linde_1982, starobinsky_1982, bardeen_1983}. The small perturbations evolve under gravitational instability as regions of space with above-average density attract matter and become denser over time. Conversely, regions of space with below-average density lose matter and become rarefied over time. The outcome is the amplification of the initial density contrast. 

Once a region reaches over-density ($\delta{\rho}/\rho \sim 1$), it breaks away from the cosmological expansion and collapses to form a dark matter halo \citep{mo_2010}. The primordial haloes are small as perturbations on the smallest scales collapse first \citep{benson_2010}. The mass and environment of the dark matter halo are a product of the evolution of the initial matter density field, or more specifically, the amplitude and pattern of the initial density perturbations, respectively. In the causal model, we loosely label this as ``initial conditions'': the `cause' of the initial haloes and their environment (a). Given that galaxies form in dark matter haloes, we associate nature with halo mass, and nurture to environment.

The ordinary baryonic matter falls into the gravitational potential well of the dark matter halo and is shock-heated to the haloes' virial temperature to produce a hot gas halo that is supported against further collapse by the pressure of the gas. Thus, halo gas mass and temperature depend on halo mass (b, c). Subsequently, the hot gas can cool through various mechanisms \citep{kauffmann_1993}, which removes the pressure support and causes the gas to sink to the centre of the gravitational potential well \citep{silk_1977, rees_1977, binney_1977, white_1991, cole_1991, lacey_1991}. If the angular momentum is conserved during the cooling process, the gas spins up as it flows inwards and forms a rotationally supported disc \citep{fall_1980, mo_1988}. 

Primarily, two factors determine the mass of cold gas in the disc: (i) the cooling rate (i.e., the mass of gas cooled per unit time) and (ii) the free-fall time (i.e., the time taken for the cooled gas to transfer from the halo to the disc) (f). The cooling rate depends on the metallicity, temperature, and density of the halo gas (d). Specifically, the temperature and density determine the ionisation state and collision rate, respectively. The free-fall time depends on the halo mass and radius (e). 

As the gas accumulates, its self-gravity dominates over that of the dark matter—and it collapses. The exact process of star formation from a self-gravitating gas cloud is unknown, but there are two theories. In the bottom-up theory, low-mass stellar cores acquire gas from the cloud in a competitive accretion process \citep{bonnell_1997}, while in the top-down theory, the gas cloud simply fragments and the sub-clouds collapse to form stars \citep{krumholz_2005}. Independent of the exact model, the star-formation rate (SFR) depends on the local density of cold gas \citep{schmidt_1959, kennicutt_1998} (g, h). This is the standard paradigm of galaxy formation. 

We remark that halo accretion also depends on environment, and generally, the halo and forming galaxy are subject to external processes (discussed in Section \ref{sec:external_processes}). Thus, galaxy formation depends not only on nature but also nurture.

\subsection{Galaxy evolution}
\label{sec:galaxy_evolution}
\begin{figure}
    \centering
\includegraphics[width=\textwidth]{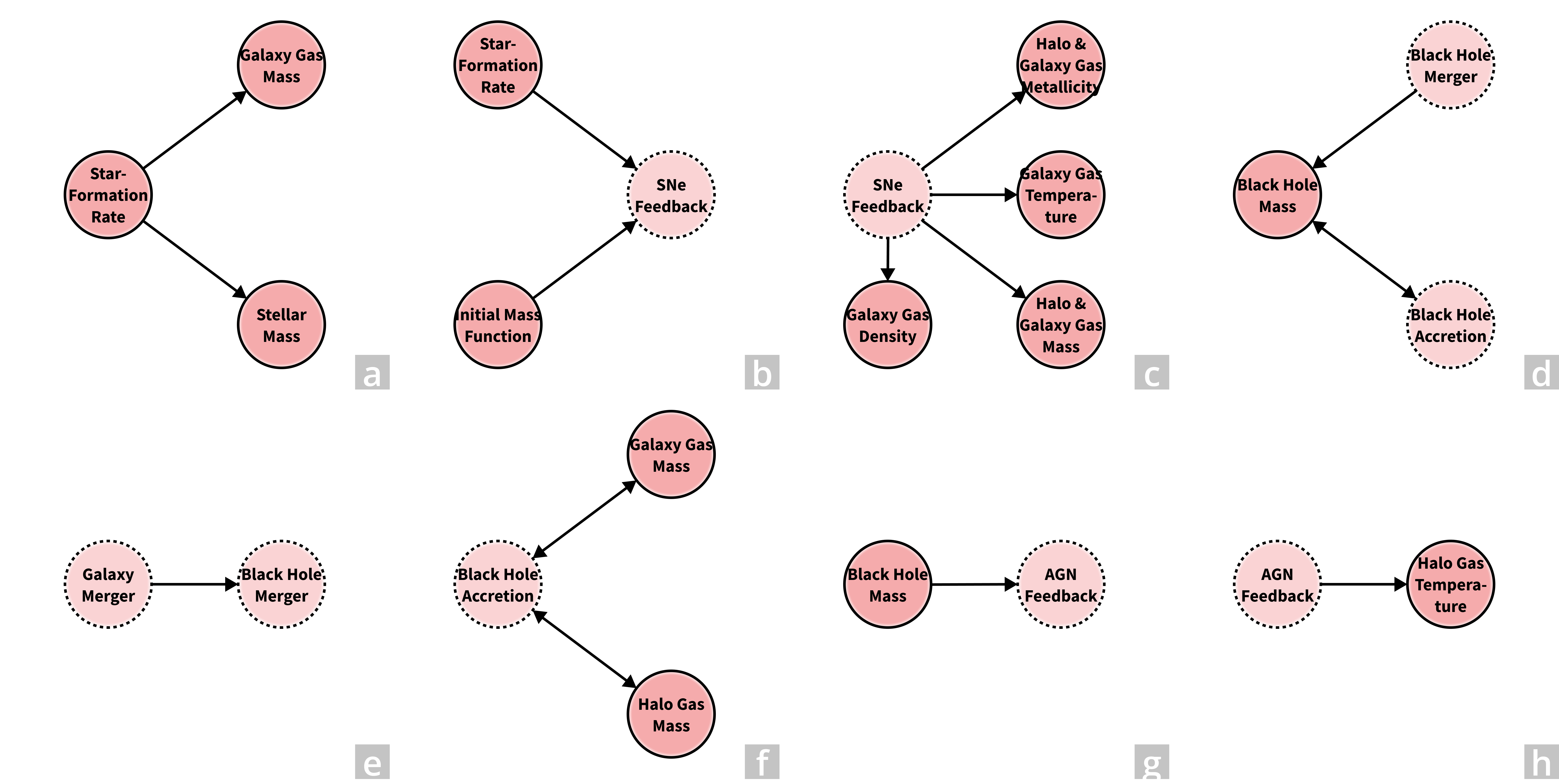}
    \caption{Mini causal models of different internal processes of galaxy evolution. The nodes are variables (solid border) or processes (dashed border), and the edges communicate the causes (with single and double arrows representing unidirectional and bidirectional causality, respectively). For visual clarity, not all the possible causal connections of a variable with others are drawn.} 
    \label{fig:mini_causal_models_galaxy_evolution_internal}
\end{figure}

\begin{figure}
    \centering
    \includegraphics[width=\textwidth]{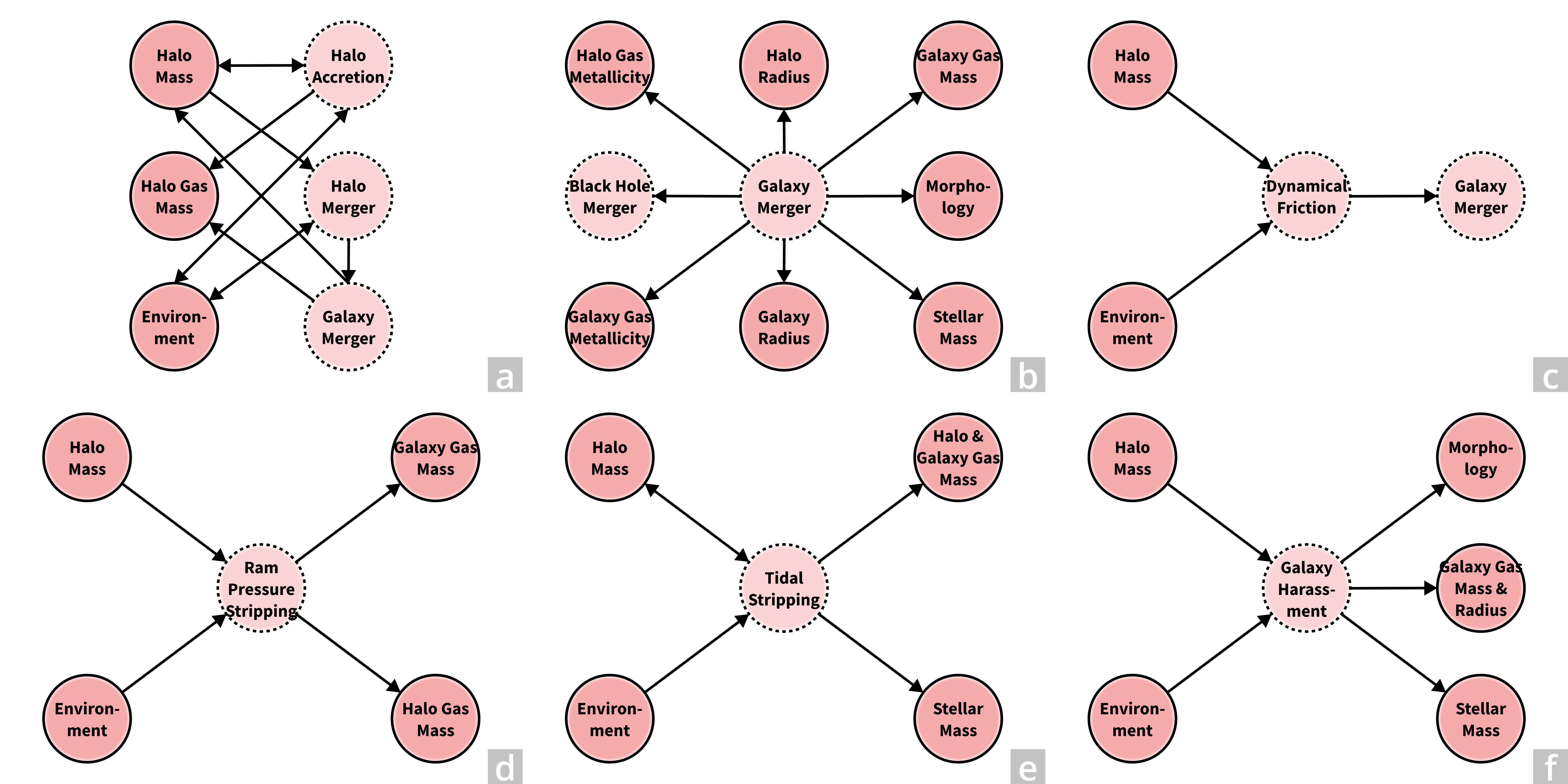}
    \caption{Mini causal models of different external processes of galaxy evolution. The nodes are variables (solid border) or processes (dashed border), and the edges communicate the causes (with single and double arrows representing unidirectional and bidirectional causality, respectively). For visual clarity, not all the possible causal connections of a variable with others are drawn.} 
    \label{fig:mini_causal_models_galaxy_evolution_external}
\end{figure}

In this section, we describe the internal and external processes that shape the evolution of galaxies. We do not attempt to model the different processes in detail or as accurately as possible because our goal is to estimate the overall causal effect of environment rather than of individual processes. Instead, we focus on conveying how the processes are related to halo mass and environment and their impact on galaxy properties, especially SFR. Figs. \ref{fig:mini_causal_models_galaxy_evolution_internal} and \ref{fig:mini_causal_models_galaxy_evolution_external} show the mini causal models of internal and external processes related to galaxy evolution, respectively.

\subsubsection{Internal processes}
\label{sec:internal_processes}
As stars form, the stellar mass of a galaxy increases, and the amount of cold gas available for future star formation decreases by construction (a). The consequence of the feedback loop between galaxy gas mass and SFR is that without further accretion of gas, a galaxy will eventually die as it exhausts its cold gas and star formation ceases.

Besides the natural evolution, feedback from massive stars can actively shape a galaxy's evolution and accelerate its demise. The most massive stars explode in a supernova at the end of their lives, and the resulting feedback \citep{larson_1974, dekel_1986} can both positively and negatively affect SFR \citep[see][for an overview]{hopkins_2014}. For example, supernova-driven galactic winds heat the interstellar medium (ISM) and eject cold gas from the disc back to the halo, or in the extreme case, out of the halo altogether, thus suppressing star formation \citep{heckmann_1990, martin_1999, scannapieco_2008}. Conversely, the blast waves may compress the cold gas to temporarily boost star formation. Supernovae (SNe) feedback also ejects material, which enriches the halo and galaxy gases. A more metal-rich halo gas increases the cooling rate (or shortens the cooling time), which may ultimately lead to increased star formation (c).

Supermassive black holes (SMBHs) are also important in the evolution of a galaxy because they are responsible for AGN feedback \citep{dekel_1986}. There are two main modes of AGN feedback: the quasar mode and the radio mode. In the quasar mode, a SMBH grows via accretion of cold gas and mergers with other SMBHs (in galaxy mergers). In the radio mode, SMBHs accrete gas directly from the halo and release a vast amount of energy, heating the halo gas and suppressing cooling \citep{croton_2006} (f, h). In both modes, AGN feedback negatively impacts star-formation activity by diminishing the cold gas. Nevertheless, like SNe feedback, there are mechanisms whereby AGN feedback can boost SFR \citep[see][for recent reviews]{fabian_2012, heckman_2014}. In the causal model, AGN feedback represents only the `output' processes.

SNe feedback depends on the initial mass function (IMF) and the SFR (b). The IMF dictates the overall fraction of stars that end up as supernovae, while the SFR determines the overall number. On the other hand, the picture for AGN feedback is far less clear as it depends on how, when, and where SMBHs form. Nevertheless, AGN feedback must scale with the mass of the SMBH \citep{soltan_1982, silk_1998} (g), which depends on accretion and merger rates, but the reverse is true for the former as well, so there is a feedback loop (d). The accretion rate depends on halo and galaxy gas masses (f), while the merger rate is broadly determined by the number of galaxy mergers (e), which as will be discussed in detail in the following section, depends on halo mass and environment. In summary, there is believed to be a causal connection between AGN feedback and star-formation activity.

Overall, there are feedback cycles between halo and galaxy gases, SFR, and feedback. For example, an increase in cold galaxy gas from enhanced cooling of hot halo gas boosts star formation. A fraction of the stars born explode in a supernova, determined by the IMF, and the resulting feedback expels and/or heats the cold gas in the galaxy, transferring it back to the halo, which in turn reduces star formation. We emphasise that while the feedback theories discussed are likely to resemble reality, the fact is that the precise mechanisms are unknown. Furthermore, it is still unclear how SMBHs form in the first place. 

In the following section, we describe the external environmental processes that shape a galaxy's evolution \citep[see][for a review]{boselli_2006}. There are many definitions of environment, but ultimately one means the mass density field. Thus, we bear this in mind to derive the causal model.

\subsubsection{External processes}
\label{sec:external_processes}
Many processes associated with environment influence galaxies, but two of the most fundamental are accretion and mergers. The halo accretion and merger rates depend on halo mass and environment. For example, a massive halo is able to accrete more matter, but environment also plays a role since it determines the amount available for accretion. Also, a massive halo in a dense environment has to compete with neighbouring haloes to attract matter. Accretion changes halo mass and environment (assuming the local mass density field), so there is a feedback loop. Undoubtedly, a feedback loop also exists between halo mass, environment, and mergers. However, unlike accretion, mergers may not affect halo mass depending on the merger type (clarified below). In summary, halo mass (i.e., nature) and environment (i.e., nurture) influence each other via accretion and mergers in a feedback loop. In this context, we define the nature versus nurture debate in this Article (a).

Mergers are broadly categorised into two types: major and minor. A major merger occurs when the progenitors are of similar masses, and in such a merger, the progenitor haloes and galaxies merge violently to form a more massive halo with a new galaxy residing at its centre. If the progenitors are disc galaxies with a mass ratio $1:1$, then the post-merger remnant resembles an elliptical \citep{toomre_1972, toomre_1977, hernquist_1992, hernquist_1993, barnes_1998, barnes_2002, cox_2006}. Later on, if the shock-heated and ejected gas from outflows cools with significant angular momentum, a disc forms, and then the post-merger remnant resembles an early-type spiral galaxy with a disc-bulge system \citep{hopkins_2009a, sparre_2016, pontzen_2017}. A galaxy merger is a cause of morphological transformation (b). 

A period of star formation activity follows a major merger if the progenitor galaxies contain large quantities of cold gas. In the short term, the influx of cold gas and/or an increase in the galaxy gas density due to interactions between galaxies trigger starbursts \citep{mihos_1994, mihos_1996, hopkins_2006, hopkins_2008a, hopkins_2008b, snyder_2011, hayward_2014, sparre_2016}. Also, the halo gas shock-heated during the merger has the opportunity to cool, leading to star formation in the long term. However, AGN feedback can prevent this from happening \citep{sanders_1998, dimatteo_2005, hopkins_2009b, treister_2012}. If the progenitor galaxies harbour SMBHs, they may merge in the process. Additionally, the same influx of gas that fuels star formation can feed the SMBH. The subsequent growth of the SMBH from the merger and accretion can consume any leftover halo and galaxy gases (in a feedback loop) and suppress the halo gas from cooling, resulting in a galaxy devoid of star formation (b).

A minor merger occurs when the progenitors are of dissimilar masses, and in such a merger, the smaller galaxy is `absorbed' by the larger galaxy. In our causal model, we have defined two types of mergers: ``halo mergers'' and ``galaxy mergers''. As the names suggest, a halo merger refers to the merger of haloes, while a galaxy merger refers to the merger of galaxies. As such, major and minor mergers are halo mergers followed by galaxy mergers in our causal model (a). We distinguish minor mergers into ``minor halo mergers'' and ``minor galaxy mergers''. In a minor halo merger, the haloes `merge' as the smaller halo orbits within the larger halo, but the galaxies may or may not. Accordingly, we refer to it as a halo merger, but not exclusively. As in simulations, we model the halo merger with the following perspective: the smaller and larger subhaloes occupy a common host halo that is a sum of its parts.

\noindent
In other words, the progenitor haloes retain their identity unless a galaxy merger follows a halo merger. Consequently, we continue referring to progenitor haloes (and the associated variables) post-merger as haloes rather than subhaloes of the host halo in the causal model.

Halo mergers are responsible for the formation of groups and clusters. In such environments, the central galaxy is the most massive and located near the centre, while the satellite galaxies orbit around it. Central galaxies reside deep in the gravitational potential well of the host halo, while satellite galaxies reside further out at different depths and distances. As a result, the environmental effect on these galaxies is asymmetrical as satellites experience most of the processes and not centrals. Satellite galaxies in these dense environments are subject to different processes that are either a consequence of, or scale with, the host halo mass.

A satellite galaxy experiences dynamical friction \citep{chandrasekhar_1943a, chandrasekhar_1943b, chandrasekhar_1943c} as it orbits within its host halo. The drag slows down the satellite, which causes it to spiral inwards and eventually merge with the central galaxy in a process called galactic cannibalism. The magnitude of dynamical friction depends on the satellite's halo mass and environment. For example, a satellite is subject to greater dynamical friction if it is more massive and occupies a larger host halo. Galactic cannibalism is the predominant environmental process that affects central galaxies (c).

Another critical process linked to environment is ram-pressure stripping (RPS; \citenum{gunn_1972}). As a satellite travels through the hot intergalactic medium (IGM) of groups and clusters, its relatively cold halo and galaxy gases encounter a hydrodynamical drag force due to the relative motion of the two fluids. If the drag force exceeds the satellites' restoring force, its cold gas is ablated. Accordingly, while environment causes RPS (assuming the amount and temperature of hot gas correlates), its extent also depends on halo mass. Evidently, the depletion of gas (hot and cold) negatively impacts the SFR, and a decline in star formation affects the visual morphology of the galaxy. More specifically, a satellite galaxy that is initially spiral may resemble a lenticular (S0) galaxy. Nonetheless, there is doubt whether the effects of RPS are negative and/or permanent. For example, there is evidence that any gas not stripped may be compressed by RPS to cause an increase in star formation activity in the disc \citep{dressler_1983, gavazzi_1995}. Additionally, the stripped gas may remain bound, to later fall back and induce starbursts \citep{vollmer_2001}. In retrospect, RPS is likely to be only partially responsible for the morphology–density and SFR–density relations \citep{abadi_1999} (d).

Groups and clusters comprise tens and upwards of hundreds of galaxies respectively, so gravitational interactions are a common occurrence. In group environments, a satellite galaxy experiences tidal forces from other galaxies. The tidal interactions can remove its cold and hot gases, stars, and dark matter via tidal stripping \citep{moore_1999}. As was the case with RPS, the effectiveness of tidal stripping depends on environment (the number of interactions is related to the density) and halo mass. Note that there is a feedback loop between tidal stripping and halo mass, so the process becomes more effective over time \citep{kampakoglou_2007} (e).

In cluster environments, satellite galaxies are typically not subject to tidal stripping and mergers (not resulting from dynamical friction) because strong gravitational interactions are infrequent due to their high velocities \citep{ghigna_1998}. Instead, they are subject to multiple, weak interactions, and the cumulative effect of such interactions is called galaxy harassment \citep{farouki_1981, moore_1996}. The high-speed encounters can impulsively heat the disc of a satellite galaxy, which pushes its stars onto elliptical orbits, and the disc transforms into a spheroidal component, thus altering its morphology \citep{moore_1998, gnedin_2003, mastropietro_2005, aguerri_2009}. At its extreme, the stars can become completely unbound, which decreases the galaxy's stellar mass \citep{smith_2010, smith_2015, bialas_2015}. The heating of the disc naturally impacts the SFR as it affects the density of the cold gas (via a change in the galaxy gas mass and/or radius). Galaxy harassment scales with the number and strength of interactions, which depends on environment, and resistance to its effects depends on halo mass, like tidal stripping (f).

The hot halo surrounding a galaxy is in constant flux as the gas condenses to form stars, and the subsequent feedback returns the cold gas back to the halo. Simply put, the hot gas acts as a reservoir for future star formation. The combination of RPS and tidal stripping can annihilate this reservoir, and without further accretion in a dense environment, a satellite galaxy eventually stops forming stars as it exhausts its fuel. This process is called strangulation \citep{larson_1980, benson_2000}. We do not have a specific node for it in our causal model because it is not a process in and of itself and is already captured by the existing variables. Also, we do not have nodes for thermal evaporation \citep{cowie_1977} and viscous stripping \citep{nulsen_1982} because they are variants of RPS. Furthermore, whilst important, variables such as colour, stellar metallicity, and luminosity are not included as they are not the subject of this study.

Finally, we have not precisely defined or modelled morphology and related processes because our target is the SFR–density relation. Nonetheless, internal dynamical effects can change the morphology of galaxies, a well-known example being the bar instability. A thin disc with high surface density is susceptible to a non-axisymmetric instability, which creates a bar-like structure. Bars can funnel gas to the central region of a galaxy \citep{athanassoula_1992}, which can fuel AGNs and star formation \citep{zurita_2004, sheth_2005}. The bar may also buckle to produce a ``pseudobulge'' \citep[see][for a review]{kormendy_2004}, which can prevent the disc from collapsing and forming stars \citep{fang_2013}. Additionally, there is a strong connection between bulges and SMBHs (\citenum{kormendy_1995, magorrian_1998, ferrarese_2000, gebhardt_2000, haring_2004}; also see \citenum{kormendy_2013}, for a review). Thus, morphology can directly and indirectly influence SFR (and vice versa). Still, detailed modelling of morphology is unnecessary to estimate the SFR–density relation because it is not a confounding factor given that it does not also impact environment. In fact, controlling for morphology may induce selection bias (Fig. \ref{fig:selection_bias}) since it is a common effect of environment and SFR.

The key findings are: (i) galaxy formation and evolution depends on nature and nurture, (ii) nature (i.e., halo mass) and nurture (i.e., environment) influence each other through accretion and mergers, and (iii) internal processes associated with nature also depend on nurture, and external processes associated with nurture also depend on nature, as both halo mass and environment determine their impact on galaxies. In conclusion, nature and nurture are heavily intertwined. Fig. \ref{fig:causal_model_galaxy_formation_and_evolution} shows the causal model of galaxy formation and evolution in its entirety with all the mini causal models and variables connected. 

\renewcommand{\thefigure}{C\arabic{figure}}% Figure counter representation
\renewcommand{\theHfigure}{C\arabic{figure}}% Hyperref figure hyperlink hook

\section{Inverse probability weighting of marginal structural models}
\label{sec:ipw_of_msms}
In this section, we describe the causal inference method we apply to disentangle nature and nurture and estimate the causal effect of enviroment on SFR: inverse probability weighting (IPW) of marginal structural models (MSMs; \citenum{robins_2000}). We begin by introducing its key component—the propensity score.

\subsection{Propensity score}
The propensity score (PS; \citenum{rosenbaum_1983}) is the conditional probability of treatment given covariates $X$,

\begin{equation}
    e(x) = p(t|x) = P(T=1|X=x).
\end{equation}

\noindent
An extension of the propensity score to continuous treatments is the generalized propensity score (GPS; \citenum{hirano_2004}),

\begin{equation}
    e(t, x) = f(t|x),
\end{equation}

\noindent
where $f(t|x)$ is the conditional probability density function (PDF). For conciseness, we refer to the GPS as simply the propensity score from hereon. Furthermore, we denote $p(t|x)$ as $f(t|x)$ in the equations even when the treatment is binary because the concept is the same. A propensity score close to zero or one means there is a low or high probability of receiving the specific treatment given covariates, respectively. Essentially, the propensity score represents the dependence of treatment on covariates. As treatment dependence correlates with confounding bias when the covariates are confounders, the propensity score can adjust for confounding to estimate causal effects. There are four known techniques to adjust using the propensity score \citep[see][for overviews]{agostino_1998, austin_2011}: 

\begin{enumerate}
    \item \textit{Matching} – Units from the treatment group are matched to their counterparts in the control group based on their propensity scores. This process makes the treatment and control groups comparable in terms of their covariate distributions, which ultimately means they are exchangeable. As explained in more detail in Appendix \ref{sec:causal_assumptions}, exchangeability ensures no confounding.
    \item \textit{Stratification} – The population is divided into distinct strata or subgroups based on the propensity score. This negates any confounding effect because within each stratum the level of confounding is similar. 
    \item \textit{Covariate adjustment} – The propensity score is included along with the treatment as a covariate in a model to predict the outcome. 
    \item \textit{IPW} – Units are weighted according to their propensity score. We discuss how this eliminates confounding bias in the following section.
\end{enumerate}

Previous studies have employed matching, stratification, and covariate adjustment to eliminate confounding bias, but just not with the propensity score. For example, the common approach of binning galaxies according to their stellar mass is a form of stratification. Also, galaxies have been matched on redshift and stellar mass when creating a treatment and control group \citep{ellison_2008, smethurst_2017, garduno_2021, sotillo_2021}, and stellar mass has been included as a variable in models \citep{teimoorinia_2016, bluck_2019, bluck_2020a, bluck_2020b, bluck_2022, brownson_2022, piotrowska_2022}. We use the IPW approach \citep[see][for an overview]{chesnaye_2022} as the other techniques are either unable or unsuitable to disentangle nature and nurture. Stratification and covariate adjustment are conditional approaches \citep{williamson_2017} and thus cannot estimate the joint effect, and there is no clear strategy with matching \citep{thoemmes_2016}. The conditional approach is to block the backdoor path between the treatment and outcome by conditioning on the confounder (as described in Section \ref{sec:confounding_bias}). The act of conditioning stops the flow of non-causal confounding association via the confounder (Fig. \ref{fig:confounding_bias}b), which means that all association is causal and the causal effect is unbiased. See Section \ref{sec:causal_effects} for definitions of the marginal and joint causal effects.

\subsection{Inverse probability weighting}
In this section, we first describe the IPW method for time-fixed treatments and then extend it to time-varying treatments.

\subsubsection{Time-fixed treatments}
IPW is a statistical technique that adjusts for confounding bias by weighting each unit with the inverse of their probability of receiving treatment, i.e., the propensity score.

\begin{equation}
    w(t) = \frac{1}{f(t|x)}.
    \label{eq:unstabilised_weight}
\end{equation}

\noindent
Intuitively, the propensity score quantifies the magnitude of confounding bias, so weighting each unit with its propensity score directly negates the influence of confounders. Specifically, the method works as follows: a unit with a high propensity score implies the treatment it received is likely given the confounders. In other words, the influence of the confounders is significant, so the unit is down-weighted to reduce its impact. Conversely, a unit with a low propensity score implies that the treatment it received is unlikely given the confounders. Crucially, such a unit is a counterfactual of units that received a different treatment, and thus it is up-weighted because it holds valuable information. In a sense, IPW is comparable to the technique of importance sampling \citep{kloek_1978}. Overall, assigning weights to each unit creates a pseudo-population in which treatment is independent of confounders, and thus IPW is a marginal approach. The marginal approach is to remove the backdoor path entirely by making the treatment independent of the confounder, as achieved experimentally with RCTs. Visually, this translates to no direct edge (i.e., arrow) from the confounder to the treatment in a DAG. By removing the backdoor path, confounding bias is eliminated altogether.

Units with specific characteristics that predispose them to a particular treatment, or from the alternative viewpoint, units subject to a treatment confined to a subpopulation, will have propensity scores close to zero or one because of the strong causal association between the covariates and treatment. Consequently, a disproportionately small fraction of units can dominate and drastically skew the causal effect. A simple solution is to truncate or trim the extreme weights from the analysis, typically at the 1st and 99th percentiles. However, a better approach is to stabilise the weights with the marginal probability of treatment $f(t)$ such that,

\begin{equation}
    w(t) = \frac{f(t)}{f(t|x)}.
    \label{eq:stabilised_weight}
\end{equation}

\noindent
Besides counteracting the effect of extreme weights, stabilised weights generally reduce the variance of causal effect estimates \citep{robins_2000}. Furthermore, when the treatment is continuous, unstabilised weights are not an option as they have infinite variance \citep{robins_2000}. For binary or discrete treatments, 

\begin{equation}
    \mathbb{E}[Y(t)] = \mathbb{E} \left[ \frac{\mathbbm{1}(T=t)Y}{f(T=t|X)} \right],
    \label{eq:ipw_estimand}
\end{equation}

\noindent
where $\mathbbm{1}$ is an indicator function that is $1$ if $T=t$ and $0$ otherwise. Thus, the ACE of a binary treatment,

\begin{equation}
    \tau = \mathbb{E}[Y(1) - Y(0)] = \mathbb{E} \left[ \frac{\mathbbm{1}(T=1)Y}{f(T=1|X)} \right] - 
\mathbb{E} \left[ \frac{\mathbbm{1}(T=0)Y}{f(T=0|X)} \right].
\end{equation}

\noindent
Using the Horvitz-Thompson estimator \citep{horvitz_1952}, 

\begin{equation}
    \hat{\tau} = \frac{1}{N} \sum \limits_{i=1}^{N}  \left( \frac{\mathbbm{1}(T_{i}=1)Y_{i}}{\hat{e}(X_{i})} - 
\frac{\mathbbm{1}(T_{i}=0)Y_{i}}{1-\hat{e}(X_{i})} \right),
\end{equation}

\noindent
where $N$ is the number of units. Therefore, ACEs of binary treatments can be directly estimated using the weights. However, this is not the case for continuous treatments as the estimand (equation (\ref{eq:ipw_estimand})) is biased for $\mathbb{E}[Y(t)]$ and is not valid \citep{hernan_2023}. For continuous treatments, a model that describes the causal relationship between the treatment and outcome is necessary. One such class of causal models are MSMs. A MSM is a model for the potential outcome under treatment, for example,

\begin{equation}
    \mathbb{E}[Y(t)] = \beta_{0} + \beta_{1}t.
\end{equation}

\noindent
Unlike `regular' models, MSMs consider the expected outcome under different treatments, which is not observable due to the ``fundamental problem of causal inference''. Nonetheless, it is possible to reliably estimate MSMs with IPW adjustment because if the causal assumptions are met, the MSM is equal to:

\begin{equation}
    \mathbb{E}[Y|T] = \beta_{0} + \beta_{1}T.
\end{equation}

\noindent
The parameters of MSMs have causal interpretations. For example, $\beta_{1}$ represents the ACE in the case of binary treatments. To summarise, weights are applied to fit a MSM to estimate ACEs of continuous treatments (and also binary treatments). As there are an infinite number of values when a variable is continuous, the goal with continuous treatments is to estimate the causal dose-response curve (CDRC), $\mu(t) = \mathbb{E}[Y(t)]$, rather than a single causal effect $\tau$. 

The weights in their current form are valid for causal effects of time-fixed treatments and marginal causal effects of time-varying treatments but may not sufficiently adjust to estimate joint causal effects of time-varying treatments. Thus, we describe the extension of the method to joint effects in the following section.

\subsubsection{Time-varying treatments}
Time-varying treatments affected by time-varying confounders necessitate adjustments to eliminate confounding bias as before. However, unlike time-fixed treatments, simply adjusting for all the confounders together in the presence of treatment-confounder feedback fails when the goal is to estimate the joint effect. Suppose we adjust for halo masses $H_{k}$ since it is the time-varying confounder in the causal model (Fig. \ref{fig:causal_model}a). Accordingly, there are no direct paths from $H_{k}$ to time-varying treatments $E_{k}$, so there is no confounding bias. But, since the same paths constitute causal pathways of $E_{k}$ to outcomes $SFR_{k}$, the joint effect now suffers from over-adjustment bias. This scenario is exactly the same as encountered with conditional approaches where the joint effect is biased whether one adjusts for confounders or not. The solution to the dilemma is simple: adjust for biases step-by-step rather than all at once. 

With joint effects, the idea is to repeat the IPW process to adjust for biases at each time point. The exact method is as follows: estimate weights for each time point and then multiply them together to construct a final weight. This strategy eliminates confounding bias without introducing over-adjustment bias, so there is no overall bias. The method creates pseudo-populations at each time point, so the principle is the same as with time-fixed treatments. The general form of stabilised weights at time point $k (=j)$ \citep{robins_2000},

\begin{equation}
    w_{j} = \prod_{k=0}^{j} \frac{f(T_{k}|\bar{T}_{k-1})}{f(T_{k}|\bar{T}_{k-1}, \bar{X}_{k})},
    \label{eq:stabilized_weights_time_varying_general}
\end{equation}

\noindent
where $T_{k}$ is the treatment at time point $k$, $\bar{T}_{k-1}$ is the treatment history up until the time point, and $\bar{X}_{k}$ is the confounder history to the time point. The numerator is the conditional PDF of the current treatment given the previous treatment history, and the denominator is the conditional PDF of the current treatment given previous treatment and confounder histories. Compared to the weights for time-fixed treatments, the treatment history is conditioned on because time-varying treatments can be, and are, confounders if the previous treatments influence the current treatment and outcome from the perspective of a point in time. As before, the weights are applied to fit a MSM to estimate joint causal effects of interest. In summary, IPW of MSMs consists of: (i) estimating weights to adjust for biases and (ii) fitting a MSM using the weights to estimate causal effects. We apply the methodology to estimate the causal effect of environment.

\renewcommand{\thefigure}{D\arabic{figure}}% Figure counter representation
\renewcommand{\theHfigure}{D\arabic{figure}}% Hyperref figure hyperlink hook

\section{Validation}
\label{sec:validation}
In this section, we verify the validity of our results by qualitatively and quantitatively checking whether the key causal assumption stated in Section \ref{sec:causal_assumptions} are met. They are exchangeability, positivity, consistency, and no interference.

The exchangeability assumption states that potential outcomes must be independent of the treatment. In other words, it must be possible to swap treatment groups without changing their potential outcomes. To achieve exchangeability, it is necessary to adjust for any confounders and thus we adjusted for halo mass, the time-varying confounder in the causal model (Fig. \ref{fig:causal_model}a). Whether or not the assumption is actually satisfied is untestable due to the possibility of unobserved confounders. We believe we have considered most (if not all) known fundamental aspects of galaxy formation and evolution to build the causal model (Fig. \ref{fig:causal_model_galaxy_formation_and_evolution}), and it is difficult to think of a physical process or variable (besides halo mass) that could causally affect both environment and SFR. Consequently, we are fairly satisfied that there is no unobserved confounding.

Positivity states that there must be a non-zero probability of receiving any treatment. In the context of this study, galaxies of all halo masses (or stellar masses, due to their correlation) must have some probability of occupying different environments to reliably estimate the causal effect of environment at any particular density. This is reasonably true according to the halo/stellar mass–environment distribution in \ref{fig:galaxy_properties_distributions}, which shows a relatively uniform halo/stellar mass coverage at different environmental densities. Though, in the lowest density environments, there is a lack of the least and the most massive haloes. To alleviate this positivity violation, we have defined the treatment grid between the 1st and 99th percentiles of the environmental density distribution, so environments at both extremes are not considered.

Consistency states that the observed outcome must equal the potential outcome under treatment. In other words, the treatment must be well-defined. The treatment in our case is the environment, which has no universal definition, and this opens up the possibility of violating the assumption. We employ the 10th nearest neighbour density, and as long as the proxy consistently measures the environmental density in varied environments and the effect of environment is similar at a particular density, we satisfy the consistency assumption.

Lastly, the no interference assumption states that the potential outcome of a unit must only depend on its treatment. This assumption is irrelevant in our case because galaxies in close proximity are roughly in the same environment, so they are subject to the same treatment. We are concerned about neighbourhood-level no interference \citep{vanderweele_2008}, which means that a galaxy's SFR must only depend on its environment and not the neighbouring environment. It is hard to imagine a physical mechanism that would result in the above being untrue, so we are reasonably confident that the no interference assumption holds. In summary, we believe the consistency and no interference assumptions hold. In the following section, we quantitatively verify exchangeability (assuming no unobserved confounders) and positivity.

\subsection{Diagnostic tests}
The critical component of our causal analysis is the weights, which we utilise to capture and adjust biases via the IPW method to estimate MSMs. Their validity directly translates to unbiased causal effects, which means they can provide clues on the satisfaction of the causal assumptions. As a result, we perform diagnostic tests on them to check exchangeability and positivity. We highlight that we conduct the tests on the weights estimated from the non-bootstrapped analysis.

The goal of IPW is to create a pseudo-population in which the treatment is independent, and this produces exchangeability as the treatment groups are comparable in terms of their covariate distributions when the treatment does not depend on anything. We assess the covariate balance using the correlation-based method of \cite{zhu_2015} to determine exchangeability. The basic premise is to determine the correlation between the confounders and treatment in the pseudo-population, and if it is minimal, then the treatment is independent, there is no confounding, and exchangeability is achieved. The exact procedure is as follows:

\begin{enumerate}
    \item Sample data with replacement from the original dataset according to the weights $w_{i}$.
    \item Compute the correlation coefficient $\rho_{m}$ between confounder $X_{m}$ and treatment $T$ in the weighted sample.
    \item Repeat the above steps $N$ times and calculate the average correlation coefficient $\bar{\rho}_{m}$.
    \item Finally, average the absolute values of all the average correlation coefficients to compute the average absolute correlation coefficient $AACC$.
\end{enumerate}

\begin{figure}
    \centering
    \includegraphics[width=0.4\textwidth]{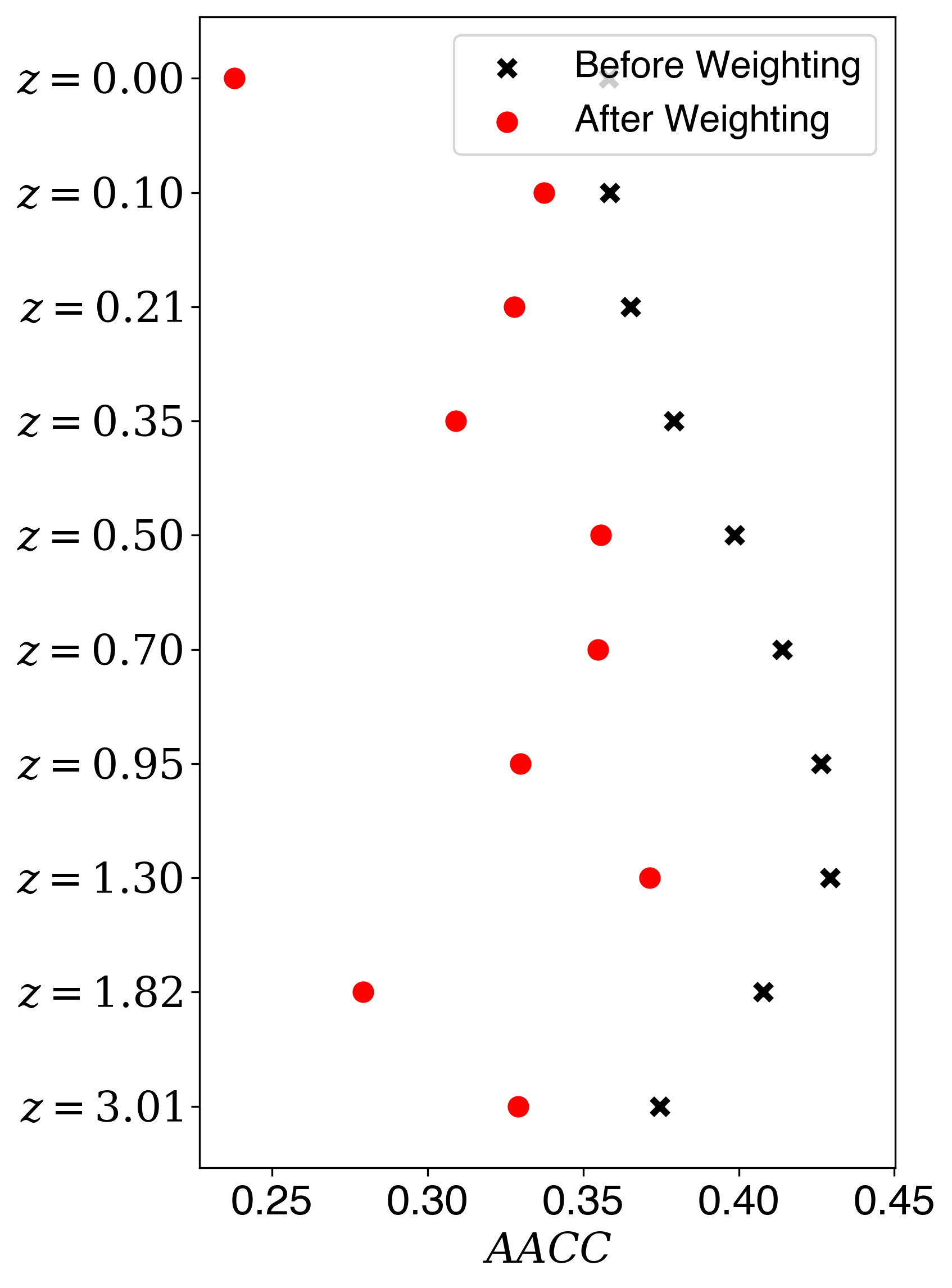}
    \caption{Average absolute correlation coefficients ($AACC$) at different redshifts in the original population before weighting and in the pseudo-population after weighting.}
    \label{fig:covariate_balance}
\end{figure}

\begin{figure}
    \centering
    \includegraphics[width=\textwidth]{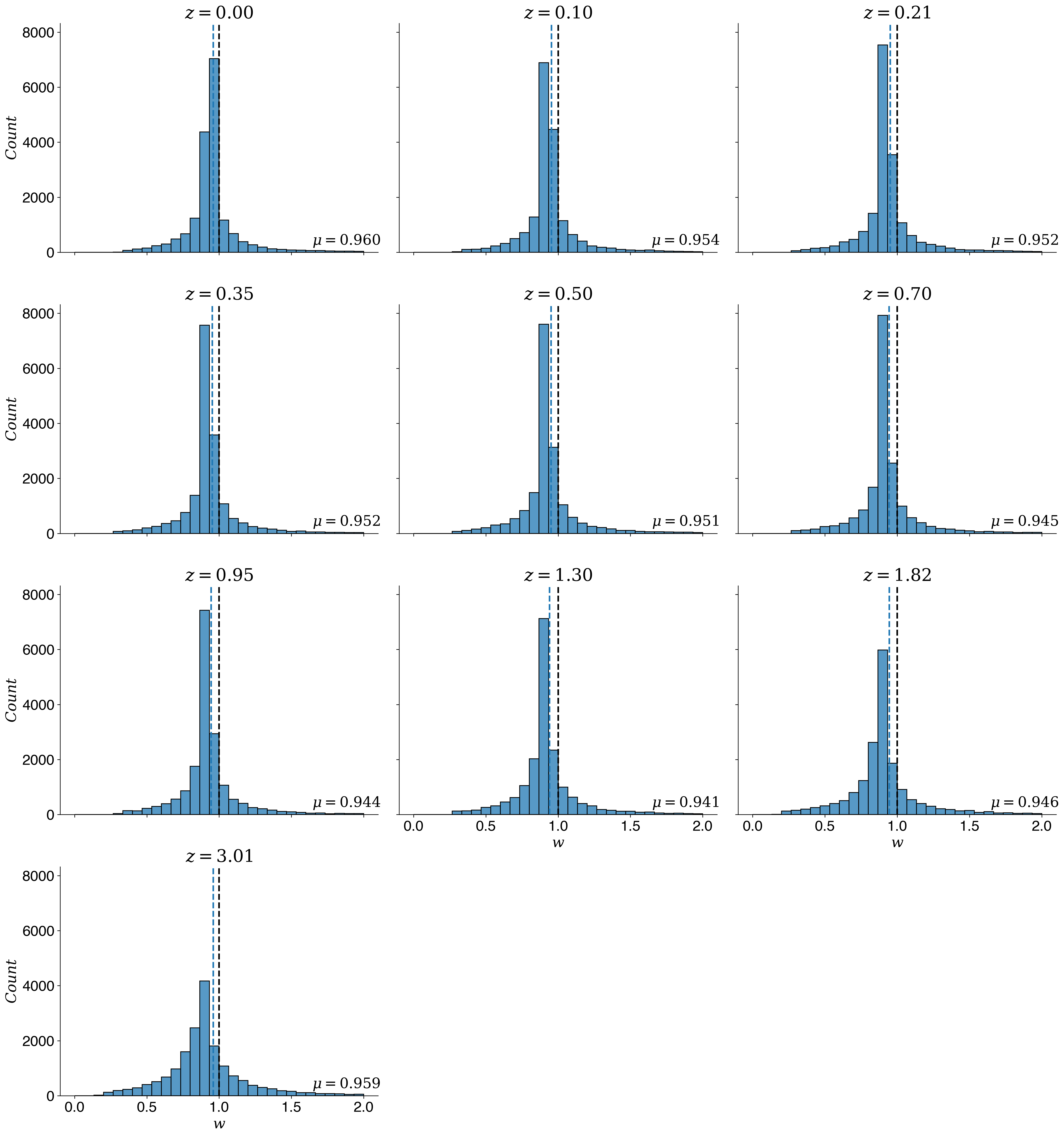}
    \caption{Weight distributions at different redshifts, assuming the causal model (Fig. \ref{fig:causal_model}a). The black and blue dashed lines indicate the reference mean ($1$) and the mean of the distribution (stated), respectively.}
    \label{fig:weight_distributions_causal_model}
\end{figure}

We assess the covariate balance at each time point. Accordingly, the data is sampled per the time-point weights, not the final product weights. For consistency, the correlation coefficients are computed with the previous environments $\bar{E}_{k-1}$ and halo masses $\bar{H}_{k-1}$ even though the confounders of the current environment $E_{k}$ and star-formation rate $SFR_{k}$ may only be the prior environment $E_{k-1}$ and halo mass $H_{k-1}$. Specifically, Kendall's tau coefficient \citep{kendall_1938} is estimated because the halo mass and environment distributions are not normal, and the relationship between them is non-linear (as can be observed in \ref{fig:galaxy_properties_distributions}). In total, $1000$ bootstrap samples are generated to calculate the average correlation coefficients.

Fig. \ref{fig:covariate_balance} shows the $AACC$ at the different redshifts in the original population before weighting and in the pseudo-population after weighting. \cite{zhu_2015} claim that there is minimal confounding when $AACC<0.1$, medium confounding when $0.1<AACC<0.3$, and large confounding when $AACC>0.55$. However, these limits are based on heuristics, and there is no theoretical $AACC$ value for exchangeability. In this case, the relative change in the $AACC$ is more important than the absolute value. As observed, there is a clear decrease in the $AACC$ post weighting across all the redshifts, which indicates that IPW has reduced confounding and improved exchangeability. 

The mean of the stabilised weights is expected to be one because the size of the pseudo-population equals that of the original population \citep{hernan_2006}. Crucially, significant deviations indicate misspecification of the weighting model, violation of positivity, or both \citep{cole_2008}. However, as is the case with the $AACC$, there is no reference value. Fig. \ref{fig:weight_distributions_causal_model} shows the weight distributions at the different redshifts. The means are close to one, so our causal model seems to be valid, and positivity is not violated. Based on the reasoning and diagnostics, the causal assumptions seem to have been met or at least not grossly violated (although it cannot be definitively proven). Thus, the results can be considered valid.

\renewcommand{\thefigure}{E\arabic{figure}}% Figure counter representation
\renewcommand{\theHfigure}{E\arabic{figure}}% Hyperref figure hyperlink hook

\section{Additional results}
\label{sec:additional_results}

\begin{figure}
    \centering
    \includegraphics[width=\textwidth]{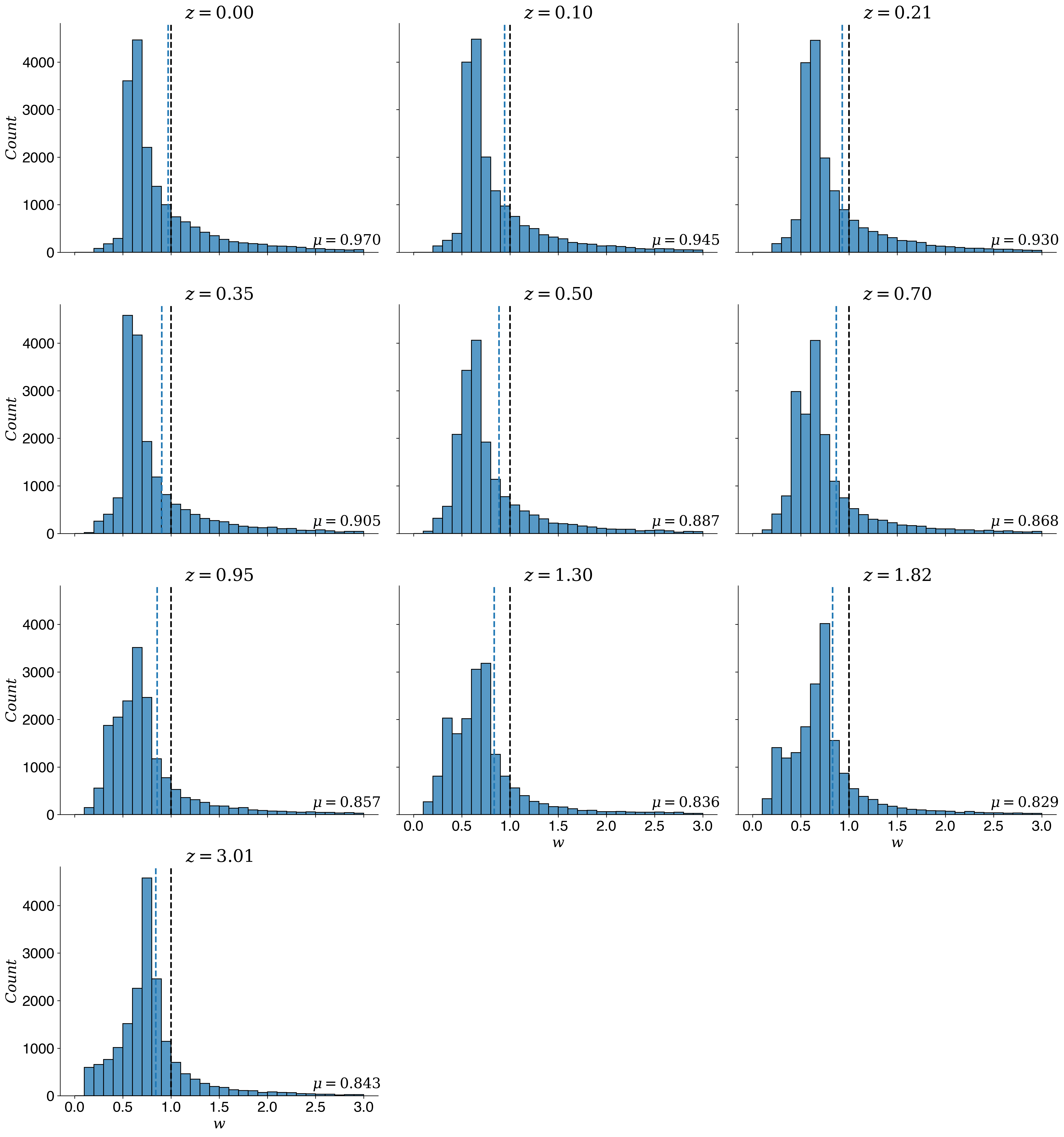}
    \caption{Weight distributions at different redshifts, assuming the traditional model (Fig. \ref{fig:causal_model}c). The black and blue dashed lines indicate the reference mean ($1$) and the mean of the distribution (stated), respectively.}
    \label{fig:weights_distributions_traditional_model}
\end{figure}

\begin{figure}
    \centering
    \includegraphics[width=\textwidth]{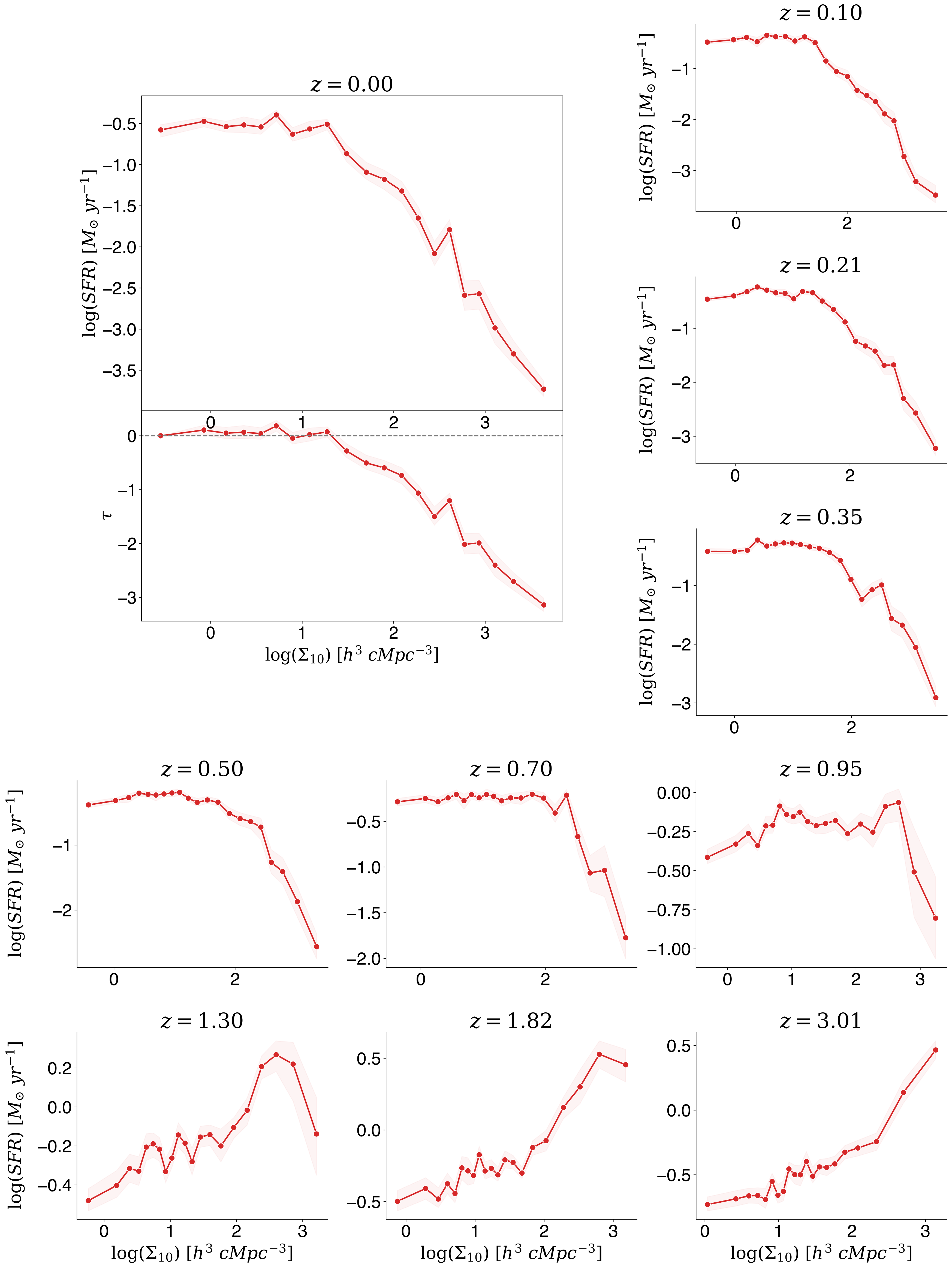}
    \caption{Causal dose-response curves (CDRCs) of the causal effects of environment on the star-formation rate (i.e., causal SFR–density relations) at $z=0$ and at different redshifts going back to $z \sim 3$, assuming the traditional model (Fig. \ref{fig:causal_model}c). Specifically, they represent the average SFR of galaxies if they inhabited the specific density environment (10th nearest neighbour density). The bottom panel of $z=0$ shows the average causal effects $\tau$ of different density environments (comparing to the lowest-density environment). The shaded regions represent the $68 \%$ confidence interval, estimated with bootstrapping.}
    \label{fig:cdrc_current_traditional_model}
\vspace{-64.19313pt}
\end{figure}




\end{appendices}

\end{document}